\documentclass{pasa}
\usepackage{graphicx}
\usepackage{multirow}
\graphicspath{{images/}}
\usepackage{aas_macros}
\usepackage{hyperref} 
\usepackage{amsmath}
\hypersetup{colorlinks,citecolor=blue,linkcolor=blue,urlcolor=blue}
\bibpunct{(}{)}{;}{a}{}{,}
\usepackage{lscape}
\usepackage{xcolor}

\newcommand{\gtsim}{\raisebox{-.5ex}{$\;\stackrel{>}{\sim}\;$}}

\newcommand{\Halpha}{\ifmmode {\rm H}\alpha \else H$\alpha$\fi}
\newcommand{\Hbeta}{\ifmmode {\rm H}\beta \else H$\beta$\fi}
\newcommand{\Hgamma}{\ifmmode {\rm H}\gamma \else H$\gamma$\fi}
\newcommand{\Hdelta}{\ifmmode {\rm H}\delta \else H$\delta$\fi}
\newcommand{\Lya}{\ifmmode {\rm Ly}\alpha \else Ly$\alpha$\fi}
\newcommand{\Lyb}{\ifmmode {\rm Ly}\beta \else Ly$\beta$\fi}
\newcommand{\HeI}{\ifmmode {\rm He}\,\textsc{i}\,\lambda5876 \else 
                  He\,\textsc{i}\,$\lambda5876$\fi}
\newcommand{\HeII}{\ifmmode {\rm He}\,\textsc{ii}\,\lambda4686 \else 
                   He\,\textsc{ii}\,$\lambda4686$\fi}

\newcommand{\nev}{[Ne\,\textsc{v}]\,}

\newcommand{\ciii}{\ifmmode {\rm C}\,\textsc{iii}] \else C\,\textsc{iii}]\fi}
\newcommand{\civ}{\ifmmode {\rm C}\,\textsc{iv} \else C\,\textsc{iv}\fi}

\newcommand{\Athena}{\textsl{Athena}}
\newcommand{\OST}{\textsl{OST}}
\newcommand{\OSS}{\textsl{OSS}}
\newcommand{\eddrat}{\ifmmode{\lambda_{\text{Edd}}} \else  $\lambda_{\text{Edd}}$ \fi}
\newcommand{\mbh}{\ifmmode{\text{M}_{\text{BH}}} \else  M$_{\text{BH}}$ \fi}
\newcommand{\mail}[1]{\href{mailto:#1}{\texttt{#1}}}
\title{The role of SPICA-like missions and the Origins Space Telescope in the quest for heavily obscured AGN and synergies with Athena}
  
\author[Barchiesi et al.]{L. Barchiesi\thanks{\mail{luigi.barchiesi2@unibo.it}}$\,\,\,^{1,2}$, F. Pozzi$^{1,2}$, C. Vignali$^{1,2}$, F. J. Carrera$^{3}$, F. Vito$^{4}$, F. Calura$^2$, L. Bisigello$^2$, G. Lanzuisi$^{1,2}$,  C. Gruppioni$^2$, E. Lusso$^{5,6}$, I. Delvecchio$^{7,8}$, M. Negrello$^{9}$, A. Cooray$^{10}$, A. Feltre$^{2,11}$, J. A. Fern\'andez-Ontiveros$^{12}$, S. Gallerani$^{4}$, H. Kaneda$^{13}$, S. Oyabu$^{14}$,  M. Pereira-Santaella$^{15}$, E. Piconcelli$^{16}$, C. Ricci$^{17,18,19}$, G. Rodighiero$^{20}$, L. Spinoglio$^{12}$, F. Tombesi$^{21,22,23}$ 
\vspace{2mm}
\affil{\textit{(Affiliations can be found after the references)}}
}%

\jid{PASA}
\doi{10.1017/pas.\the\year.xxx}
\jyear{\the\year}

\begin{document}

\begin{frontmatter}
\maketitle
\begin{abstract}
In the BH-galaxy co-evolution framework, most of the star-formation (SF) and the black hole (BH) accretion is expected to take place in highly obscured conditions. The large amount of gas and dust absorbs most of the UV-to-soft-X radiation and re-emits it at longer wavelengths, mostly in the IR. Thus, obscured Active Galactic Nuclei (AGN) are very difficult to identify in optical or X-ray bands, but shine bright in the IR. Moreover, X-ray background (XRB) synthesis models predict that a large fraction of the yet-unresolved XRB is due to the most obscured (Compton thick, CT: N$_{\text{H}}\ge 10^{24} \,$cm$^{-2}$) of these AGN. In this work, we investigate the synergies between  putative IR missions (using \textit{SPICA}, proposed for ESA/M5 but withdrawn in October 2020, and Origins Space Telescope, \OST, as `templates') and the X-ray mission \textsl{Athena}, which should fly in early 2030s, in detecting and characterizing AGN, with a particular focus on the most obscured ones. Using an XRB synthesis model, we estimated the number of AGN and the number of those which will be detected in the X-rays by \textsl{Athena}. For each AGN we associated an optical-to-FIR spectral energy distribution (SED) from observed AGN with both X-ray data and SED decomposition, and used these SEDs to check if the AGN will be detected by \textit{SPICA}-like or \OST{} at IR wavelengths. We expect that, with the deepest \textsl{Athena} and \textit{SPICA}-like (or \OST)  surveys, we will be able to photometrically detect in the IR more than $90\,\%$ of all the AGN (down to L$_{2-10\text{keV}} \sim 10^{42}\,$erg/s and up to $z \sim 10$) predicted by XRB synthesis modeling, and we will detect at least half of them in the X-rays. The spectroscopic capabilities of the \OST{} can provide $\approx 51\,000$ and $\approx 3400$ AGN spectra with R= 300 at $25-588\,\mu$m in the wide and deep surveys, respectively, the last one up to $z\approx 4$. \textsl{Athena} will be extremely powerful in detecting and discerning moderate- and high-luminosity AGN, allowing us to properly select AGN even when the mid-IR torus emission is ``hidden'' by the host-galaxy contribution. We will constrain the intrinsic luminosity and the amount of obscuration for $\sim 20\,\%$ of all the AGN (and $\sim 50\,\%$ of those with L$_{2-10\text{keV}} > 3.2 \times 10^{43}\,$erg/s) using the X-ray spectra provided by \textsl{Athena WFI}. We find that the most obscured and elusive CT-AGN will be exquisitely sampled by \textit{SPICA}-like mission or \OST{} and that \textsl{Athena} will allow a fine characterization of the most-luminous ones. This will provide a significant step forward in the process of placing stronger constraints on the yet-unresolved XRB and investigating the BH accretion rate evolution up to very high redshift ($z \ge 4$).
\end{abstract}

\end{frontmatter}

\section*{Preface}
 
The articles of this special issue focus on some of the major scientific questions that a future IR observatory will be able to address. We adopt the {\it SPace Infrared telescope for Cosmology and Astrophysics (SPICA)} design as a baseline to demonstrate how to achieve the major scientific goals in the fields of galaxy evolution, Galactic star formation and protoplanetary disks formation and evolution. The studies developed for the \textit{SPICA} mission serve as a reference for future work in the field, even though the mission proposal has been cancelled by ESA from its M5 competition. \par
The mission concept of {\it SPICA} employs a 2.5\,m telescope, actively cooled to below $\sim$ 8\,K, and a suite of mid- to far-IR spectrometers and photometric cameras, equipped with state of the art detectors \citep{roelfsema18}. In particular the {\it SPICA} Far-infrared Instrument (SAFARI) is a grating spectrograph with low (R\,$\sim200-300$) and medium (R\,$\sim3000-11\,000$) resolution observing modes instantaneously covering the $35-210\,\mu$m wavelength range. The {\it SPICA} Mid-infrared Instrument (SMI) has three operating modes: a large field of view ($10' \times 12'$) low-resolution $17-36\,\mu$m imaging spectroscopic (R\,$\sim50-120$) mode  and photometric camera at $34\, \mu$m (SMI-LR), a medium resolution (R\,$\sim1300-2300$) grating spectrometer covering wavelengths of $18-36\, \mu$m (SMI-MR) and a high-resolution echelle module (R\,$\sim29\,000$) for the $10-18\, \mu$m domain (SMI-HR). Finally, B-BOP, a large field of view ($2'.6 \times 2'.6$), three channel, ($70\, \mu$m, $200\, \mu$m and $350\,\mu$m) polarimetric camera complements the science payload.\par

\section{Introduction}
One of the main open issue in astrophysics is the role of Supermassive Black Holes (SMBH) in shaping the galaxies and the influence that the star formation (SF) has on the properties of the Active Galactic Nuclei (AGN). The scaling relations between the mass of the SMBH and the stellar mass (or luminosity) of the spheroid in which it is hosted \citep{Kormendy95} or the spheroid velocity dispersion
\citep{Ferrarese00, Gebhardt00} suggest a tight link between star-formation activity in the spheroidal components of galaxies and SMBH growth. The BH-galaxy co-evolution paradigm states that an intense phase of SF is triggered by a wet merger, at least for the most luminous and massive systems \citep{silk98, dimatteo05, lamastra13}. A fraction of the gas reservoir of the galaxy is funneled towards the SMBH and turns on the AGN activity. Hydrodynamical simulations (e.g. \citealt{dimatteo05, hopkins06, Barai18}), semianalytic models (e.g. \citealt{menci08}) and observations (e.g. \citealt{tombesi15, feruglio15, smith19}) suggest a scenario where AGN provide feedback that regulates the SF in the host galaxies, thus originating the observed correlations. Such link between SMBH growth and galactic star formation history is supported by the striking similarity of the evolution of the SMBH accretion rate (BHAR) and that of the star formation rate (SFR) densities, especially at high redshifts where the SF mostly occurs in the spheroidal components \citep{Shankar09, Fiore17, gruppioni20}. The evolutions of the SFR and the BHAR follow a parallel evolution, showing both a peak of activity around $z \approx 1-3$, known as the ``cosmic noon'' \citep{madau14, heckman14, vito18}. However, the mechanisms responsible for this similarity are not well understood and the co-evolution between galaxy and SMBH is still a matter of debate. Investigating the BH-galaxy co-evolution requires sizable samples of galaxies and AGN over a broad range of redshift and luminosities, to characterize the accretion and SF histories and disentangle the two contributions within the same source as accurately as possible.\par
To study the link between star formation and black hole accretion we need, in particular, to investigate the epoch in which the bulk of these growths seems to take place (i.e. the cosmic noon). However, the high quantity of gas and dust that fuels both processes absorbs most of the energy emitted by stars and accreting SMBHs and re-emits it at longer wavelengths, mostly in the IR. The search for and the characterization of these objects in the optical and UV is challenging due to the heavy reddening that affects them \citep{rowan-robinson97, hughes98}. A solution to this problem is to study indirectly the primary emission, by measuring the dust-reprocessed radiation in the IR. A study at such wavelengths allows us to observe the physical processes at work in the obscured regions and, therefore, to derive the SFR density (SFRD) and the BHAR density (BHAD) if we are capable of properly separating them. A direct measure of the dust-obscured SF activity is possible with space-based IR observatories, which do not need to be corrected by dust attenuation.\par
 The mid- to far-IR photometric observations in deep fields \citep{delvecchio14, schreiber15} have provided extremely reliable determination of the SFRD and the BHAD up to redshift $\sim 3-4$ \citep{gruppioni13, magnelli13, delvecchio14} using \textsl{Herschel PACS} data (observing at 100 and $160\,\mu$m; \citealt{poglitsch10}) and have detected high redshift galaxies (up to $z > 6$; e.g., \citealt{riechers13, rowan-robinson16}) using \textsl{SPIRE}, observing at 250, 350 and $500\,\mu$m; \citealt{griffin10}). However, the deepest cosmological surveys performed by Herschel at high-z have detected only the most luminous galaxies (L$_{\rm IR} >10^{12}\,L_{\odot}$ at $z\ge 3$; \citealt{gruppioni13}) and suffer from the uncertainties of not being able to accurately disentangle the AGN from the SF emissions, as both contributes to the mid- and far-IR emission. Even in the case when Spectral Energy Distribution (SED) fitting allows us to separate the two components, the results usually depend on the adopted technique and modeling. Therefore, the initial phase of the BH-galaxy co-evolution paradigm has been, so far, the most elusive and difficult to track. 
According to the co-evolution scenario, this heavily obscured phase should be characterized by the presence of a strong starburst and an obscured AGN, both giving a significant contribution to the mid- and far-IR emission (e.g. \citealt{page04, vignali09, Lapi18}). 
\par
Compton-thick (CT) AGN, i.e. highly-obscured AGN with column density $\log{(\text{N}_{\text{H}}/\text{cm}^{-2})} \ge 24$, are also required to explain the spectra of the extragalactic X-ray background (XRB): their flat spectrum is needed to reproduce the hump observed at $20-30\,$keV in the XRB (e.g. \citealt{comastri95, gilli07, ballantyne11}). However, the fraction of CT-AGN derived from XRB synthesis models ranges from $10\,\%$ to $30-50\,\%$ \citep{treister09, gilli07, ueda14, ananna19} and suffers from large uncertainties due to degeneracies between several model parameters. Although the expected number of CT-AGN is large, their detection and identification beyond the local universe are challenging and, even in the local universe, their observed fraction ($\sim 5-10\,\%$, e.g. \citealt{burlon11, ricci15}) is significantly lower than models predictions, but X-rays detection bias may be responsible of such low numbers \citep{burlon11}. \par 
To select unobscured or mildly-obscured AGN, one of the most effective tools is the observation in the X-ray band, thanks to the radiation coming directly from the AGN innermost regions. However, this selection loses effectiveness with the increasing column density (at N$_\text{H} \ge 10^{23}\,\text{cm}^{-2}$ the soft continuum is severely depressed) and, even in deep fields, only a limited fraction of the most obscured AGN has been revealed in the X-ray (e.g. \citealt{tozzi06, lanzuisi13, marchesi16, delmoro17}). Moreover, the BHAD estimation from the X-ray luminosity is strongly dependent on the bolometric correction ($k_{bol}$), that allows to estimate the ``total'' bolometric power ($\text{L}_{bol}$) of the AGN from the AGN emission in a specific band, but suffers from significant uncertainties in the $k_{bol}-\text{L}_{bol}$ relation (e.g. \citealt{hopkins07, lusso10, lusso12, duras20}). Likewise for obscured SF galaxies, the IR is an effective band in detecting these obscured AGN, thanks to the energy absorbed by the obscuring material and re-emitted at sub-mm wavelengths (e.g. \citealt{pozzi07}). The effectiveness of identifying highly obscured AGN by selecting bright mid-IR sources with faint optical or near-IR emission  has been shown in the past by \textit{Spitzer} (e.g. \citealt{houck05, weedman06, polletta08}) and \textit{WISE} (e.g. \citealt{mateos12,assef13}).\par

To be able to photometrically identify highly-obscured and CT-AGN, both mid-IR and far-IR instruments are necessary, as we need to distinguish between the emission from hot dust (T>100 K), mostly due to reprocessed AGN emission from the dusty torus and galactic dust, and the emission from warm/cold dust (T<100 K) dominated by the reprocessed SF emission.

A future IR cryogenic cooled space telescope may be able to deliver all of this. We investigated two different IR mission, using as `templates' the proposed, now withdrawn, ESA \textit{SPICA} mission and the NASA concept study \textsl{Origins Space Telescope} (\OST). A \textit{SPICA}-like mission (described extensively in section~\ref{sec:spica_ins}) with a mid-IR spectrophotometer  will be able to sample the band where the peak of the CT-AGN emission is up to $z \approx 6$ and provide also low-resolution spectra, while far-IR photometric cameras will cover the wavelength range where the SF is the main contributor, thus helping in separating the two emission mechanisms. Detected sources may be followed-up with SPICA SAFARI-like instruments to provide low resolution spectra in the mid- and far-IR wavelength range up to $z \approx 3.5$. In these spectra high ionization lines, such as \nev at $14.32\,\mu$m and $24.32\,\mu$m, $[\text{Ne}\,\textsc{vi}]$ at $7.7\,\mu m$ and $[\text{O}\,\textsc{iv}]$ at $25.89\,\mu m$, would be signatures of AGN presence and their fluxes could be used as proxy of the AGN bolometric luminosities (e.g. \citealt{spinoglio92, garcia16, fernandez16}). Alternatively an highly capable grating mid- and far-IR spectrometer, as the \OSS{} in the \OST{} (see section~\ref{sec:oss_ins}), will be able to cover instantaneously both the hot and warm/cold dust wavelength regime, thus enabling us to distinguish between the torus and the host-galaxy emission and to estimate, in a reliable way, the AGN bolometric power, the host SF and the BH accretion rate, without the need of additional constraints from other optical-IR instruments or surveys.

To confirm the CT nature of sources, X-ray spectroscopy has been the most reliable method so far. Unfortunately, with the current deepest X-ray surveys, we have only scratched the surface of the obscured AGN population at $z \sim 1-4$. Moreover, the identification of an AGN as CT depends strongly on the adopted analysis techniques and modeling, with different techniques giving sometimes different classifications of the same sources (e.g. \citealt{castellomor13}). The best results have come from X-ray deep and ultradeep surveys (e.g. COSMOS, \citealt{elvis09, civano16}, CDF-S, \citealt{giacconi02, luo08, xue11, xue17}) that allowed to detect CT-AGN even at high redshift, but with the disadvantage of having small fields, thus low source statistics. \textit{XMM-Newton} and \textit{Chandra} surveys unveiled only up to few tens of candidates of CT-AGN (e.g. \citealt{comastri11, brightman12, georgantopoulos13, lanzuisi17}), while an hard X-ray mission like \textit{NuSTAR} has difficulties in detecting objects beyond $z\sim 1$, due to its sensitivity (\citealt{lansbury17b,lansbury17a},  the notable exceptions of CT-AGN at $z\sim 2$ detected by \citealt{delmoro14, zappacosta18}).\par
A breakthrough in selection and identification of CT-AGN is expected with the new ESA X-ray observatory \textsl{Athena}, due for launch in early 2030s. \textsl{Athena}, with  its  combination  of angular resolution, field of view and  collecting area  at  $\sim 1\,$keV, is the ideal instrument to perform X-ray surveys and will be more than two orders of magnitude faster than \textit{Chandra} or \textit{XMM-Newton} \citep{nandra13}.  The excellent spectral capabilities of \textit{Athena WFI} for surveys will yield samples of the most heavily obscured AGN up to redshifts $z=4$ and up to 100 times larger than is currently possible. Since $\text{L}_{\text{X-ray}} \gtrsim 10^{42}\,$erg$\,$s$^{-1}$ are commonly associated with AGN activity, \textsl{Athena} will also be invaluable in quickly identifying moderate- and high-luminous AGN, especially in cases where we can not distinguish between the torus mid-IR emission and the host-galaxy contribution from SED fitting analysis.\par
The planned launch date for \textsl{Athena} is $\gtrsim 2030$. Having, in the same years, a new IR mission dedicated (also) to survey high-$z$ galaxy and AGN would be an extraordinary opportunity to study obscured accretion and to characterize AGN up to high redshift. A possible example of such mission would have been \textit{SPICA}, but also the NASA concept \OST{} for an $4.5\,$K cooled IR space observatory, which has a proposed launch date of $\sim 2035$ and similar survey strategy and wavelength coverage as \textit{SPICA}. Having data at X-ray and IR wavelengths will be the key to accurately reconstruct the accretion luminosity across cosmic time up to very high redshift ($z \approx 6$). The search for the elusive CT-AGN may decisively benefit from the synergies between \textsl{Athena} and future IR mission: while a \textit{SPICA}-like telescope and the \OST{} should be able to effectively detect the majority of them (even those too obscured to be detected in X-rays), the X-ray spectra provided by \textsl{Athena} will be invaluable to confirm their CT-nature, even in paucity of photons, for a large fraction of them.\par
The goal of this paper is to study the synergies between \textsl{Athena} and future IR surveys in searching and identifying AGN, with a particular focus on the most obscured ones. We will show how the search for these elusive CT-AGN, very difficult to identify and characterize with the current instruments, can benefit from a combined use of IR and X-ray data. In Section~\ref{sec:instruments} we give a brief description of the \textit{SPICA} and \OST{} instruments and planned surveys and a comparison with the \textsl{Athena} surveys. The simulation of the intrinsic AGN prediction obtained from the XRB synthesis models are presented in Section~\ref{sec:xrb}, while the number of those which will be detected by \textsl{Athena} are reported in Section~\ref{sec:athena}. In Section~\ref{sec:sed} we present the optical-to-far-IR SED used to predict the AGN emission and their detectability with \OST{} and \textit{SPICA}-like mission. The prediction of the number of AGN that will be detected by the IR surveys, as well as the synergies with \textsl{Athena} that we can exploit to study them, are reported in Section~\ref{sec:spica}. In Section~\ref{sec:conclusion}, we present our conclusions. \par 
Throughout this paper, we adopt the following cosmological parameters: H$_0 = 70\, \text{km}\, \text{s}^{-1}\, \text{Mpc}^{-1}$, $\Omega_{\text{M}} = 0.3$ and $\Omega_{\Lambda} = 0.7$ \citep{spergel03}.

\section{Instruments and Surveys}\label{sec:instruments}
\subsection{SPICA}\label{sec:spica_ins}
\textit{SPICA} was a joint European-Japanese project to develop a new generation cryogenic infrared space telescope, with the early design dating back to more than two decades ago (e.g. \citealt{nakagawa98,nakagawa14,swinyard09}). Here we discuss the \textit{SMI} and \textit{B-BOP} instruments as it were designed during the ESA M5 selection process and a \textit{SPICA}-like survey strategy. \par

\subsubsection{SMI}
\textit{SMI} is composed of the four channels, SMI-CAM, LR, MR and HR. In this paper, we consider \textit{SMI-CAM}  and \textit{SMI-LR}, that work simultaneously. \textit{SMI-CAM} is a mid-IR slit viewer photometric camera that covers the $30-37\,\mu$m range and provides $34\,\mu$m broad-band images with a field of view of $10'\times 12'$ (with the exclusion of the four slits used by the spectrometer). \textit{SMI-LR} is a wide field-of-view multi-slit prism spectrometer, composed of four long slits of $10'$ in length and $3.7''$ in width. It has a resolution of R = $50-120$ and allows spectroscopic surveys in the $17-36\,\mu$m wavelength range. The \textit{SMI-CAM} provides photometric scientific data, and is invaluable in accurately determining the position of the slits on the sky for pointing reconstruction in creating spectral maps. We refer to \citet{kaneda17} for further instrument specifications and survey strategies. 

\subsubsection{B-BOP}
The B-BOP polarimetric instrument allows simultaneous imaging observations in three bands, centered at $70\,\mu $m, 200$\,\mu $m  and 350$\,\mu $m, over an instantaneous field of view from $ 1.8'\times 1.8'$ to $ 2.7'\times2.7'$ at FWHM resolutions of $6''$, $17''$ and $30''$, respectively.  B-BOP is two to three orders of magnitude more sensitive than current or planned far-IR/submillimeter polarimeters, e.g. \textit{SHARP} \citep{li08}, \textit{HAWC+ (SOFIA)} \citep{dowell98}, \textit{BLAST-TNG} \citep{galitzki14}. It provides wide-field $70-350\,\mu$m polarimetric images in Stokes Q and U of comparable quality (in terms of resolution, signal-to-noise ratio, and both intensity and spatial dynamic ranges) to \textsl{Herschel} images in Stokes I.
More details about the B-BOP instrument are in \citet{rodriguez18} and \citet{andre19}.

\subsubsection{Surveys with SMI}\label{sec:survey_smi}
Hereafter, we consider two reference blind spectroscopic surveys:
\begin{itemize}
\item a \textbf{1deg$^2$ Ultradeep survey} with a total observational time of $600\,$hr, an SMI-CAM $5\,\sigma$ sensitivity of $\approx 3\,\mu$Jy and an SMI-LR $5\,\sigma$ sensitivities (high-background case) of $\approx 50\,\mu$Jy at $20\,\mu$m, $\approx 110\,\mu$Jy at $30\,\mu$m and a line sensitivity of $\approx 9.2\times10^{-20}\,$W$/$m$^2$ (this survey configuration corresponds to the \citealt{kaneda17} deep survey);
\item a \textbf{15$\,$deg$^2$ Deep survey} with a total observational time of $600\,$hr, an SMI-CAM $5\,\sigma$ sensitivity of $\approx 13\,\mu$Jy and an $5\,\sigma$ SMI-LR sensitivities (high-background case) of $\approx 160\,\mu$Jy at $20\,\mu$m, $\approx 380\,\mu$Jy at $30\,\mu$m and a line sensitivity of $\approx 33\times10^{-20}\,$W$/$m$^2$ (this survey has the same depth of the \citealt{kaneda17} shallow survey, although over a larger area).
\end{itemize}

\subsection{The \textsl{Origins Space Telescope}}\label{sec:oss_ins}
The \OST{} \citep{battersby18, meixner19} is a concept study for a Far-Infrared Surveyor mission, the subject of one of four science and technology definition studies supported by NASA for the 2020 Astronomy and Astrophysics Decadal Survey. The \OST{} is being designed with the aim of covering a large area of the sky, thus allowing to search for rare objects at low and high redshifts.\par
\OST{} is composed of a $5.9\,$m-diameter telescope with a \textsl{Spitzer}-like architecture, actively cooled down to $4.5\,$K. It is designed with three onboard instruments: the \textsl{Origins Survey Spectrometer} (\OSS) to cover the $25-588\,\mu$m wavelength range instantaneously at R$=300$, \textsl{MISC-T} (\textit{Mid-Infrared Spectrometer Camera Transit}) a $2.8-20\,\mu$m transit spectrometer and the \textsl{Far-infrared Imager and Polarimeter (FIP)}, that delivers imaging or polarimetry at $50\,\mu$m and $250\,\mu$m. In this paper, we take into consideration primarily the \OSS{} instrument, as it is perfectly suited for detecting and characterizing AGN in blind surveys.
\subsubsection{\textsl{OSS}}
The \textsl{Origins Survey Spectrometer} is a R=300 spectrometer covering the full $25-588\,\mu$m wavelength range instantaneously, using six logarithmically-spaced grating modules. True 3D spectral mapping is performed thanks to the 30 to 100 spatial beams that each module couples. Higher spectral resolution can be achieved by inserting into the light-path two mirrors that divert the light into interferometer optics allowing a $R=43\,000$ resolution, keeping the full spectral range but working only on single pointing (i.e. losing the survey capabilities). Even higher $R=300\,000$ spectral resolution can be achieved using an insertable etalon, further restricting the FoV and reducing the spectral range to $100-200\,\mu$m. As in this paper we focus on detecting AGN in surveys, we take into consideration only the base R=300 spectroscopy mode.\par
The six \OSS{} bands cover the $25-44$, $42-74$, $71-124$, $119-208$, $200-350$, $336-589\,\mu$m wavelength ranges, with beam sizes of 1.41, 2.38, 4.0, 6.78, 11.3, 19.0 arcsec and  $1\,$hr $5\,\sigma$ sensitivities (R=300) of 22, 28, 40, 104, 104, 338$\,\mu$Jy. With the current \OSS{} specs, an improvement of more than a factor $1\,000$ in sensitivity is expected with respect to already flown far-IR observatories (\textsl{SOFIA}, \textsl{Herscel-SPIRE}, and \textsl{Herschel-PACS}). We refer to \citet{meixner19} for further instrument specifications.

\subsubsection{Survey with \textsl{OSS}}
For \OST{} we consider two blind blank-field spectroscopic surveys as described in \citet{meixner19}:
\begin{itemize}
\item a \textbf{0.5$\,$deg$^2$ deep survey} with an observational time of $\approx 1000\,$hr, a R=4, $5\,\sigma$ sensitivity of 4.5, 5.8, 8.4, 21.8, 21.8, 70.9$\,\mu$Jy (for channel 1 to 6) and a R=300 $5\,\sigma$ sensitivity of 39, 50, 72, 189, 189, 614$\,\mu$Jy;
\item a \textbf{20$\,$deg$^2$ wide survey} with an observational time of $\approx 1000\,$hr, a R=4, $5\,\sigma$ sensitivity of 30, 39, 56, 145, 145, 473$\,\mu$Jy (for channel 1 to 6) and a R=300 $5\,\sigma$ sensitivity of 262, 336, 483, 1260, 1260, 4094$\,\mu$Jy.
\end{itemize}
\subsection{\Athena{}}

\Athena{} (Advanced Telescope for High ENergy Astrophysics \footnote{\url{https://www.the-athena-x-ray-observatory.eu/}} \citealt{barcons17}) is the X-ray observatory mission selected by ESA, within its Cosmic Vision programme, to address the \textit{Hot and Energetic Universe} scientific theme. It is the second L(large)-class mission within that programme and is due for adoption in late 2022 and launch in early 2030s.

\Athena{} will consist of a single large-aperture grazing-incidence X-ray telescope, utilizing a novel technology (High-performance silicon pore optics) developed in Europe, with 12~m focal length and 5 arcsec Half Energy Width (HEW) on-axis angular resolution, degrading gradually to less than 10~arcsec at 30~arcmin off-axis \citep{bavdaz18}. There will be two instruments in the focal plane. One is the Wide Field Imager (WFI)\footnote{\url{http://www.mpe.mpg.de/ATHENA-WFI/}} \citep{meidinger18} providing simultaneous sensitive wide-field imaging and spectroscopy (FWHM$\leq 170$~eV at 7~keV) and high count-rate capability ($>$90\% throughput and $<$1\% pile-up for 1~Crab) over the 0.2 to 15~keV energy range. This is achieved through two sets of Silicon-based detectors using DEPFET Active Pixel Sensor technology: the Large Detector Array is a mosaic of 2$\times$2 arrays spanning a $\sim 40 \times 40\,\text{arcmin}^2$ Field of View (FoV) oversampling the PSF by more than a factor of two and the Fast Detector is a single array optimised for high count rate applications.

The other instrument is the X-ray Integral Field Unit (X-IFU)\footnote{\url{http://x-ifu.irap.omp.eu/}} \citep{barret18} delivering simultaneous spatially resolved (5~arcsec pixels) high-resolution X-ray spectroscopy (FWHM$<$2.5~eV below 7~keV) over a limited field of view ($\sim$5~arcmin equivalent diameter) over the 0.2 to 12~keV energy range, with high count-rate capability (10~eV spectral resolution at 1~Crab intensities with low pile-up and $>$50\% throughput). This performance is based on a large format array of superconducting molybdenum-gold Transition Edge Sensors coupled to absorbers made of gold and bismuth, cooled at $\sim$90~mK inside a nested set of cryostats. 

\subsubsection{Surveys with \textsl{Athena}}

Some of the core science objectives of \Athena{}\footnote{The full list of science requirements can be found under \url{https://www.cosmos.esa.int/web/athena/study-documents}} (e.g. finding distant evolved groups of galaxies at $z>2$, complete the census for high-redshift and heavily obscured AGN) require performing surveys of the X-ray sky with the WFI. These surveys are expected to take a significant part of the observation time during the four years of the nominal mission lifetime. At the time of writing, the nominal number of pointings and exposure times (the so-called ``Tier 2 post-CORE strategy'') includes an ``ultradeep'' layer (4$\times$1400~ks), a ``deep'' layer (3$\times$980~ks$+$7$\times$840~ks) and a ``shallow'' layer (106$\times$84~ks) (see Table \ref{tab:survey}). Each pointing covers $\sim 0.4$~deg$^2$, so the total area of each layer is 1.6, 4 and 42.5~deg$^2$, respectively. The exact places and geometries (single pointings vs. mosaics) of those layers will of course be determined at a later stage by the international astronomical community through the guaranteed time and open time, but they are expected to include regions of the sky with substantial multi-wavelength coverage. An initial estimate is encapsulated in the \Athena{} Mock Observing Plan (MOP).\par

\subsection{Surveys comparison}
\textsl{Athena} WFI, \OST{} \OSS{}, and \textit{SPICA SMI} have planned survey with different layers of depth and area. In this work we compared the different layers, matching them to have similar areas as much as possible. As we compared the expected number of sources per deg$^2$ for all the surveys and investigated the fraction of sources that may be detected, the exact area coverage is not critical. We investigate the capabilities of the observatories considering a \textbf{DEEP} survey (comparing the \textit{SPICA} ultradeep survey with the \textsl{Athena} ultradeep survey and the \OST{} deep survey), and a \textbf{WIDE} survey (composed of the \textit{SPICA} deep survey, the \textsl{Athena} shallow survey and the \OST{} wide survey). Table~\ref{tab:survey} summarizes the main parameters of all the \textit{SPICA}, \OSS{} and \textsl{Athena} surveys, while Table~\ref{tab:survey_name} reports the naming scheme used in this work.

\begin{table*}
\caption{Parameters of the \textit{SPICA}, \textsl{Athena} and \OST{} surveys. The time per field $t_{\text{field}}$ are pure integration times, without over-heads. The sensitivities are computed at $5 \sigma$. The \textsl{SPICA} R=150 sensitivities are reported at $20\,\mu$m. The B-BOP sensitivities refer only to the $70\,\mu$m channel, while those for the \OST{} refer to channel 1 and channel 6. In this table, the \textsl{Athena} survey strategy is simplified, as the deep survey encompasses also the ultradeep pointings and the shallow survey will comprehend also the deep pointings. The \textsl{Athena} sensitivities refer to the $2-10\,$keV band. The surveys investigated and the naming scheme used in this work are summarized in Table~\ref{tab:survey_name}.}
\centering
\begin{tabular}{@{}ccccccc@{}}
\hline\hline
 \multicolumn{2}{c}{Instrument} & Survey & $t_{\text{field}}$ (ks) & Sensitivity ($\mu$Jy) & $t_{\text{tot}}$ (hr) & Area (deg$^2$) \\
\hline%
\multirow{4}*{\textit{SPICA}}& SMI & \multirow{2}*{ultradeep} & 67.86 & 3 (R=4), 50 (R=150) & 605 & \multirow{2}*{1} \\
 &   B-BOP & & - & 60 & 70 &  \\
 &  SMI & \multirow{2}*{deep}  & 5.22 & 13 (R=4), 160 (R=150) & 482 & \multirow{2}*{15} \\
 &   B-BOP & & - & 100 & 350 &  \vspace{0.2cm}\\
\multirow{2}*{\textsl{OST}} & \multirow{2}*{\textsl{OSS}} & deep & - & $4.5-79$ (R=4), $39.4-614$ (R=300) & 1000 & 0.5  \\
 & & wide & - & $30-470$ (R=4), $260-4090$ (R=300) & 1000 & 20  \vspace{0.2cm}\\ 
 \multirow{3}*{\textsl{Athena}} & \multirow{3}*{\textsl{WFI}} & ultradeep & 4$\times$1400 & $\sim 3\times 10^{-17}\,$erg/s/cm$^2$ & $\sim 1600$ & 1.6  \\
 & & deep & 3$\times$980$+$7$\times$840 & $\sim 4\times 10^{-17}\,$erg/s/cm$^2$  & $\sim 2500$ & 4\\
 & & shallow & 106$\times$84  & $\sim 1.2\times 10^{-16}\,$erg/s/cm$^2$ & $\sim 2500$ & 42.5\\
\hline\hline
\end{tabular}
\label{tab:survey}
\end{table*}

\begin{table}
\caption{Reference surveys used in this work. The surveys composing the reference surveys were matched in similar area coverage.}
\begin{tabular}{cccc}
\hline\hline
Instrument & Survey & Area (deg$^2$) & ref. survey \\
\hline
SPICA & ultradeep & 1.0 & \multirow{3}*{DEEP 1$\,$deg$^2$} \\
OST & deep & 0.5 & \\
Athena & ultradeep & 1.6 & \\
\hline
SPICA & deep & 15 & \multirow{3}*{WIDE 15$\,$deg$^2$} \\
OST & wide & 20 & \\
Athena & shallow & 42.5 & \\
\hline\hline
\end{tabular}
\label{tab:survey_name}
\end{table}

\section{Simulation of the intrinsic AGN counts from XRB synthesis model}\label{sec:xrb}

We have estimated the `intrinsic' AGN number as a function of the redshift $z$, 2-10~keV intrinsic X-ray luminosities $\text{L}_{\text{x}}$ and intrinsic column density $\text{N}_{\text{H}}$ based on the \citet{gilli07} X-ray background synthesis model. We have started by dividing the $(z,\text{L}_{\text{x}},\text{N}_{\text{H}})$ space in bins, with 21 $z$ bins $\in [0,10]$, 18 $\text{L}_{\text{x}}$ bins with $\text{L}_{\text{x}}\in [10^{42},1.5\times10^{48}]$~erg/s and 6 $\text{N}_{\text{H}}$ bins with $\log(\text{N}_{\text{H}}/\text{cm}^{-2})\in [20,26]$. We, then, make use of the software POMPA\footnote{\url{http://www.bo.astro.it/~gilli/counts.html}} to get the $\log{N(>S)}$ source density in each of those bins, with $S$ being the 2-10~keV source flux.\par
From the $N(>S)$ we also obtained the total intensity of each bin (energy received per unit area, unit time and unit sky area). Instead of the composite power law plus reflection spectral model in \citet{gilli07}, we have adopted here a torus model from \citet{brightman11} with photon index $\Gamma=1.9$,  aperture angle 30~deg and inclination 80~deg. We have also included an additional scattering component modelled with a power-law with the same photon index and normalization equal to one percent of that of the primary emission.

In addition, we also included ``normal'' galaxies using the galaxy $\log N- \log S$ curve of \citet{lehmer12} and a power-law spectral shape with $\Gamma=1$ (the median value for the 332 sources with spectral slope not based on upper limits in \citealt{lehmer12}) and $S\in[5\times 10^{-20}, 5\times 10^{-16}]$~erg~cm$^{-2}$~s$^{-1}$.\par
Physical fluxes are converted to counts using the latest (at the time of writing) set of FoV-averaged without filter WFI matrices\footnote{\url{http://www.mpe.mpg.de/ATHENA-WFI/response\_matrices.html}} using \texttt{xspec}\footnote{\url{https://heasarc.gsfc.nasa.gov/xanadu/xspec/}} \citep{xspec}. 

\section{Predictions of \textsl{Athena} source counts}\label{sec:athena} 

When observed by the WFI on board \textsl{Athena}, all the sources above are seen against a ``background'' which includes several components\footnote{\url{https://www.mpe.mpg.de/ATHENA-WFI/public/resources/background/WFI-MPE-ANA-0010_i7.1_Preparation-of-Background-Files.pdf}}{\color{blue}$^,$}\footnote{Exact matrix used: \url{athena_wfi_rib2.3_B4C_20190122_wo_filter_FovAvg.rsp}}:
\begin{itemize}
\item{Particle background:} due to cosmic rays impacting the detector and the surrounding structure. It is assumed to contribute $6\times 10^{-4}$~cts/keV/s/arcmin$^2$ with a flat power-law spectrum characterized by a spectral index of $\Gamma=0$.

\item{Diffuse Galactic X-ray background:} it is assumed to be uniform and to have the spectral shape and intensity from \citet{mccammon02}, which includes two thermal components, one unabsorbed representing emission from the Local Hot Bubble and one with foreground absorption by our own Galaxy representing emission from the hot Galactic halo, in \texttt{xspec} parlance: \texttt{apec$_1$ + phabs (apec$_2$)} with $kT_1$=0.099~keV and $kT_2$=0.225~keV and respective \texttt{apec} normalizations of 1.7$\times 10^{-6}$ and 7.3$\times 10^{-7}$, corresponding to 1~arcmin$^2$; the column density for the Galactic absorption is assumed to be the same as for extragalactic sources (see below).

\item{XRB:} due to the integrated emission of the above AGN and galaxy populations. It is assumed to have intensity and power-law spectral shape from \citet{mccammon02} but with $\Gamma=1.45$ and normalisation $10^{-6}$~photons~s$^{-1}$cm$^{-2}$keV$^{-1}$, again normalized to 1~arcmin$^2$.

\item{Stray light:} X-ray photons from the two previous components, coming from outside the FoV but impacting the WFI after a single reflection on the mirror. We have modelled it by averaging the contribution of the two previous components over the full WFI FoV and fitting it with two thermal components and a power-law, with the same temperatures and photon index but with free relative normalizations.
  
\end{itemize}

All extragalactic components are assumed to undergo absorption by gas in our own Galaxy with hydrogen column density of $\text{N}_{\text{H}}=1.8\times 10^{20}\,\text{cm}^{-2}$, which is the average Galactic column density of the estimated survey pointings in the MOP and coincides with the foreground column density in \citet{mccammon02}.

Depending on the $(z,\text{L}_{\text{x}},\text{N}_{\text{H}})$ and the exposure time, a fraction $f_{det}$ of the sources in each of the bins, described above, will be detected, contributing to resolve a fraction of the full intensity of the extragalactic XRB, thus lowering the remaining (unresolved) background. For each of the exposure time values of the survey layers, we have determined the resolved fraction iteratively starting from an initial fiducial value of 80\% and $f_{det}$ is also determined for each bin in each iteration. To decide whether a source is detected above the background we have used Cash statistic $C$ \citep{cash79}: calling $T$ the total source+background counts, $B$ the background counts and $t$ the exposure time, the mode (most probable value) of the source counts is $\hat s=\max(0,T-B)$. For a 5$\sigma$ detection for a single free parameter ($s$) we required that the Cash statistic improves by $\Delta C=25$ between $s=0$ (i.e. no source) and $s=\hat s$. For each bin, 100 sources are simulated at the bin centre and $f_{det}$ is the fraction of those sources that are detected.  It turns out that the resolved fraction is around 80\% and varies a few percent for exposure times above $\sim$10~ks. In order to estimate the source+background and the background counts we have assumed a circular extraction area of radius 5.7~arcsec, which is the WFI FoV-weighted average of the \Athena{} mirror HEW.

\subsection{\textsl{Athena} source characterization}
By their own nature, X-ray detectors work in photon-counting mode. One of the consequences of this is that they all are "Integral Field Units", providing spectra for all detected sources. For \Athena{}, this ranges from the superb $\sim$2.5~eV resolution of X-IFU to the modest (but good for a Si-based detector) $170\,$eV resolution at $7\,$keV of WFI. We have simulated 100 spectra at the "centres" of each of the $1\leq z \leq 4$, $10^{44}\leq \text{L}_{\text{x}} (\text{erg/s}) \leq 5\times 10^{45}$, $24\leq \log(\text{N}_{\text{H}}/\text{cm}^{-2})\leq 26$ bins and fitted them. If both the fitted luminosity and the column density, obtained from the spectral fitting, are within 30\% of the input values, we say that we have ``characterized'' the source. Finally, by dividing the number of characterized sources by 100 (the number of spectra we simulated) we have estimated the fraction of characterized sources in each bin.

\section{AGN and host galaxy SED}\label{sec:sed}
To compute the AGN and host galaxy expected fluxes at different redshifts in the \textit{SPICA} and \OST{} bands, we need a proper set of SEDs to model the AGN as a function of different parameters: the intrinsic rest-frame $2-10$ keV luminosity $\text{L}_{\text{x}}$, the equivalent hydrogen column density $\text{N}_{\text{H}}$ and the redshift $z$.\par
We started with the AGN and host galaxy SED compilation from \citet{lanzuisi17}. They used a sample of $2\,333$ X-ray selected (either \textit{XMM-Newton} or \textit{Chandra}) AGN in the COSMOS field \citep{scoville07} with at least 30 X-ray counts from \citet{lanzuisi13, lanzuisi15}, \citet{marchesi16} and then selected only those with a $>3\sigma$ IR detection in one of the \textit{Herschel} bands. The X-ray spectral analysis provided us with column density $\text{N}_{\text{H}}$ and absorption-corrected $2-10\,$keV luminosity for each source (see \citealt{lanzuisi17} and references therein for more details). The final sample was composed of 692 sources X-ray and FIR detected, all with both X-ray spectral properties and SED decomposition. The SEDs were obtained by \citet{delvecchio14,delvecchio15}, following the recipes described in \citet{berta13}, using the \texttt{sed3fit} code (with a ``smooth'' torus model), which allowed us to disentangle the AGN and host-galaxy contribution using three components -- stellar emission, AGN torus emission and SF-heated dust emission -- with photometric points from the UV to sub-mm. We excluded from our sample the sources with 2-10 keV rest-frame luminosities $\log{(\text{L}_{\text{x}}/\text{erg/s})}< 42\,$, as SF galaxies without AGN should not be able to exceed this threshold. Moreover, to overcome the degeneracy between the AGN and the SF contribution during the SED fitting, we selected only the SEDs with AGN significance $S_{AGN}\ge99\%$, evaluated through the F-test between the best-fit reduced $\chi^2$ with and without the AGN component (see \citealp{delvecchio14}). Thus, the final sample is composed of 422 AGN in the redshift range $0 < z \leq 4$, 2-10 keV rest-frame luminosities in the $42 \leq \log{(\text{L}_{\text{x}}/\text{erg/s})} \leq 45.5$ range and host-galaxy masses $7.9 \le \log{\text{M}/M_{\odot}} \le 12.2 $. 
\par
With the aim of obtaining template SEDs representative of AGN with different intrinsic rest-frame $2-10$ keV luminosity and obscuration, we created four $\text{L}_{\text{x}}$ bins and four $\text{N}_{\text{H}}$ bins. The binning was chosen to maximize the number of sources in each bin, whilst maintaining a good sampling of the  $\text{L}_{\text{x}}$ and $\text{N}_{\text{H}}$ parameter space. The binning used and the number of SEDs in each bin are shown in Table~\ref{tab:binning}. Two of the most extreme bins had no sources in them; for these bins we chose to use SEDs randomly extracted from the two nearest bins (one with the same $\text{L}_{\text{x}}$ and one with the same $\text{N}_{\text{H}}$). The binning choice was also driven by the binning we adopted to obtain the number of expected AGN (see Sec~\ref{sec:xrb}) and we used the same binning edges, although we merged some of the less populated bins.\par
Throughout this paper, we assumed a bolometric correction of $k_{bol}=L_{bol}/L_{\text{x}}$, eq~\ref{eq:kbol}, from \citet{lusso12}.
\begin{equation}
\label{eq:kbol}
\begin{split}
\log{(\text{L}_{\text{x}}/L_{\odot})}=0.230 x+0.050 x^2+0.001 x^3 +1.256 \\
\text{with}\,\, x=\log{(L_{bol}/L_{\odot})}-12
\end{split}
\end{equation}

In this work, we will refer to the predicted AGN in the simulated redshift range $0<z<10$ with $20 \le \log{(\text{N}_{\text{H}}/\text{cm}^{-2})}<26$ and $42 \le \log{(\text{L}_{\text{x}}/\text{erg/s}}) <48.2$ ($43 \le \log{(L_{bol}/\text{erg/s}}) <49.7$), thus covering the whole simulated parameter space, as all the AGN. The naming scheme used for referring to AGN with different properties (namely, N$_{\text{H}}$ and luminosity) is reported in Table~\ref{tab:namingscheme}.

\begin{table}
\caption{Naming scheme used in this work for referring to AGN with different amounts of obscuration and X-ray luminosity. As reported in section~\ref{sec:sed}, we used the \citet{lusso12} bolometric correction (eq~\ref{eq:kbol}) to compute $\text{L}_{\text{bol}}$.}
\label{tab:namingscheme}
\centering
\begin{tabular}{cccccc}
\hline
\multicolumn{2}{c}{$\log{(\text{N}_{\text{H}}/\text{cm}^{-2})}$} &  \\
\multicolumn{2}{c}{$20-22$} & unobscured \\
\multicolumn{2}{c}{$22-24$} & obscured \\
\multicolumn{2}{c}{$24-26$} & CT-AGN \vspace{0.2cm}\\
\hline
$\log{(\text{L}_{\text{bol}}/\text{erg s}^{-1})} $ & $\log{(\text{L}_{\text{x}}/\text{erg s}^{-1})} $ &  \\
$43.0-44.6$ & $42.0-43.5$ & low-luminosity \\
$44.6-49.7$ & $43.5-48.2$ & high-luminosity \\
\hline
\end{tabular}
\end{table}

\begin{table}
\caption{Number of binned sources, using the compilation of \citet{lanzuisi17} as explained in section~\ref{sec:sed}, in each 2-10 keV rest-frame luminosity $\text{L}_{\text{x}}$ and amount of obscuration $\text{N}_{\text{H}}$ bin. The used $\text{N}_{\text{H}}$ bins are: $20 \le \log\, (\text{N}_{\text{H}}/\text{cm}^{-2}) \le 22$; $22 < \log\, (\text{N}_{\text{H}}/\text{cm}^{-2}) \le 23$; $23 < \log\, (\text{N}_{\text{H}}/\text{cm}^{-2}) \le 24.18 $; $24.18 < \log\, (\text{N}_{\text{H}}/\text{cm}^{-2}) \le 26$. We refer to the sources in the first two $\text{L}_{\text{x}}$ bins as low-luminosity ones and to those in the other two as high-luminosity ones. We consider the sources in the first $\text{N}_{\text{H}}$ bin as unobscured AGN, those in the second and third bins as obscured and those in the last $\text{N}_{\text{H}}$ bin as CT-AGN. For the two low-luminosity CT bins we had no SED available and we chose to use the SED of the nearest bins.}
\label{tab:binning}
\centering
\begin{tabular}{rcccc}
\hline
$\log{(\text{N}_{\text{H}}/\text{cm}^{-2})}$ & 21.5 & 22.5 & 23.5 & 24.5 \\	 
$42.0\le \log{(\text{L}_{\text{x}}/\text{erg/s})}  <42.9 $ & 21 & 14 &  5 & - \\
$42.9\le \log{(\text{L}_{\text{x}}/\text{erg/s})}  <43.5 $ & 43 & 54 & 10 & - \\
$43.5\le \log{(\text{L}_{\text{x}}/\text{erg/s})}  <44.2 $ & 48 & 76 & 53 & 3 \\
$44.2\le \log{(\text{L}_{\text{x}}/\text{erg/s})}  <48.2 $ & 23 & 25 & 38 & 6 \\
\hline
\end{tabular}
\end{table}

\section{Predictions of IR source counts}\label{sec:spica} 
In Sec~\ref{sec:xrb}, we predicted the number of expected AGN (per deg$^2$) as function of $z$, $\text{L}_{\text{x}}$ and $\text{N}_{\text{H}}$. 
For each $z$, $\text{L}_{\text{x}}$ and $\text{N}_{\text{H}}$ bin we randomly extracted 20 template SEDs, from the corresponding bins in Section~\ref{sec:sed}), and assigned them a weight equal to one-twentieth of the total number of predicted AGN. We measured their total (AGN+host) flux densities in the SMI $34\mu$m and B-BOP $70\mu$m, $200\mu$m $350\mu$m bands (for \textit{SPICA}) and in the six \OSS{} bands (for \OST). We obtained the number sources detectable at $5\,\sigma$ by summing the weights of the sources with flux larger than the survey sensitivity.\par
In case of detection, we also differentiated if we were primarily detecting either the AGN or the host galaxy, on the basis of which of the two components had the highest flux in the band we were considering. In the case of a \textit{SPICA} SMI $34\mu$m detection, we considered whether we could also detect the source with the SMI-LR mode (as the SMI-CAM photometric channel operates at the same time with the SMI-LR spectroscopic channel). We used, as SMI-LR sensitivity, a mean value between the sensitivities at the boundaries of the SMI-LR wavelength range. For \OST{} \OSS{} we always considered both the detection in photometric mode (R=4) and in spectroscopic mode (R=300).\par
We iterated this whole process forty times (different SEDs were extracted at each iteration), each time obtaining an estimate for the number of detected sources, AGN/torus and spectroscopic detection. We computed the median of these numbers and used the 16$^{\text{th}}$-84$^{\text{th}}$ percentiles as uncertainties. Extracting 20 SEDs for each bin (and not one SED for each of the expected detection in that bin) allowed us to save computational time while maintaining a result resolution of $5\%$. We chose to iterate the process forty times to have good statistical significance of the median and of the 16$^{\text{th}}$-84$^{\text{th}}$ percentiles.\par
In conclusion, from the IR simulations, we derive, for each bin, the percentage of sources photometrically detected, spectroscopically detected, and for which it is the AGN emission to be primarily detected. From the X-ray simulations, we obtained the total number of AGN in each bin, as well as the fraction of sources photometrically detected and spectroscopically characterized using \textsl{Athena WFI}. In each bin, we computed the fraction of sources detected both in the IR and in the X-rays, as the minimum between the detection fraction in the two wavelength ranges. Although, in some bins, there may be sources detected in the IR and not in the X-rays, alongside those detected in the X-rays and not in the IR, usually the detection fraction in one of the wavelength ranges is much higher than the other, and we can safely consider the minimum as a reliably estimate of the fraction of sources detectable with both instruments. \par
\begin{figure*}
  \centering
  \resizebox{0.99\hsize}{!}{\includegraphics{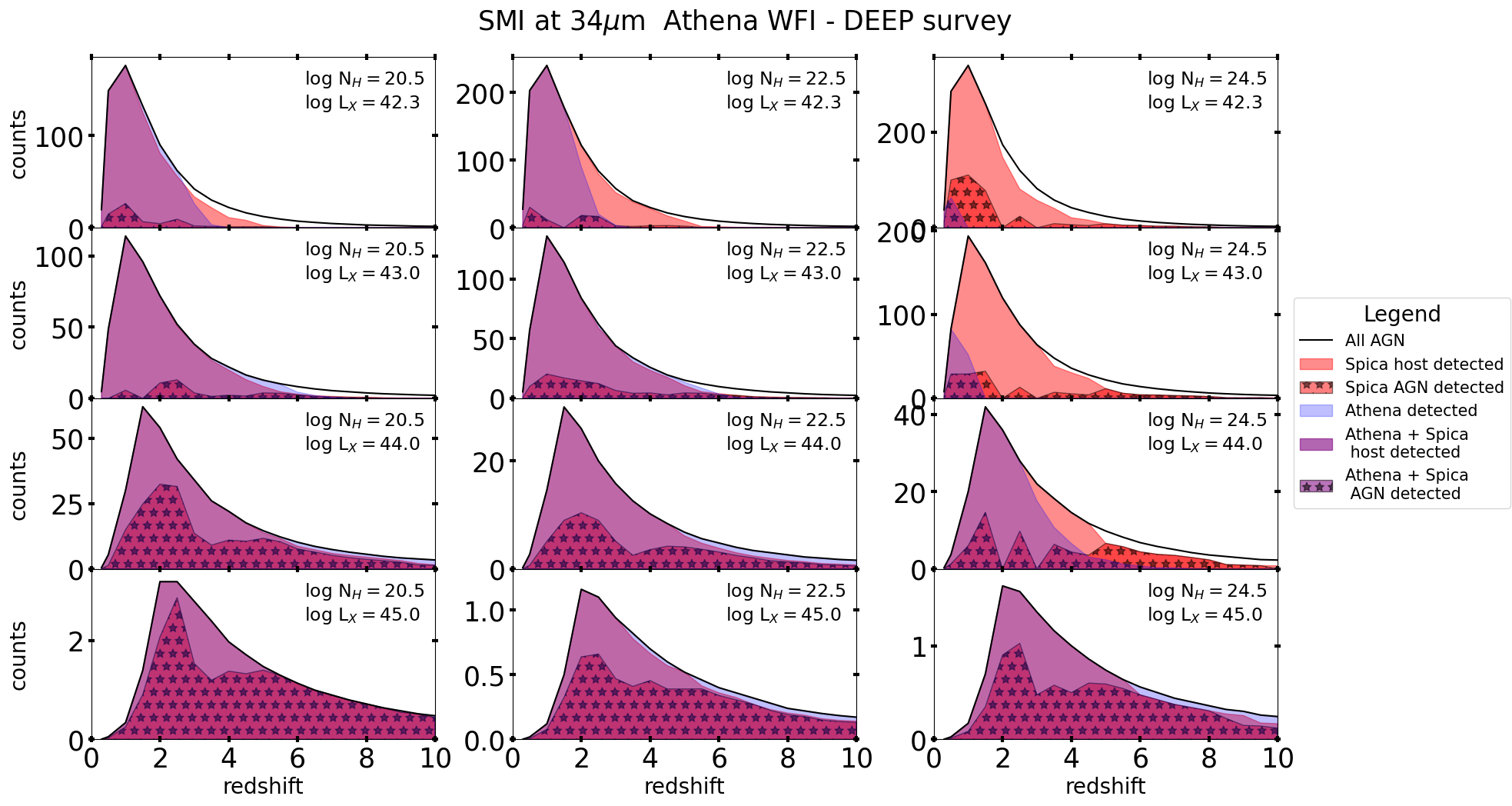}}
  \caption{Number of AGN expected per deg$^2$ and per $\Delta z =1$ for a \textit{SPICA}-like DEEP survey. The black lines are the total number of expected AGN, the red areas represent those which can be detected with the \textit{SPICA} SMI-CAM at 34$\mu$m, the blue areas those which can be detected in the X-rays with the \textsl{Athena WFI}, the purple areas are the AGN that will be detected both by \textit{SPICA} SMI-CAM and \textsl{Athena}. The red uniform area represents the sources detected by \textit{SPICA} in which the main component is due to the host-galaxy emission, while for those represented with the starry red area the AGN is the main contributor to the detected emission. The columns refer to AGN with different amount of obscuration ($20 \le \log\, (\text{N}_{\text{H}}/\text{cm}^{-2}) \le 21$; $22 < \log\, (\text{N}_{\text{H}}/\text{cm}^{-2}) \le 23$; $24.18 < \log\, (\text{N}_{\text{H}}/\text{cm}^{-2}) \le 25$, from left to right), the rows to different AGN luminosity ($42.0\le \log{(\text{L}_{\text{x}}/\text{erg/s})}  <42.3 $, $42.9\le \log{(\text{L}_{\text{x}}/\text{erg/s})}  <43.2 $, $43.9\le \log{(\text{L}_{\text{x}}/\text{erg/s})}  <44.2 $, $44.9\le \log{(\text{L}_{\text{x}}/\text{erg/s})}  <45.2 $, from top to bottom). For forty times we extracted 20 SED for each bin, measured the flux in each \textit{SPICA} band and compared them with the $5\,\sigma$ sensitivities to compute the number of detectable sources. The median of these forty values are the numbers used in this figure, while the $84^{\textit{th}}-16^{\textit{th}}$ percentiles are used as uncertainties (not reported in the figure).}
  \label{fig:smi34UD} 
\end{figure*}

\begin{figure*}
  \centering
  \resizebox{0.95\hsize}{!}{\includegraphics{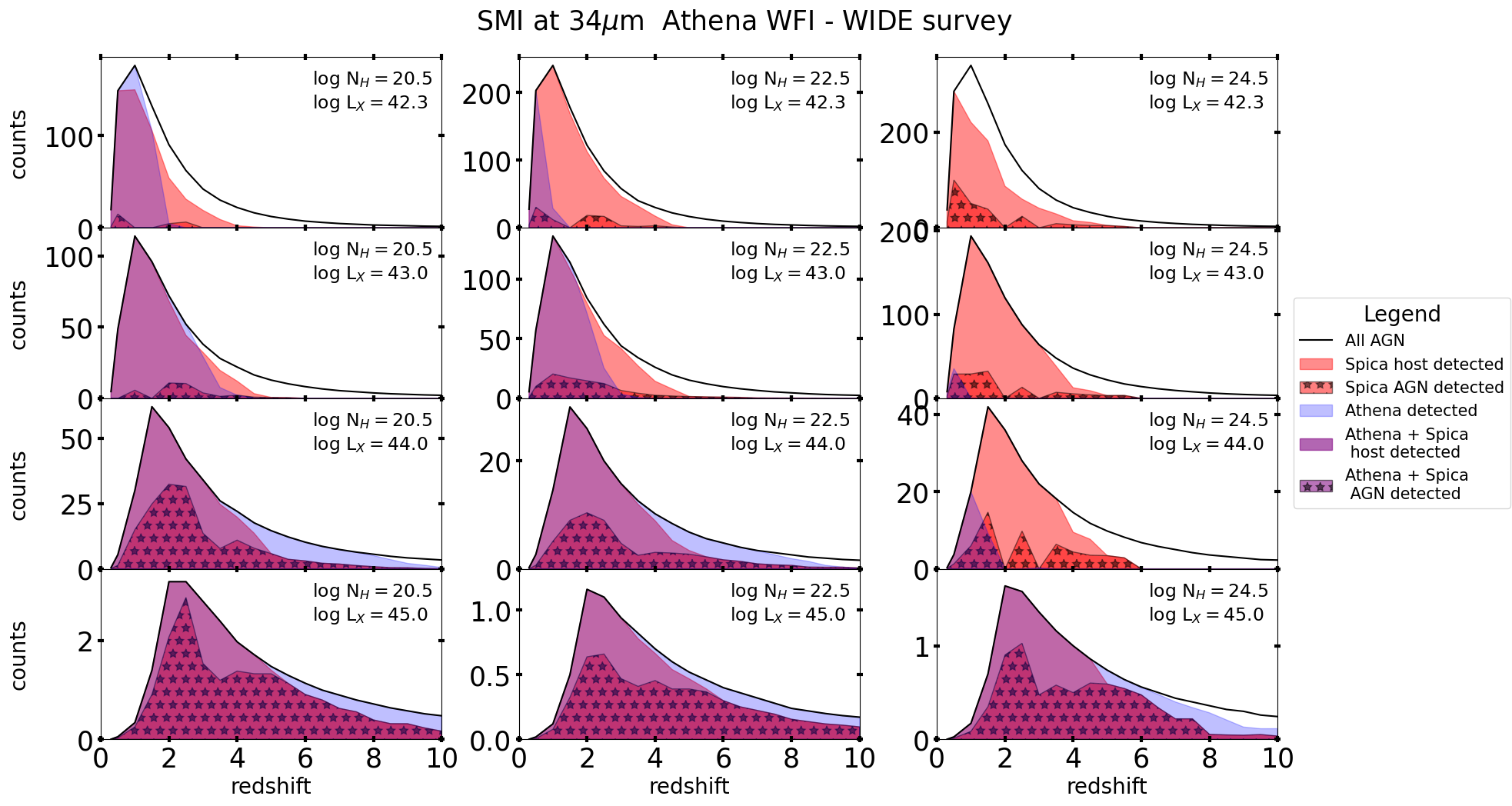}}
  \caption{Number of AGN expected per deg$^2$ and per $\Delta z =1$ for a \textit{SPICA}-like DEEP survey. The lines and areas are coded as in Figure~\ref{fig:smi34UD}.}
  \label{fig:smi34D_84} 
\end{figure*}

\begin{figure*}
  \centering
  \resizebox{0.99\hsize}{!}{\includegraphics{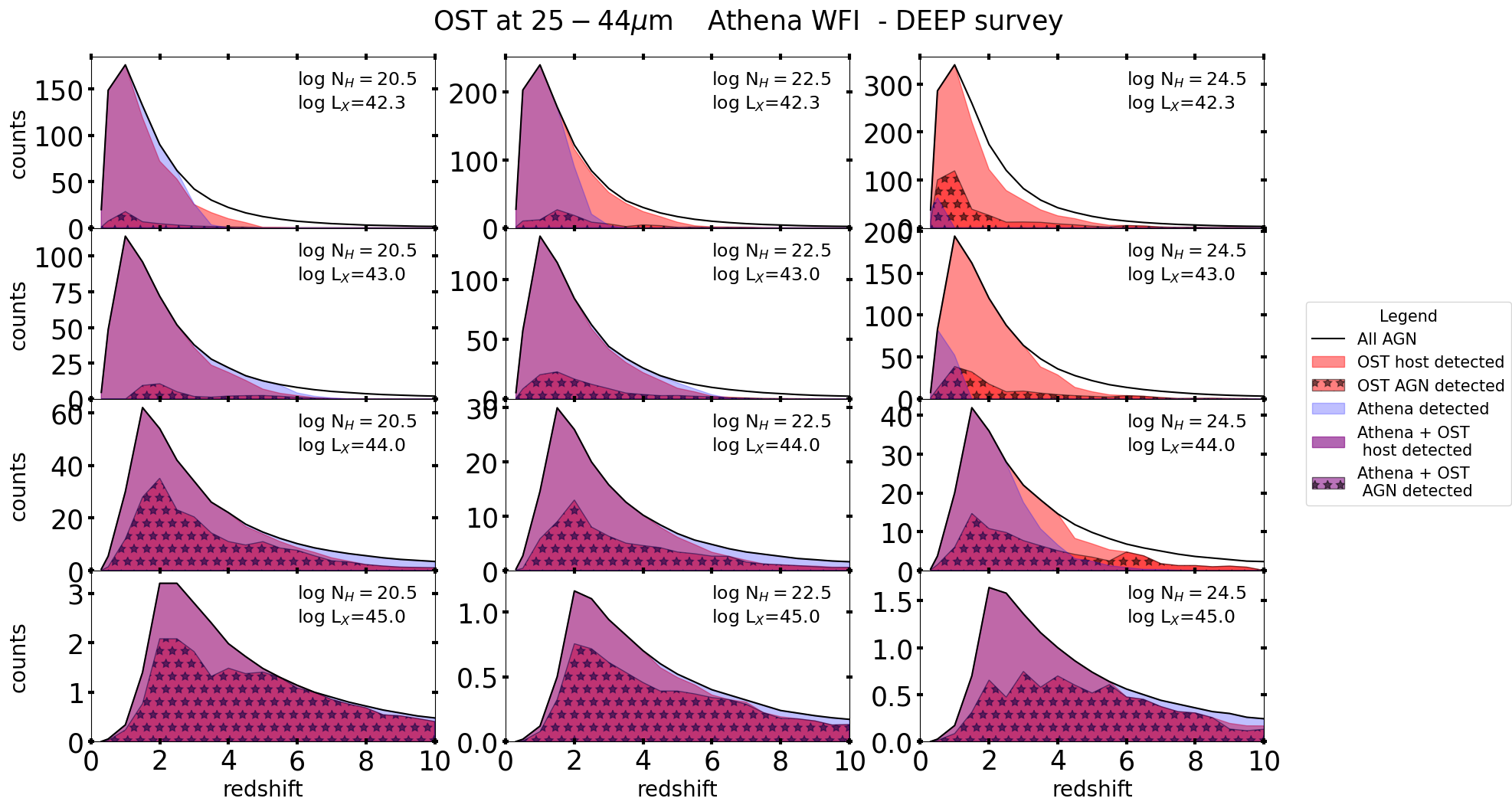}}
  \caption{Number of AGN expected per deg$^2$ and per $\Delta z =1$ for a \OST{} \OSS{} DEEP survey. The black lines are the total number of expected AGN, the red areas represent those which can be detected in the band 1 by \OST{} \OSS{}, the blue areas those which can be detected in the X-rays with the \textsl{Athena WFI}, the purple areas are the AGN that will be detected both by \OSS{} and \textsl{Athena}. The red uniform area represents the sources detected by \OSS{} in which the main component is due to the host-galaxy emission, while for those represented with the starry red area the AGN is the main contributor to the detected emission. The columns refer to AGN with different amount of obscuration ($20 \le \log\, (\text{N}_{\text{H}}/\text{cm}^{-2}) \le 21$; $22 < \log\, (\text{N}_{\text{H}}/\text{cm}^{-2}) \le 23$; $24.18 < \log\, (\text{N}_{\text{H}}/\text{cm}^{-2}) \le 25$, from left to right), the rows to different AGN luminosity ($42.0\le \log{(\text{L}_{\text{x}}/\text{erg/s})}  <42.3 $, $42.9\le \log{(\text{L}_{\text{x}}/\text{erg/s})}  <43.2 $, $43.9\le \log{(\text{L}_{\text{x}}/\text{erg/s})}  <44.2 $, $44.9\le \log{(\text{L}_{\text{x}}/\text{erg/s})}  <45.2 $, from top to bottom). For forty times we extracted 20 SED for each bin, measured the flux in each \OSS{} band and compared them with the $5\,\sigma$ sensitivities to compute the number of detectable sources. The median of these forty values are the numbers used in this figure, while the $84^{\textit{th}}-16^{\textit{th}}$ percentiles are used as uncertainties (not reported in the figure).}
  \label{fig:ost1deep} 
\end{figure*}

\begin{figure*}
  \centering
  \resizebox{0.95\hsize}{!}{\includegraphics{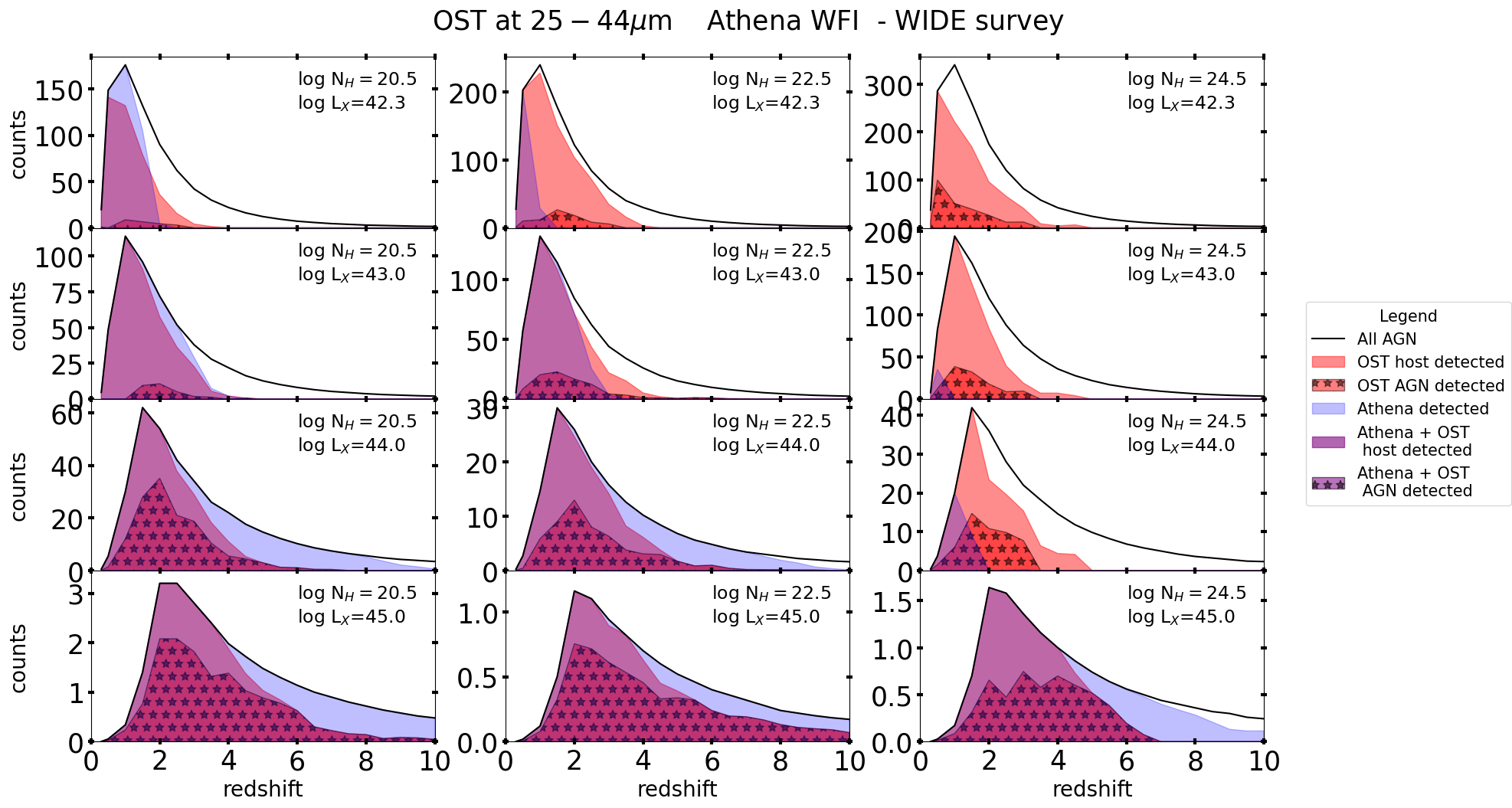}}
  \caption{Number of AGN expected per deg$^2$ and per $\Delta z =1$ for a \OST{} \OSS{} WIDE survey. The lines and areas are coded as in Figure~\ref{fig:ost1deep}.}
  \label{fig:ost1wide} 
\end{figure*}

       
We summarize the main results of our prediction in Tables~\ref{tab:recap1} and \ref{tab:recap2}. Figures~\ref{fig:smi34UD}, \ref{fig:smi34D_84}, \ref{fig:ost1deep}, and \ref{fig:ost1wide} illustrate the number of photometric detections per deg$^2$ for \textit{SPICA-SMI} DEEP, \textit{SPICA-SMI} WIDE, \OSS{} DEEP, and \OSS{} WIDE, respectively. All the figures also report the number of sources detected by \textsl{Athena WFI}.  The black lines are the total number of AGN expected, the red areas represent those which can be detected using \textit{SPICA} SMI-CAM at 34$\mu$m (or band 1 of \OSS), the blue areas those which can be detected in the X-rays by \textsl{Athena}, the purple areas are the AGN that can be detected both by \textit{SPICA} SMI-CAM (or \OSS) and by \Athena. The red uniform area represents the IR detected sources in which the main component of emission is due to the host-galaxy, while for those represented with the starry red area, the AGN is the main contributor to the detected emission. Figures~\ref{fig:smi34UDLR} and \ref{fig:ost1deepLR} have a similar color code, but with the dashed areas representing the sources for which we will have spectrocopic detection in a DEEP survey with a \textit{SPICA} SMI-LR-like instrument or with the \OSS{}. Finally, figure~\ref{fig:oss_spica_comp_deep} (figure~\ref{fig:oss_spica_comp_wide}) shows a comparison of the number of sources that we will be able to detect in a DEEP (WIDE) survey with \textit{SPICA} and \OST{}, with the color code indicating the expected number of photometric bands in which we may detect the source (considering SMI-CAM and B-BOP for \textit{SPICA} and the six \OSS{} bands for the \OST{}). \par  

\begin{table*}
\caption{ Percentage of all the AGN, CT-AGN, CT-AGN at $z \le 4$, and CT-AGN at $z \le 2$  photometrically detected with various configurations of instruments and surveys. SMI refers to \textsl{SPICA SMI-CAM}. OSS refers to the sources detected with the \OST{} \OSS{} in photometric mode (R=4) in at least one of the bands, while the number within the parenthesis to those with detection in all the six \OSS{} bands. SMI AGN  and OSS AGN refer to the direct detection of AGN emission (thus the cases where the AGN is more luminous than the host-galaxy in the considered band) for respectively the \textsl{SPICA SMI-CAM} and the \OST{} \OSS{} instruments. WFI refers to the source photometrically detected by \textsl{Athena WFI}, while SMI+WFI (OSS+WFI) refers to the sources with both \textsl{SMI-CAM} (\OSS) detection and \textsl{Athena} photometric detection.}
\centering
\begin{tabular}{cccccccccccc}
\hline\hline
 & Survey & SMI & SMI AGN & OSS & OSS AGN & WFI & SMI+WFI & OSS+WFI \\
 & & \% & \% & \% & \% & \% & \% & \% \\
\hline
 \multirow{2}*{All AGN} & DEEP & 88$\,\pm\,3$ & 22$_{-6}^{+8}\,$ & 95$_{-2}^{+1}\,$ (84$\,\pm\,3$) & 21$_{-7}^{+6}\,$ ($<1$) & 52 & 51 & 51 (50) \\
                        & WIDE & 72$\,\pm\,5$ & 16$\,\pm\,6$ & 83$_{-4}^{+3}\,$ (60$_{-5}^{+6}\,$) & 15$\,\pm5$ ($<1$) & 29 & 27 & 28 (25)  \vspace{0.2cm}\\
 \multirow{2}*{CT-AGN} & DEEP & 87$\,\pm\,3$ & 		& 96$\,\pm\,1$ (84$\,\pm\,4$) &  			   & 20  & 20 & 20 (20) \\
			& WIDE & 71$\,\pm\,5$ &		& 84$\,\pm\,3$ (59$_{-6}^{+5}\,$) &			   & 4   & 4  & 4 (4)  \vspace{0.2cm}\\
 \multirow{2}*{CT $z\le4$} & DEEP & 94$\,\pm\,2$ & 		& 99$\,\pm\,1$ (89$_{-3}^{+4}\,$) &  			   & 22  & 22  & 22 (22) \\
			& WIDE & 80$\,\pm\,5$ &		& 93$_{-3}^{+2}\,$ (68$_{-7}^{+6}\,$) &		   & 5   & 5  & 5 (5)  \vspace{0.2cm}\\	
 \multirow{2}*{CT $z\le2$} & DEEP & 99$\,\pm\,1$ & 		& 99$\,\pm\,1$ (92$_{-3}^{+2}\,$) &  			   & 28  & 28  & 28 (28) \\
			& WIDE & 84$_{-4}^{+5}$ &		& 98$\,\pm\,1$ (78$\pm 6\,$) &			   & 6   & 6  & 6 (6)  \\						
\hline\hline
\end{tabular}
\label{tab:recap1}
\end{table*}

\begin{table*}
\caption{Percentage of all the AGN, CT-AGN, CT-AGN at $z \le 4$, and CT-AGN at $z \le 2$  spectroscopically detected with various configurations of instruments and surveys. SMIsp refers to \textsl{SPICA SMI-LR}. OSSsp refers to the sources detected with the \OST{} \OSS{} in spectroscopic mode (R=300) in at least one of the bands, while the number within the parenthesis to those with detection in all the six \OSS{} bands. WFIsp refers to the source spectroscopically detected by \textsl{Athena WFI}, while SMIsp+WFI (OSSsp+WFI) refers to the sources with both \textsl{SMI-LR} (\OSS) spectroscopic detection and \textsl{Athena} photometric detection.}
\centering
\begin{tabular}{ccccccc}
\hline\hline
 & 			Survey  & SMIsp 		&  OSSsp 				& WFIsp & SMIsp+WFI & OSSsp+WFI \\
 & & \% & \% & \% & \% & \% \\
\hline
 \multirow{2}*{All AGN} & DEEP & 38$\,\pm\,6$ 	& 81$\,\pm\,3$ (55$\,\pm\,6$)		& 20 & 28 & 47 (38) \\
                        & WIDE & 18$\,\pm\,5$ 	& 53$\,\pm\,5$ (20$_{-4}^{+5}\,$)	&  6 & 10 & 22 (11)   \vspace{0.2cm}\\
 \multirow{2}*{CT-AGN} & DEEP & 37$\,\pm\,6$  	& 82$_{-2}^{+3}\,$ (55$_{-7}^{+6}\,$) & 1 & 16 & 20 (18)  \\
                        & WIDE &17$\,\pm\,5$  	& 54$\,\pm\,6$ (22$_{-6}^{+5}\,$)	&  <1 & 3 & 4 (4)    \vspace{0.2cm}\\
 \multirow{2}*{CT $z\le4$} & DEEP & 43$_{-7}^{+8}\,$	& 92$_{-7}^{+6}\,$(63$\pm\,7\,$)	& 2 & 18 & 22 (21) \\
                        & WIDE & 18$\,\pm\,5$		& 63$_{-6}^{+7}\,$(26$_{-6}^{+7}\,$)	&  <1 & 4 & 5 (4)   \vspace{0.2cm}\\
 \multirow{2}*{CT $z\le2$} & DEEP & 55$_{-7}^{+9}\,$	& 98$\pm\,1\,$(75$_{-7}^{+6}\,$)	& 2 & 22 & 28 (27) \\
                        & WIDE & 20$\pm\,6\,$ 	& 77$_{-6}^{+5}\,$(34$_{-7}^{+8}\,$)	&  <1 & 5 & 6 (6)  \\

\hline\hline
\end{tabular}
\label{tab:recap2}
\end{table*}

\subsection{Photometric detections}\label{sec:photom} 
Considering the DEEP survey we will be able to detect in the IR $\gtsim 90\,\%$ of all the AGN, approximately half of them will have photometric detection both in the IR and in the X-ray bands; this synergy will allow us to identify the source as an AGN and will help in placing better constraints to its properties. By a DEEP survey, working jointly, a IR cryogenic observatory (like \textit{SPICA} or \OST{}) and \textsl{Athena} can detect all the AGN up to $z=4$, \textit{de-facto} completely covering the ``cosmic noon'' and bringing important insights about the AGN density evolution at higher redshifts. At the same time, the bulk of the intrinsic X-ray emission is produced at the ``knee'' of the luminosity function ($\text{L}_{\text{x}}\sim 5 \times 10^{44}\,$erg/s for $z \sim 1-4$, \citealt{aird10}) and these sources will be easily detected and characterized by \textsl{Athena} up to $z \sim 3$. Moreover, \textsl{Athena} will be of paramount importance to identify as AGN even sources for which the IR torus emission is diluted and hidden behind the more powerful host-galaxy contribution. The combined used of data in both bands will be fundamental in obtaining a more robust estimate of the AGN bolometric power. In particular, on the one hand, obtaining the AGN bolometric power with only X-ray data requires to assume an AGN X-ray bolometric correction, thus introducing large uncertainties in the derived values, due to our limited knowledge of the $k_{bol}-\text{L}_{bol}$ relation \citep{hopkins07, lusso10, lusso12, duras20}. On the other hand, using only an instrument similar to \textit{SPICA} SMI-CAM, it will not be possible to distinguish between the AGN and the galaxy emission, as \textit{SPICA SMI-CAM} can measure only the total (AGN+host-galaxy) emission. \par
Considering a WIDE survey, the fractions of sources that will be detected are lower of $\sim 20\,\%$ with respect to the DEEP survey, but the (at least) fifteen times wider area will allow us to detect at least 10 times more sources. Similarly to the DEEP, the WIDE survey may allow us to perfectly cover the ``cosmic noon'' with the advantage of a ten-folded statistic (at least). In a WIDE survey, \textsl{Athena WFI}  may loose a significant fraction of the most obscured AGN or of the low-luminosity ones. However, a \textit{SPICA}-like (or \OST-like) mission should be able to effectively recover these sources, although we may need an effective way (e.g. \textit{SPICA-SAFARI} or \OST{} spectroscopic follow-up, see sec~\ref{sec:spectros} and \ref{sec:sedfitting}) to characterize their AGN properties (intrinsic bolometric luminosity and amount of obscuration). For this shallower survey it is still valid our statement that, for most of the detected sources at $\log{(\text{N}_{\text{H}}/\,\text{cm}^{-2})}\le 23$, $ \log{(\text{L}_{\text{x}}/\text{erg/s})}  >43$ and $z \le 3$ (and higher $z$ for higher L$_{\text{x}}$), \textsl{Athena} detections would provide essential evidence of the AGN nature of those sources. \par
Focusing on \OST, the six spectrophometric bands of \OSS{} will give us a good coverage at and beyond the ``cosmic noon'' even in photometric mode. Moreover, thanks to the capabilities of \OSS, for the majority of the sources detected by \textsl{Athena}, we will have six photometric points (and maybe up to six spectra) in the mid- and far-IR.\par

\subsection{Spectroscopic detections}\label{sec:spectros} 
We estimated that, exploiting \textsl{Athena WFI} spectra, we will be able to recover the intrinsic AGN luminosity and obscuration (with $30\,\%$ uncertainties) for $\approx 20\,\%$ of all the AGN and up to $\approx 50\,\%$ of the high-luminosity ones. Even low-count X-ray spectra will allow reliable $\text{N}_{\text{H}}$ characterization, as for obscured and CT-AGN, we expect simple spectra with a flat continuum and a possibly strong iron K$\alpha$ line. Furthermore, an X-ray detection at a level of $\sim 2\times 10^{-16}$~erg cm$^{-2}$ s$^{-1}$ in the $2-10$~keV band at $z \ge 1$, or simply any detection at $z \gtrsim 2.3$ in the DEEP survey, imply X-ray luminosities $\gtrsim 10^{42}\,$erg/s in that band, revealing almost unequivocally an AGN. \par
\begin{figure*}
  \centering
  \resizebox{0.93\hsize}{!}{\includegraphics{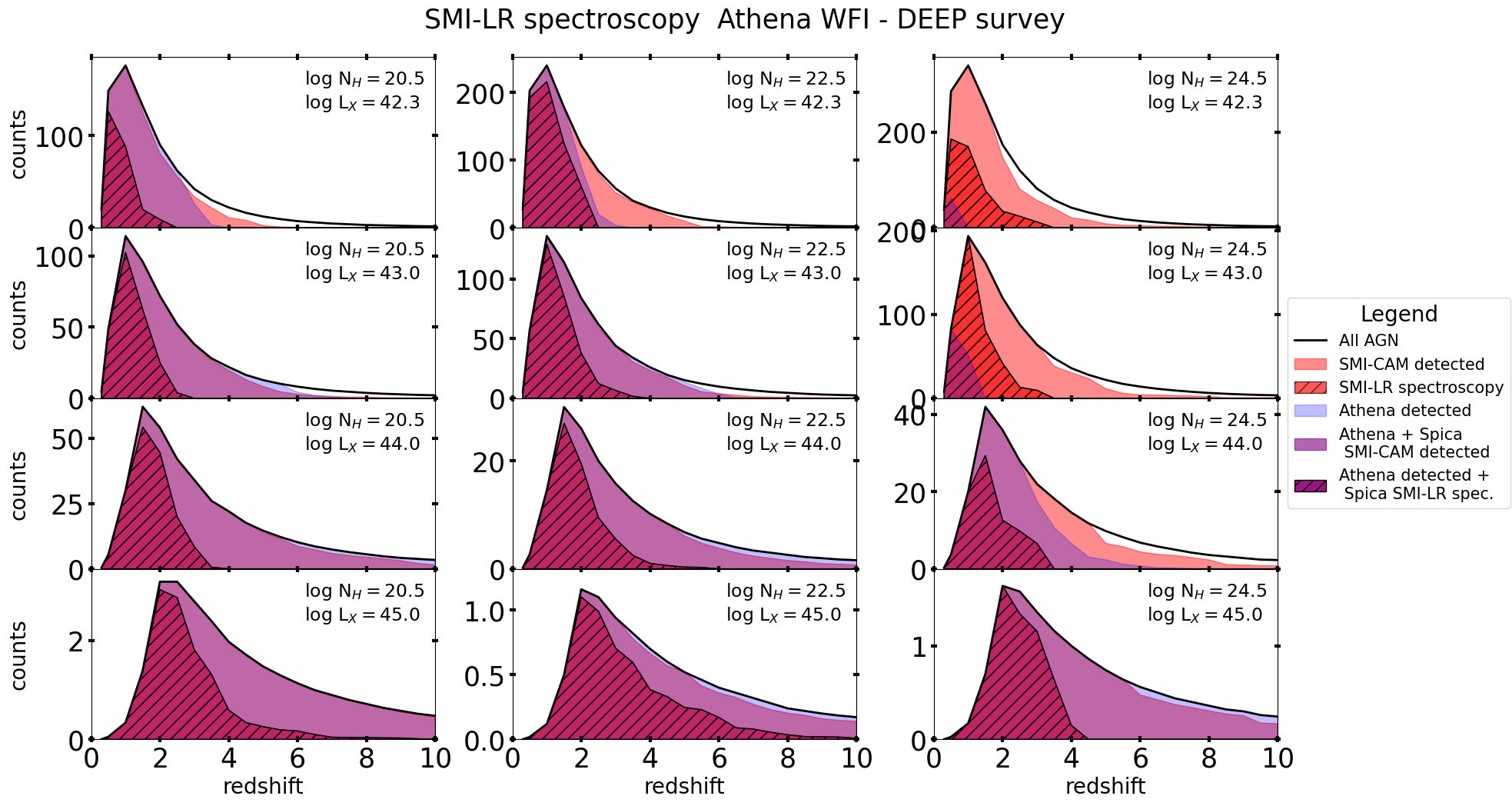}}
  \caption{Number of AGN expected per deg$^2$ and per $\Delta z =1$ for a \textit{SPICA}-like DEEP survey. The black lines are the total number of AGN expected, the red areas represent those which can be detected with the \textit{SPICA} SMI-CAM at 34$\mu$m, the blue areas those which can be detected in the X-rays by \textsl{Athena SMI}, the purple areas are the AGN that will be detected both by \textit{SPICA} SMI-CAM and \textsl{Athena}. The dashed areas represent the sources which could be spectroscopically detected with SMI-LR ($5\,\sigma$ detection with a resolution R$=50-120$ in the $17-36\,\mu$m wavelength range). The columns refer to AGN with different amount of obscuration ($20 \le \log\, (\text{N}_{\text{H}}/\text{cm}^{-2}) \le 21$; $22 < \log\, (\text{N}_{\text{H}}/\text{cm}^{-2}) \le 23$; $24.18 < \log\, (\text{N}_{\text{H}}/\text{cm}^{-2}) \le 25$, from left to right), the rows to different AGN luminosity ($42.0\le \log{(\text{L}_{\text{x}}/\text{erg/s})}  <42.5 $, $42.9\le \log{(\text{L}_{\text{x}}/\text{erg/s})}  <43.2 $, $43.9\le \log{(\text{L}_{\text{x}}/\text{erg/s})}  <44.2 $, $44.9\le \log{(\text{L}_{\text{x}}/\text{erg/s})}  <45.2 $, from top to bottom).}
  \label{fig:smi34UDLR} 
\end{figure*}
\begin{figure*}
  \centering
  \resizebox{0.93\hsize}{!}{\includegraphics{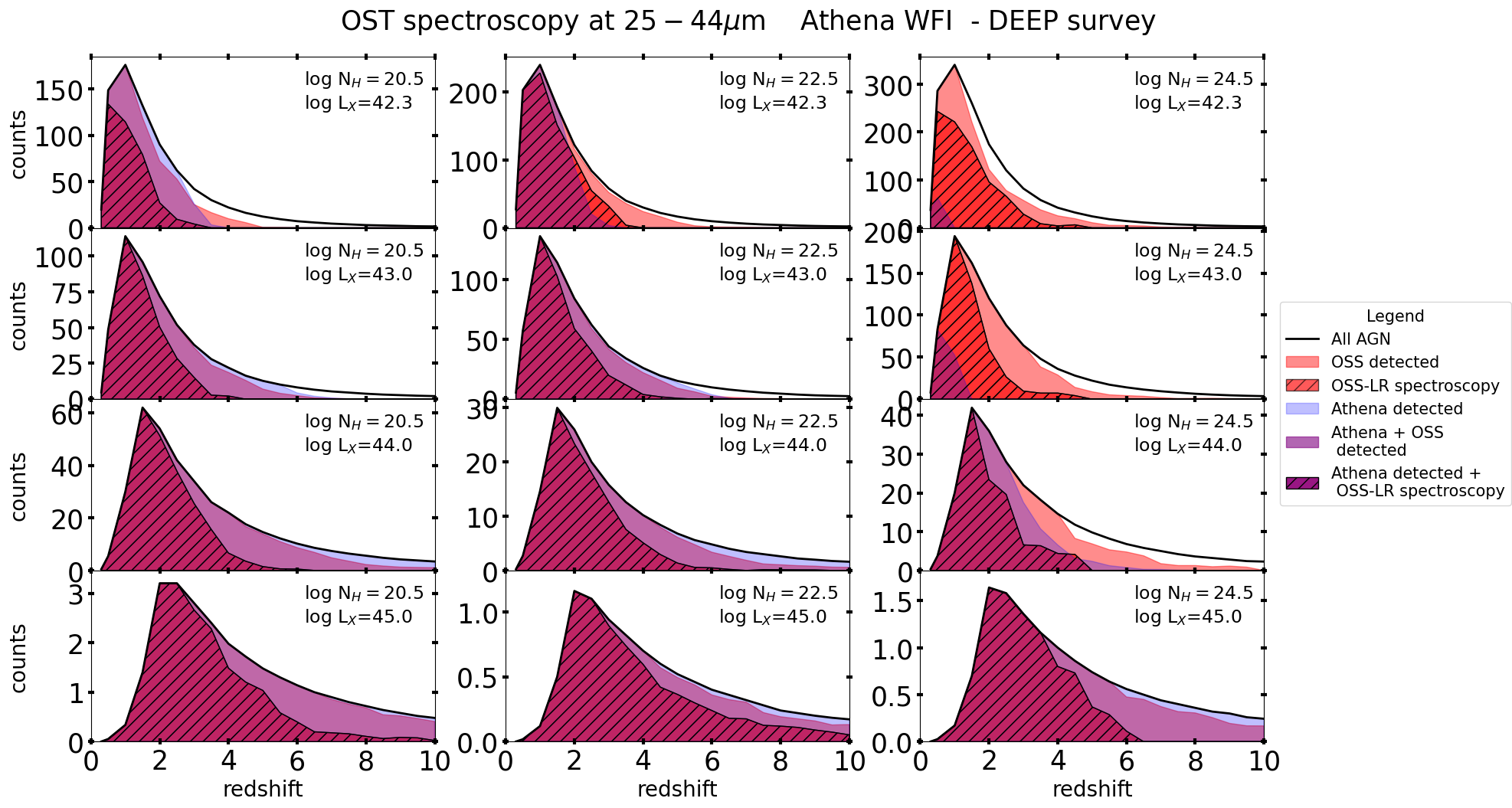}}
  \caption{Number of AGN expected per deg$^2$ and per $\Delta z =1$ for a \OSS{} DEEP survey. The lines and areas are coded as in Figure~\ref{fig:smi34UDLR}, but with the red areas referring to the band 1 of the \OST{} \OSS{} instrument.}
  \label{fig:ost1deepLR} 
\end{figure*}

Considering a SMI-LR-like instrument (figure~\ref{fig:smi34UDLR}), we find that with a DEEP survey we will be able to detect spectroscopically more than half of all the AGN within $z=2$. For the high-luminosity objects, we expect to have some spectroscopic detections even up to $z\approx 4$. Considering a WIDE survey, the fraction of AGN that we should be able to detect spectroscopically drops considerably. For the sources for which a fast low-resolution spectroscopic characterization will not be possible, we may make use of photometric data (such as those produced by SMI-CAM- or B-BOP-like instruments) and of SED-fitting technique (see section~\ref{sec:sedfitting}) to disentangle the AGN and SF emissions.\par 
Considering the \OSS, we find that, with a DEEP survey, we will be able to detect spectroscopically between $55\,\pm 6\,\%$ and $81\,\pm 3\,\%$ of all the AGN, depending on the \OSS{} band considered. Thus, we will have R=300 full IR spectra for more than half of all the predicted AGN at $z\le 10$. For the high-luminosity objects, we expect to have some spectroscopic detections even up to $z > 4$. In the WIDE survey, the fraction of AGN that we should be able to detect spectroscopically lowers of $\approx 30\,\%$ with respect to the DEEP survey.\par

Several important galaxy and AGN mid-IR spectral features fall in the wavelength range of a \textit{SPICA SMI-LR}-like (or \OSS) instrument and can be of fundamental importance in both characterizing AGN and host-galaxy properties and measuring the redshift. In particular, we expect AGN-related high-ionization lines\footnote{The high-ionization potential assures us that the emission can not be associated with star-formation activity and has to be, instead, linked to AGN activity}, such as [Ne$\,\textsc{v}$]$14.3\mu$m and $24.3\mu$m, [O$\,\textsc{iv}$]$25.9\mu$m (e.g. \citealt{tommasin08, tommasin10, feltre16}), but also SF-related Polycyclic Aromatic Hydrocarbons (PAHs) features (e.g. \citealt{leger89}) and the $9.7\mu$m and $18\mu$m silicate features. The $9.7\mu$m feature is typically associated with type 1 AGN when observed in emission and with type 2 when it is in absorption (with the few notable exceptions of the type 2 quasars of \citealt{sturm06, teplitz06}, which show the silicate feature in emission); moreover, more obscured AGN usually have deeper silicate features, with depths that correlate with the X-ray derived $\text{N}_{\text{H}}$ (e.g. \citealt{wu09,shi06}, but see also \citealt{Goulding12}). \par

Considering the AGN spectroscopic properties, the most evident mid-IR AGN lines ([Ne$\,\textsc{v}$]$14.3\mu$m and $24.3\mu$m, [O$\,\textsc{iv}$]$25.9\mu$m) are redshifted out of a \textit{SMI-LR} spectral range at modest redshift ($z < 1$), and would require follow-ups with either \textit{SPICA-SAFARI} or \textit{SMI-MR} to be properly detected. In particular, exploting 5 hours pointed observations with a \textsl{SPICA SAFARI}-like instruments, for low-luminosity (high-luminosity) unobscured AGN the [Ne$\,\textsc{v}$]$24.3\mu$m and [O$\,\textsc{iv}$]$25.9\mu$m lines can be studied up to $z \approx 1$ and $z \approx 2$ ($z \approx 3$ and $z \approx 4$). The same lines can be detected up to $z \approx 1.5$ and $z \approx 2.5$ ($z \approx 4$ and $z \geq 4$) for low-luminosity (high-luminosity) obscured AGN \citep{spinoglio17, spinoglio21}. For the majority of the sources, the brightest and most recognizable features may be due to the galaxy emission, in particular the dust continuum and the PAH emission lines (e.g. \citealt{lutz98, hollenbach99, yan07, fadda10, spinoglio21}). However, for most cases, we expect the $9.7\mu$m silicate feature, to be easily recognizable. The silicate feature can be helpful in estimating the AGN amount of obscuration (e.g. \citealt{vignali11, lacaria19}) and can be entirely within a SMI-LR-like instrument wavelength range up to $z \approx 3$.  This is particularly important for the CT-AGN, for which, with the exception of the most luminous ones ($\text{L}_{\text{x}} \approx 10^{44}$ erg/s for $z<3$), it will be difficult to obtain the AGN properties using only X-ray spectroscopy. Thus, the analysis of both X-ray and IR spectroscopic data will play a fundamental role in discerning their characteristics.\par
\begin{figure}
  \centering
  \resizebox{\hsize}{!}{\includegraphics{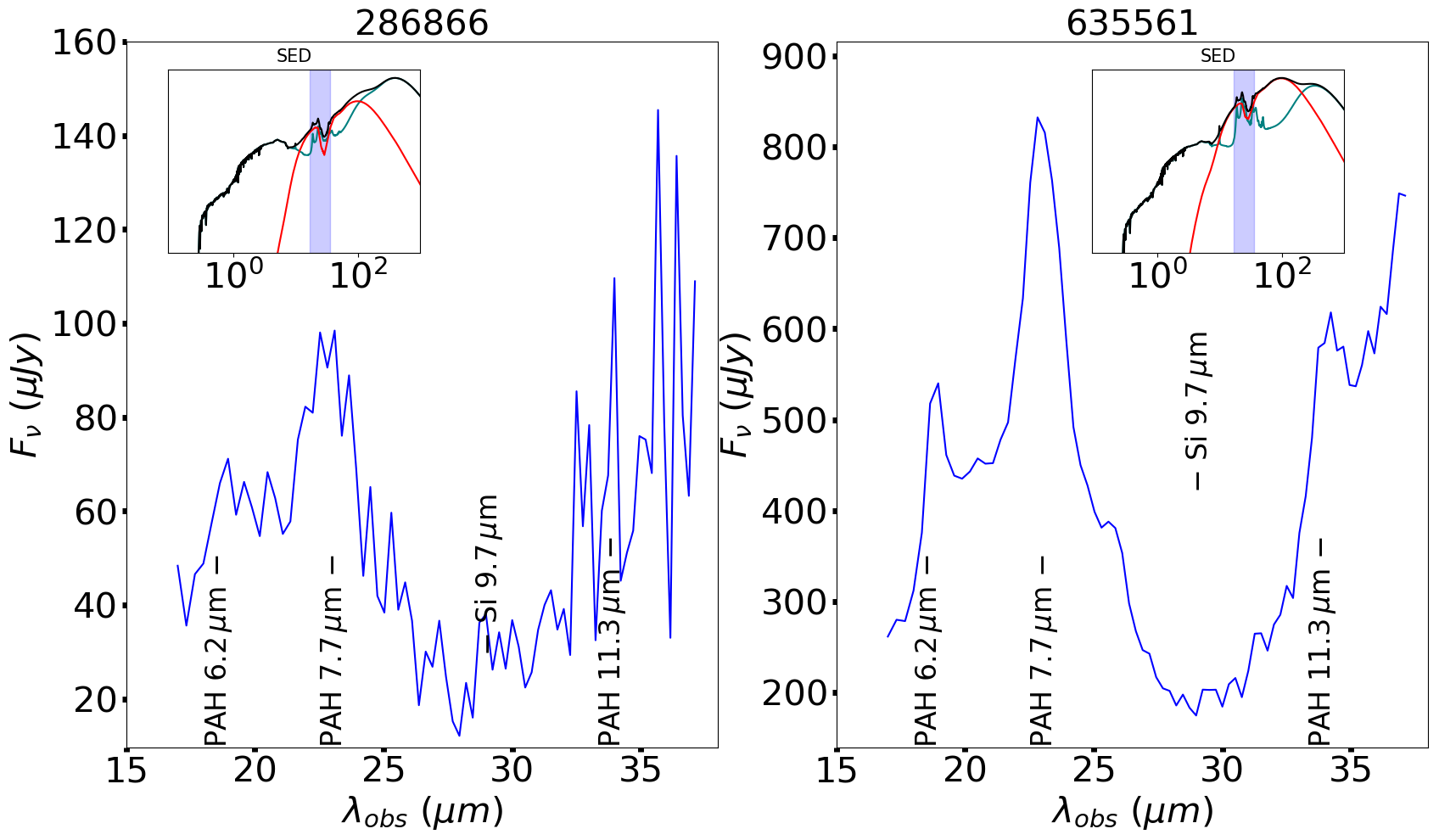}}
  \caption{Simulated SMI-LR spectra for a \textit{SPICA}-like DEEP survey. Left panel is for $\log{ (\text{L}_{\text{x}}/\text{erg/s})}=43.9$, $\log{ (\text{N}_{\text{H}}/\text{cm}^{-2})}=23.4$ (i.e. moderate-luminous obscured) AGN at $z=2$; right panel for $\log{ (\text{L}_{\text{x}}/\text{erg/s})}=44.3$, $\log{ (\text{N}_{\text{H}}/\text{cm}^{-2})}=24.1$ (i.e. high-luminous) CT-AGN $z=2$. We used the \citet{lanzuisi17} SEDs (inset in the upper left of the plots, where the red line is the AGN component, the blue one is related to the host-galaxy and the black one is the total AGN+galaxy emission) as spectral templates and added a white noise with amplitude based on the LR continuum sensitivity expected for a DEEP survey. The spectra were sampled with a resolution of $R=50$ at $17\,\mu$m and R$=120$ at $36\,\mu$m.}
  \label{fig:spettri} 
\end{figure}
\begin{figure}
  \centering
  \resizebox{\hsize}{!}{\includegraphics{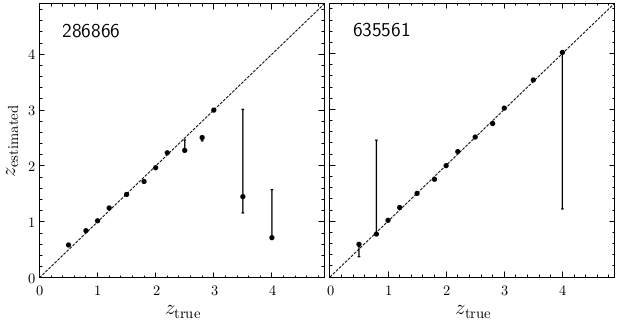}}
  \caption{Redshift estimate from simulated \textit{SMI-LR}-like spectra using a modified version of the PAHFIT code (Negrello et al. in prep). Left panel is for $\log{ (\text{L}_{\text{x}}/\text{erg/s})}=43.9$, $\log{ (\text{N}_{\text{H}}/\text{cm}^{-2})}=23.4$ moderate-luminous obscured AGN; right panel for $\log{ (\text{L}_{\text{x}}/\text{erg/s})}=44.3$, $\log{ (\text{N}_{\text{H}}/\text{cm}^{-2})}=24.1$ high-luminous CT-AGN. We find that we should be able to effectively recover the redshift up to $z \approx 3-4$; at $z>4$, there are too few strong PAH features in the \textit{SMI-LR} wavelength range to allow a proper determination of the redshift.}
  \label{fig:spectra} 
\end{figure}

One of the advantages of using a \textit{SPICA SMI}-like (or \OSS-like) instrument is that it can provide, at the same time, photometry and spectroscopy, and it can quickly produce low-resolution spectroscopic data for a large number of sources. X-ray spectral analysis is, so far, the best way to characterize the AGN (e.g. to determine the intrinsic luminosity), but it requires a redshift determination. Fortunately, there is a good prospect that redshifts can be determined solely from X-ray spectra (e.g. \citealt{simmonds18, peca21}). We simulated mid-IR spectra to evaluate the possibility of measuring the redshift from a \textit{SMI-LR}-like spectra. In particular, we used a code that exploits the most prominent PAH features to recover the redshift. This implies that it is most suited for sources with sufficient star-formation to produce strong PAH features and low- or moderate-luminosity AGN, as the equivalent width of the PAH features is greatly reduced when the AGN luminosity is high \citep{voit92,genzel98,odowd09}. We chose ten sources among the $\sim 500$ of our sample and use their SEDs as spectral templates. The ten sources were chosen so that in our sample there were un-obscured, obscured and CT-AGN, high- and low- luminosity sources, AGN- and host-dominated sources, with emission and absorption $9.7\,\mu$m features. For each template, we simulated sixteen spectra in the redshift range $0.5 \le z \le 6$ (as we can see from Fig~\ref{fig:smi34UDLR}, at $z>6$ the fraction of spectroscopically detected sources is very small). The simulated spectra were created accordingly to the latest \textit{SMI} specifications. We used a spectral resolution of $R=50$ at $17\,\mu$m and R$=120$ at $36\,\mu$m and assumed a linear regime between these two values. We used the LR continuum sensitivity (see section~\ref{sec:survey_smi}), rescaled to the Ultradeep survey, as standard deviation of the Gaussian noise that we included in the spectra. Figure~\ref{fig:spettri} shows two of the simulated spectra. Finally, we exploit a modified version of the PAHFIT code \citep{smith06} that uses a power-law continuum, several PAH emission lines and silicate absorption, to fit the spectra and provide the source redshift (Negrello et al. in prep). As shown in figure~\ref{fig:spectra}, we find that we should be able to effectively recover the redshift in case of moderate and luminous type 2 AGN up to $z \sim 3-4$. However, the code currently does not use AGN silicate $\,9.7\mu$m emission feature in finding the $z$ and is proned to mis-interpret it as a PAH feature when it is strong; however, the code is still able to recover the redshift in case of faint AGN silicate emission line, as long as the silicate line is not mistakenly interpreted as a PAH feature. In particular, excluding three sources with strong Si$\,9.7\mu$m emission features and one source with very low signal-to-noise ratio, we recovered the redshifts with a median error $ |{z_{best}-z_{true}|\,/\,(1+z_{true})} =0.02$. In conclusion, a \textit{SPICA SMI}-like instrument can provide us with low-resolution spectra for more than $6\,000$ AGN in a DEEP survey and, in the cases of sources with prominent host-galaxy PAH lines and low AGN activity (i.e. unable to destroy these lines), measures of their spectroscopic redshifts. Moreover, it should be possible to obtain an estimate of the AGN obscuration from the depth of the silicate $\,9.7\mu$m absorption feature (\citealt{wu09,shi06} had shown that there is a correlation between these two quantities), that may provide support to the outcomes of X-ray spectroscopy in terms of obscuration (especially for low-statistics obscured AGN).\par

Focusing on the \OSS{} capabilities, this instrument is able to exploit the above mentioned mid-IR lines, but with the advantages of higher spectral resolution and larger wavelength range coverage than a \textit{SPICA}-like instrument. Lines as the [Ne$\,\textsc{v}$]$14.3\mu$m and $24.3\mu$m, [O$\,\textsc{iv}$]$25.9\mu$m, which are redshifted out of \textit{SMI-LR} spectral range at modest redshift ($z < 1$), thus requiring follow-ups with other instruments, can be detected up to very high-redshift, their detection being a matter of instrument sensitivity and not of limited wavelength range. Fine structure lines of O, C, Ne, S, N, Fe, Ar, and Si are keys to probe neutral and ionized gas phases and can be used to estimate redshift, amount of gas, its ionization, presence of outflows and the contribution of SF and AGN to the emission \citep{meixner19, mordini21}. The \OSS{} DEEP survey will allow us to detect the PAH up to $z\approx4$ for galaxies with SFR $\sim 10 \text{M}_{\odot}/\text{yr} $ and over $z>6$ for those with SFR $\sim 100 \text{M}_{\odot}/\text{yr} $ \citep{meixner19}. Regarding the estimate of the BHARD, the high sensitivity and large spectral range of the \OSS{} survey will allow the detection of the [Ne$\,\textsc{v}$] line (whose flux correlates with the AGN intrinsic emission), as well as of the [O$\,\textsc{iv}$] and  [Ne$\,\textsc{ii}$] line, whose line flux ratio can be used as a diagnostic of AGN power (e.g., \citealt{genzel98, lutz03, armus06, armus07, mordini21}). \citet{gruppioni16} well calibrated the relations between [Ne$\,\textsc{v}$] (or [O$\,\textsc{iv}$]) and AGN bolometric emission in samples of local galaxies. \citet{bonato19} explored the spectroscopic survey capabilities of the \OSS{} and find out that, in a DEEP survey, it can detect emission lines associated to SF up to $z \approx 8.5$ and those of AGN emission, in particular [Ne$\,\textsc{v}$]$24.3\mu$m and [O$\,\textsc{iv}$]$25.9\mu$m, up to $z \approx 3$ and $z \approx 5$, respectively.

\subsection{SED fitting}\label{sec:sedfitting} 
Working jointly, an IR cryogenic observatory, like \textit{SPICA} or \OST{}, and \textsl{Athena} can detect a huge fraction of the the AGN up to $z=10$. The combined used of data from both wavelength bands will be fundamental, as with only IR photometric detections, it will not be possible to distinguish between the AGN and the galaxy emission, as \textit{SPICA SMI-CAM} or \OSS{} can measure only the total (AGN+host-galaxy) emission, this in the lucky case of having already identified the source as an AGN. In this regard, a spectrometer similar to \textit{SPICA} SMI-LR or \OSS{} may surely help, but we will not be able to spectroscopically detect all the AGN, as the spectroscopic recovery fraction decreases rapidly for $z>2$. The additional use of other photometric data (such as those of B-BOP-like instruments, or from all the \OSS{} bands), along with SED decomposition technique, could be used to overcome this limitation. Even in photometric mode, having six far-IR detections of \OST{} will be invaluable in disentangling the AGN and host-galaxy contribution: indeed, these bands cover both the regime where the torus has its peak emission and the regime where the SF is the main contributor to the SED. But we expect that the real game-changer in obscured AGN studies will come from a multi-wavelength approach.\par
 The use of \textsl{Athena} data, optical and near-IR photometric datapoints (in fields where these are already available, such as COSMOS and the Euclid Deep Fields), mid-IR and far-IR photometric instruments (similar to those designed for \textit{SPICA}) can provide a never-reached-before coverage of all the AGN and galaxy emission up to high redshift. In addition, with follow-up pointed spectroscopic observations of these targets (photometrically detected, but without a clear separation between galaxy and AGN components) we will be able to distinguish the AGN and host-galaxy emissions using fine-structure lines and PAH features up to $z \sim 4$ \citep{spinoglio21}. Therefore, we expect to be able to improve our knowledge of the AGN $k_{bol}-\text{L}_{bol}$ relation, similarly to what was done in \citet{lusso12}, but with a larger sample and to higher redshift. This should enable us to extend the $k_{bol}-\text{L}_{bol}$ dynamical range and, having more sources, to reduce its uncertainties.\par

\begin{figure}
  \centering
  \resizebox{0.95\hsize}{!}{\includegraphics{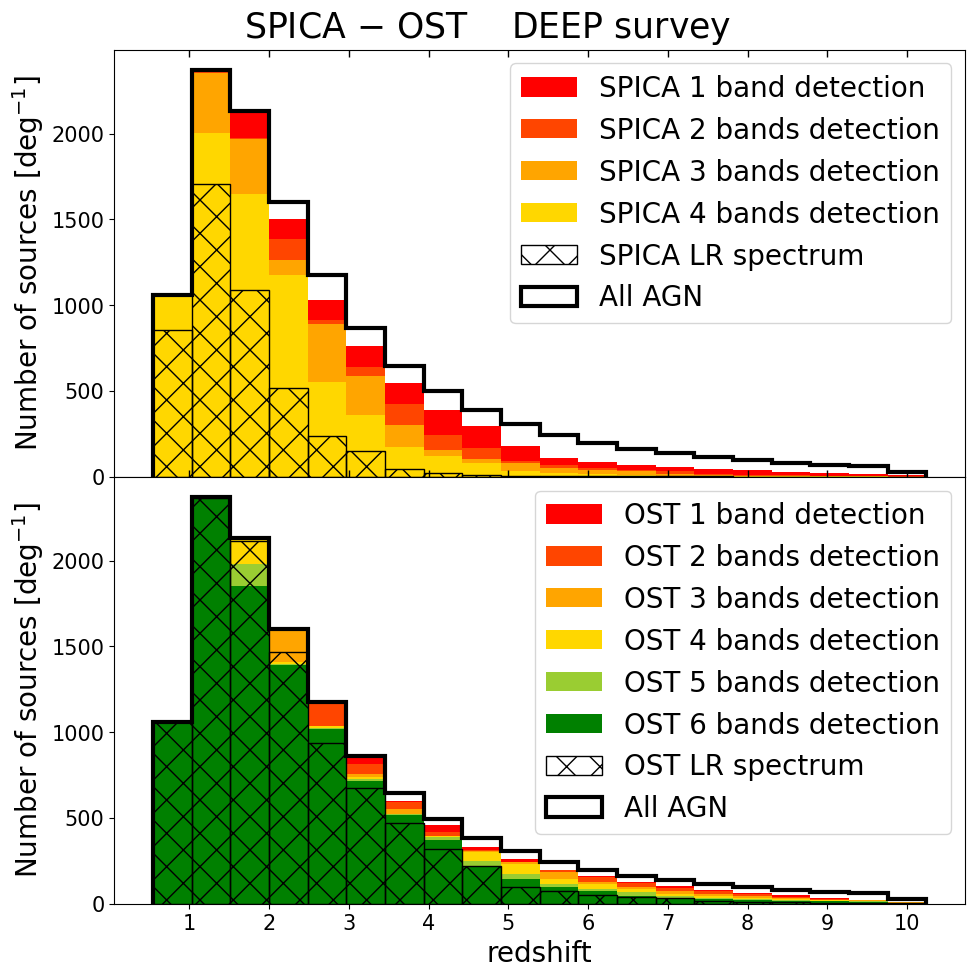}}
  \caption{Distribution of the number of AGN expected per deg$^2$ for a \textit{SPICA}-like (\textit{top panel}) and for the \OST{} \OSS{}  (\textit{bottom panel}) from a DEEP survey. The color code indicates our predictions for the number of photometric detections. For the \textit{SPICA} panel we took into consideration the SMI-CAM and the three B-BOP channel, while the dashed area represents the sources which can be spectroscopically detected with a SMI-LR-like instrument ($5\,\sigma$ detection with a resolution R$=50-120$ in the $17-36\,\mu$m wavelength range). For the \OST{} panel, we consider the six \OSS{} bands in photometric mode (R=4), while the dashed line indicates the sources for which we can have R=300 spectra in at least one of the \OSS{} bands.}
  \label{fig:oss_spica_comp_deep} 
\end{figure}

\begin{figure}
  \centering
  \resizebox{0.95\hsize}{!}{\includegraphics{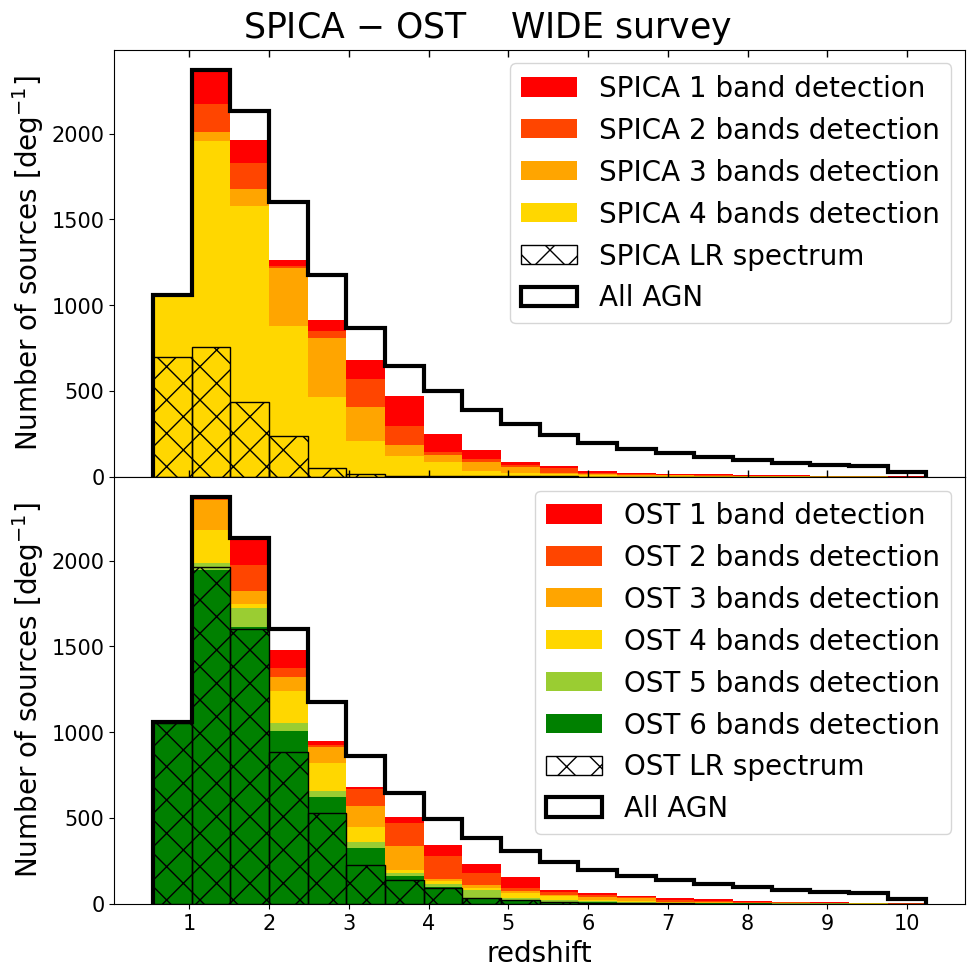}}
  \caption{Distribution of the number of AGN expected per deg$^2$ for a \textit{SPICA}-like (\textit{top panel}) and for the \OST{} \OSS{}  (\textit{bottom panel}) from a WIDE survey. The lines and areas are coded as in Figure~\ref{fig:oss_spica_comp_deep}.}
  \label{fig:oss_spica_comp_wide} 
\end{figure}

In figure~\ref{fig:oss_spica_comp_deep}, we show the comparison of the total number of AGN detected by a \textit{SPICA}-like observatory and by the \OST{} \OSS{} for a DEEP survey. The color code indicates the number of bands for which we can have photometric detection, considering SMI-CAM and the three B-BOP channel for \textit{SPICA} and the six \OSS{} bands for \OST{}. For $z<2$ both mission are able to detect almost all the sources in at least one band. While at $z>2$ \textit{SPICA} begins to lose a fraction of the sources, \OSS{} continues to detect them all up to $z\approx 3.5$.\par
Moreover, having a sufficient number of photometric points, would allow us to break the SF-AGN emission degeneracy and to estimate the AGN properties using SED-fitting decomposition techniques (jointly with additional instruments at different wavelengths) when spectra are not available. Although X-ray photometry and/or low-resolution spectroscopic data may not be used in SED fitting (but see X-CIGALE, \citealt{yang20}), X-ray information could be crucial to select the templates to fit in the IR. \par
 
\begin{figure}
  \centering
  \resizebox{\hsize}{!}{\includegraphics{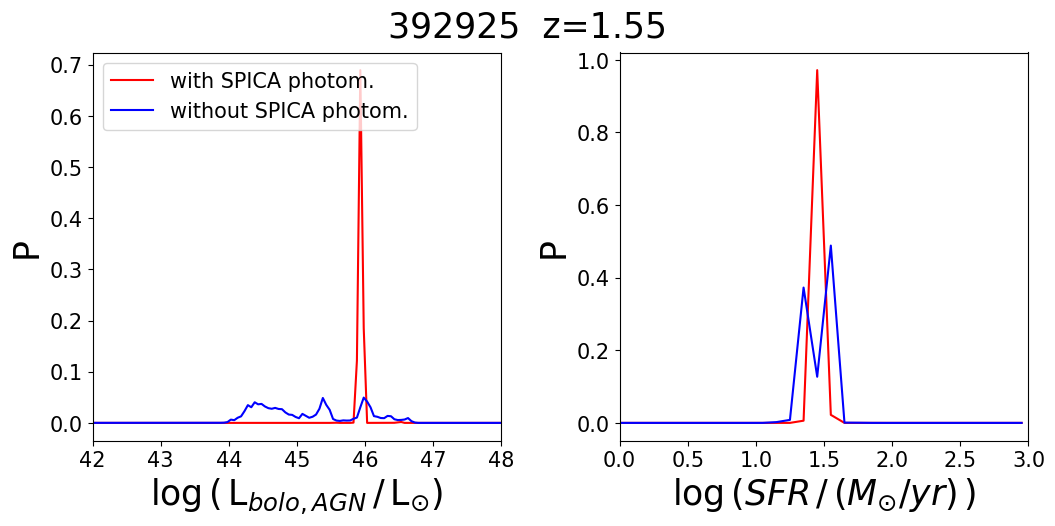}}
  \caption{SED-fitting probability distribution functions (PDFs) for AGN 392925 ($\log{ (\text{L}_{\text{x}}/\text{erg/s})}=45.0$, $\log{ (\text{N}_{\text{H}}/\text{cm}^{-2})}=24.1$) at $z=1.55$. \textit{Left panel}: AGN bolometric luminosity; \textit{right panel}: SFR in the last 0.1 Gyr. The blue lines are obtained from the SED-fitting using 33 photometric bands from optical to far-IR (in particular, from IR telescope \textit{Spitzer} and \textit{Herschel}). The red lines are the PDF obtained using the same 33 photometric bands plus the four simulated photometric observations of \textit{SPICA SMI-CAM}, \textit{B-BOP1}, \textit{B-BOP2} and \textit{B-BOP3}. The use of the additional \textit{SPICA} photometric points provides better constraints on the AGN bolometric power and, overall, allows us to properly disentangle the AGN and the SF emission.}
  \label{fig:pdf_improvement}
\end{figure}

As a quick test for this statement, we simulated \textit{SPICA}-like observations and used \texttt{sed3fit} to recover the stellar mass, SFR and AGN bolometric luminosity. In particular, we selected a random sample of seventy sources among the $\approx 550$ of the \citet{lanzuisi17} compilation that we used in this work and exploit \texttt{sed3fit} with up to 33 \citet{laigle16} photometric data (from optical to sub-mm including \textsl{Spitzer} and \textsl{Herschel}, although not all the photometric points were available for all the sources) to obtain the AGN, host-galaxy and total SED. We used the total SED to estimate the \textit{SPICA SMI} and \textit{B-BOP} fluxes, added these four photometric points to the previous 33 and used all of them in new runs of \texttt{sed3fit}. We focused on the width of the $84^{\textit{th}}-16^{\textit{th}}$ percentile of the probability distribution functions (PDF) from \texttt{sed3fit} to check for the improvement provided by the use of \textit{SPICA}-like data. We found that for, respectively, 50 and 45 of the 70 sources investigated we have narrower AGN bolometric luminosity and SFR PDFs. The median SF relative error (i.e. $0.5\times(SFR_{84^{\textit{th}}}-SFR_{16^{\textit{th}}})/SFR_{best}$) improves from $11\,\%$ to $1\,\%$ and that of the AGN L$_{bol}$ from $61\,\%$ to $29\,\%$.
The improvements in the stellar mass are smaller, since \texttt{sed3fit} estimates this parameter exploiting the optical wavelength band, therefore we get only a minor advantage in using more IR data. The major improvements come from the cases where the AGN-SF degeneracies are stronger; as we can see from Figure~\ref{fig:pdf_improvement}, the additional mid- and far-IR points provided by a \textit{SPICA}-like observatory help in disentangling the two contributions and allow us to obtain better estimates of both the host-galaxy and AGN parameters (e.g. stellar mass, SFR, AGN L$_{bolo}$).

\subsection{CT-AGN}
\begin{figure*}[h]
  \centering
  \resizebox{0.75\hsize}{!}{\includegraphics{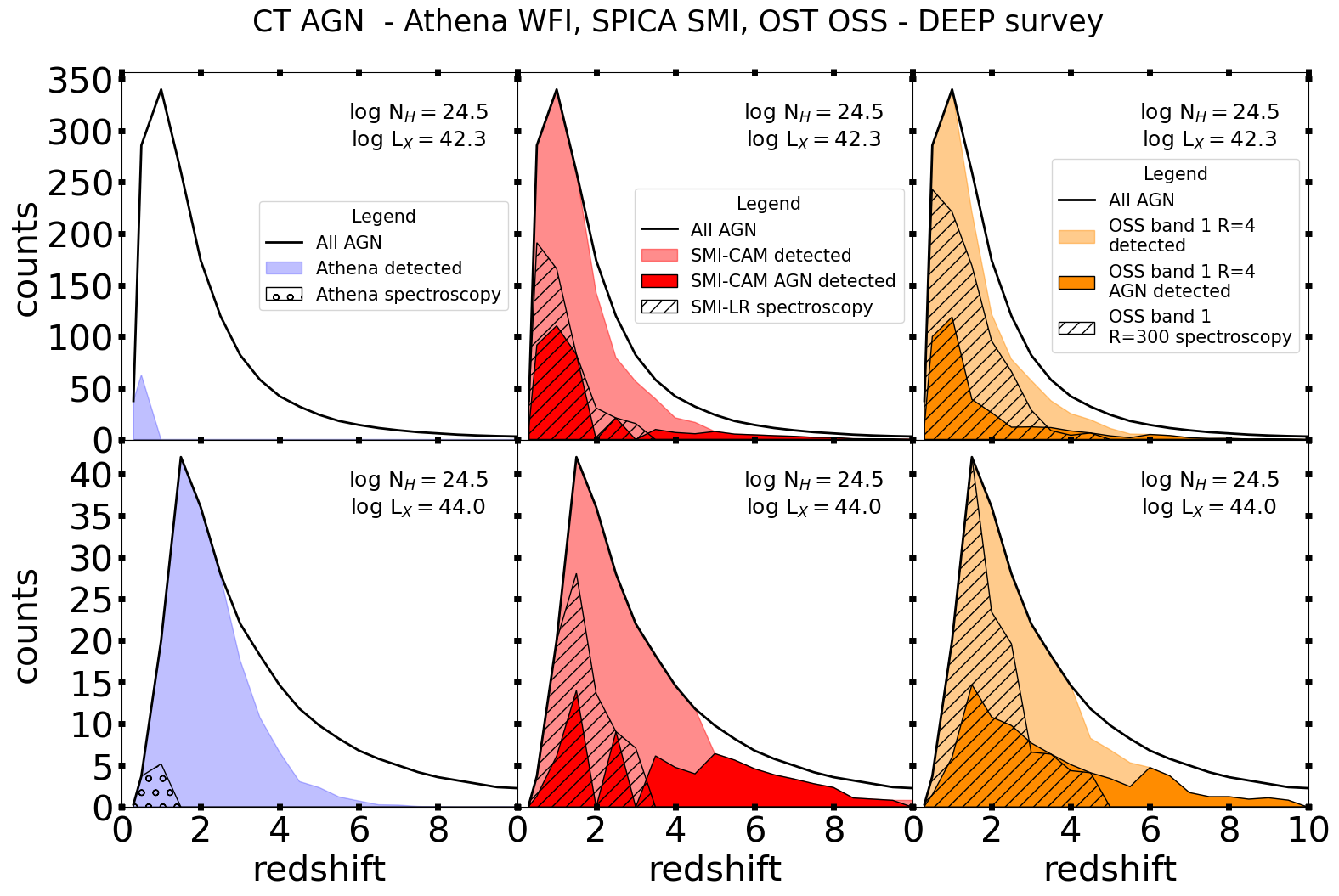}}
  \caption{Number of CT AGN expected per deg$^2$ and per $\Delta z =1$ from a DEEP survey with \textsl{Athena} (\textit{left panels}), with a \textit{SPICA}-like observatory (\textit{middle panels}), and with the \OST{} \OSS{}(\textit{right panels}). The black lines are the total number of expected CT-AGN, the blue areas rapresent those which can be detected in the X-ray by \textsl{Athena WFI} and the dotted areas those for which we will be able to recover the $\text{L}_{\text{x}}$ and $\text{N}_{\text{H}}$ (with $30\,\%$ uncertanties) using \textsl{Athena WFI} spectroscopy; the red areas represent those which can be detected in the IR with the \textit{SPICA} SMI-CAM at 34$\mu$m. The light red areas indicate that the host galaxy has the largest contribution to the flux in the SMI-CAM detection, while the dark red areas that the AGN is the dominant component. The dashed areas indicate that we will be able to have also mid-IR spectroscopic $5\,\sigma$ detection with \textit{SPICA} SMI-LR with a resolution R$=50-120$ in the $17-36\,\mu$m wavelength range. The light orange areas are the sources for which we expect detection in the \OSS{} band 1 ($25-44\,\mu$m) in photometric mode (R=4), the dark orange area those for which the AGN is the dominant component of the detected emission. The dashed areas refer to AGN for which we will have enough flux to have low-resolution (R=300) spectra in \OSS{} band 1. For the \OST{}, we showed only the prediction for the band 1 of the \OSS, but the instrument has five other bands (up to $588\,\mu$m), for which we predict a larger number of AGN detected both in photometry and in spectrometry. The upper row refers to AGN with luminosity $42.0\le \log{(\text{L}_{\text{x}}/\text{erg/s})}  <42.5 $, while the lower to those with $43.9\le \log{(\text{L}_{\text{x}}/\text{erg/s})}  <44.2 $.}
  \label{fig:CT_AGN} 
\end{figure*}

The search for CT-AGN is of fundamental importance in view of the still open questions of the missing sources responsible for the XRB and the galaxy co-evolution paradigm. About the former issue, X-ray background synthesis models predict a putative population of CT-AGN challenging to detect with current facilities \citep{gilli07, gilli13, harrison16}. In addition, the black-hole-galaxy co-evolution paradigm states that the scarcely known first phases of BH (and galaxy) growth should be associated with very obscured AGN activity and obscured SF, therefore these elusive AGN are of the foremost importance to search for (e.g. \citealt{silk98, dimatteo05, lamastra13}).\par
The adopted model for the XRB predicts that low-luminosity CT-AGN form the bulk of the very obscured AGN population. However, due to their large obscuration, it is difficult to detect them with X-ray instruments, while a \textit{SPICA}-like (or \OST) observatory should be able to effectively sample them with a detection fraction as high as $90 \%$ in the DEEP survey and $70\,\%$ in the WIDE survey. \par
As we can see from Figure~\ref{fig:CT_AGN}, low-luminosity obscured AGN are difficult to detect even with the planned \textsl{Athena} facility. In fact, we estimated that \textsl{Athena} will be able to detect (light blue)  more than half of all the high-luminosity CT-AGN, while the major part of the CT-AGN will be missed ($\sim 20\,\%$ detection fraction), being low-luminosity sources ($\sim 4260$ of the 5200 CT-AGN per deg$^2$ have $\log{(\text{L}_{\text{x}}/\text{erg$\,$s$^{-1}$})} < 43.5$). The characterization of these sources with \textsl{Athena WFI} (dotted area) will be challenging due to their low statistics: we may be able to recover $\text{L}_{\text{x}}$ and $\text{N}_{\text{H}}$ only for $\sim 6\,\%$ of the high-luminosity ones and $<1\,\%$ for the low-luminosity. Sufficiently powerful IR surveys will be capable of detecting the reprocessed emission due to accretion at IR wavelengths. For example, a \textit{SPICA}-like DEEP survey is capable of detecting (red areas) $\approx 90\,\%$ of the CT-AGN and provide low-resolution spectra (shaded areas) for one third of them. The even more powerful \OST{} will allow us to detect photometrically (R=4) almost all the CT-AGN and to have R=300 spectra for at least half of them. Spectroscopic follow-up with other IR instruments will be fundamental in identifying CT-AGN with photometric detection but without a clear detection with \textsl{SMI-LR}-like or \OSS{} instruments. Low-luminosity (high-luminosity) CT-AGN can be spectroscopically identified with a \textit{SAFARI}-like instrument up to $z\sim 1.5$ ($z\sim 2.5$) and $z\sim 2.5$ ($z > 4$) via the [Ne$\,\textsc{v}$]$24.3\mu$m and the [O$\,\textsc{iv}$]$25.9\mu$m lines with 5 hours observations \citep{spinoglio21}. Longer exposures (>10 hours) will allow to identify CT-AGN beyond $z \sim 4$. Despite the fact that the X-ray detection of these CT objects will be difficult, the \textsl{Athena} contribution to their study remains essential. In fact, only X-ray spectra (for individual sources or at least stacked X-ray data in case of no detection) can confirm the AGN nature of these sources (i.e., a flat X-ray continuum plus a strong iron line are clear indicators of heavy obscuration).\par
\begin{figure}
  \centering
  \resizebox{\hsize}{!}{\includegraphics{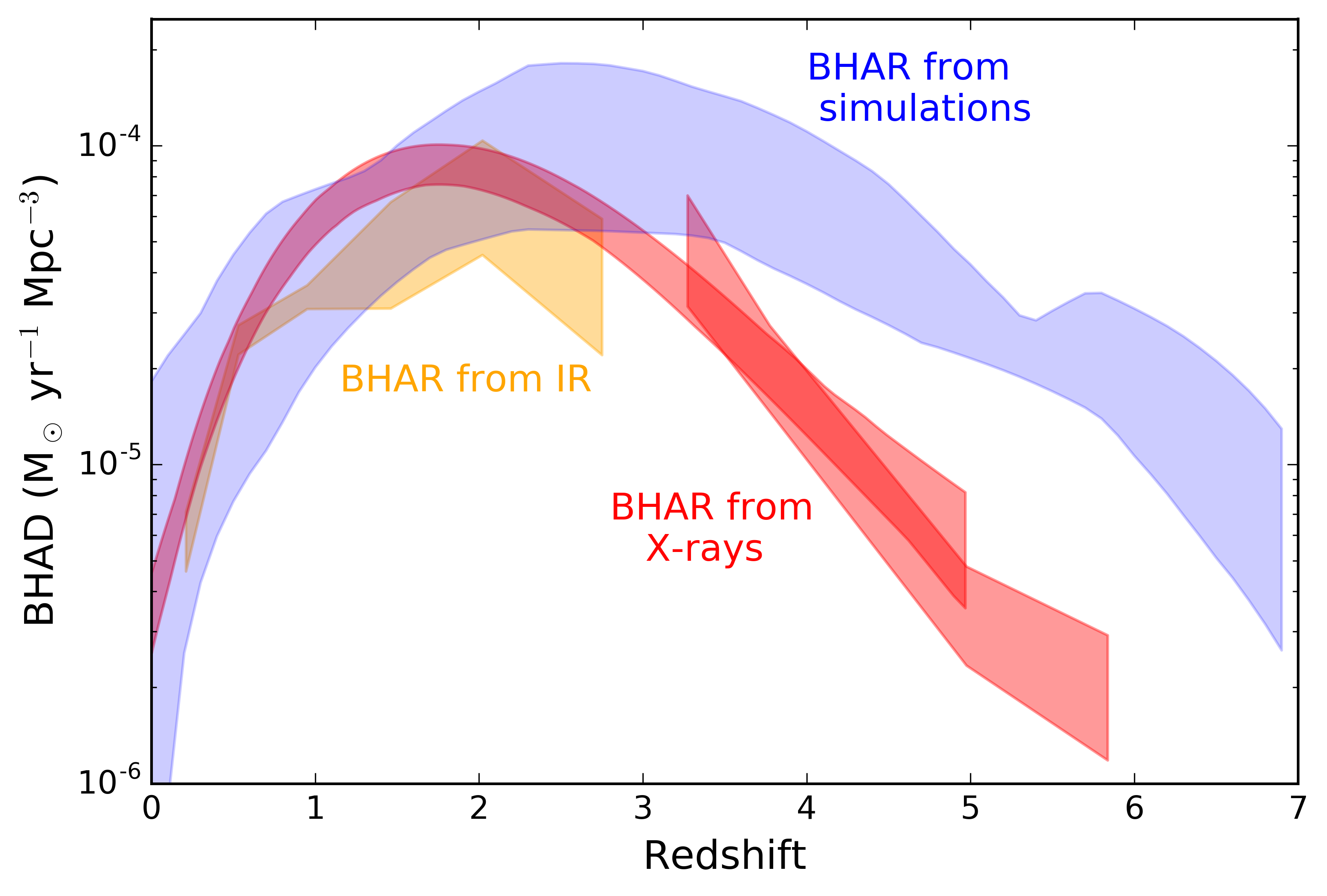}}
  \caption{Redshift evolution of the BHAD adapted from \citet{vito18}. The blue area is a theoretical BHAD curve obtained from \citet{volonteri16, sijacki15, shankar13}. The red areas are the X-ray-derived BHADs, obtained from observations, from \citet{aird15, ueda14, vito14}. The yellow area is the IR-derived BHAD from \citet{delvecchio14}. \textit{SPICA}-like or \OST{} missions will allow to sample the BHAD at higher $z$ than actual IR survey and will be able to detect very obscured CT-AGN, missed by X-ray surveys, that should contribute to the total BHAD.}
  \label{fig:BHAD} 
\end{figure}

For the most luminous ones, the combined use of \textsl{Athena} X-ray spectroscopy, mid-, and far-IR spectrophotometry (with SMI-CAM- and B-BOP-like instruments or with the \OSS) will provide never achieved before insights of the intrinsic AGN emission and bolometric power and opportunities to sample them at redshift higher than the cosmic noon. As shown in Figure~\ref{fig:BHAD}, at present, there are tensions between BHAD traced by X-rays, and those obtained by simulations, especially at $z > 2$. At  $z \sim 4$, the BHAD expected from simulations \citep{aird15,vito18} are $\sim 5$ times higher than those computed via X-ray surveys, while actual IR-surveys do not go beyond $z \sim 3$. Different arguments, mostly driven by observations in the IR \citep{gruppioni16, bisigello20}, predict a higher number of deeply obscured AGN, as for now undetected in the X-ray, that may alleviate this tension.\par

Exploiting all the capabilities of new generation IR cryogenic observatories (i.e. fast LR spectrophometry to detect and characterize low-redshift CT-AGN, spectroscopic follow-ups to identify those with only photometric detection and multi-wavelength SED-fitting exploting already observed deep fields to separate the AGN and host-galaxy emission) will allow us to discover the deeply obscured CT-AGN population up to very high redshift ($z > 4 $). The presence of such an obscured population does not violate the limit imposed by the spectral energy density of the XRB, since these sources will provide an almost negligible contribution in the X-ray band \citep{comastri15}. The updated picture, in terms of obscured accretion over cosmic time, that the IR missions discussed in this paper will lay out is fundamental in providing inputs to the simulations, which are still affected by considerable uncertainties and depend strongly on the adopted assumptions (e.g., \citealt{gonzalez11, jaacks12, sijacki15, shankar13, volonteri16, thomas19}).

\section{Conclusions}\label{sec:conclusion}
In this work we have investigated the capabilities of IR deep and wide surveys to detect and characterize AGN, as well as the advantages resulting by the synergies with the future X-ray observatory \textsl{Athena}. We used XRB synthesis models to predict the total number of AGN as a function of intrinsic X-ray luminosity, amount of obscuration and redshift. A sample of more than 500 AGN from the COSMOS field with both X-ray spectra and optical-to-FIR SED-fitting has been used to investigate the AGN mid- and far-IR emission and to predict the fraction of AGN that will be detected by instrument similar to \textit{SPICA SMI}, \textit{B-BOP} and the \OST{} \OSS. We compared the estimated total number of AGN with the fraction of those that will be detected by \textsl{Athena}, by a \textit{SPICA}-like observatory and by the \OST, with the goal of enlightening the synergies between X-ray and IR surveys in terms of detection and characterization of AGN (in particular, of the most obscured ones) up to high redshift (i.e. $z\sim 10$) and down to low luminosity (i.e. $\log{(\text{L}_{\text{x}}/\text{erg/s})} \sim 42$). In the IR bands, we also differentiated if we were primary detecting either the AGN or the host-galaxy emission. Moreover, we investigated if we were able to spectroscopically identify and characterize the AGN emission with \textit{SMI-LR}-like instruments or exploiting the full R=300 spectral resolution of the \OSS. In particular, we put emphasis on the elusive CT-AGN, a large fraction of those being missed by the current X-ray surveys, and that will require deep IR surveys for their detection and \textsl{Athena} spectra for their characterization. Our main results are the following:
\begin{enumerate}

\item An \OST{} WIDE survey will be able to photometrically detect $\sim 60\,\%$ of all the AGN in all the \OSS{} bands and more than $\sim 80$ in at least one of the bands, while the DEEP survey will detect at least $95\, \%$ of all the AGN at $z \le 10$ in at least one of the \OSS{} bands and more than $\sim 80\,\%$ in all of them.

\item Considering  an \textsl{Athena} and \textit{SPICA}-like DEEP survey, we estimate that we can have both IR and X-ray detections for half of all the AGN and for more than the $80\,\%$ of the high-luminosity ($\log{(\text{L}_{\text{x}}/\text{erg/s})}\ge 43.5$) AGN; for these sources just the \textsl{Athena} detection will assure us of the presence of AGN activity, even when the mid-IR torus contribution is diluted into the host-galaxy emission. The \textsl{Athena WFI} spectral capabilities should also provide X-ray spectra, hence a reliable way to characterize the AGN, for thousands of AGN. This will allow us to estimate the AGN bolometric power and, having such a large number of sources with data in both the wavelength bands, to improve the AGN L$_{\text{mid-IR}}-\text{L}_{\text{x}}$ and $k_{bol}-\text{L}_{bol}$ relations. Moreover, the use of low-resolution spectra (such as those produce by a SMI-LR-like spectrometer), or alternatively, of mid- and far-IR photometric points (available with SMI-CAM- and B-BOP-like instruments) and SED-fitting decomposition technique, will help in constraining the AGN properties even in case of low-statistic X-ray spectra, like those we expect for very obscured and CT-AGN. However, even for sources with low photon statistics, the detection of a flat continuum and/or a strong iron emission line in the \textsl{Athena WFI} spectra will be invaluable clues to identify these sources as very obscured AGN.

\item The sinergies of \textsl{Athena} with the \OST{} are similar to those with a \textit{SPICA}-like mission, but with the advantages of having up to six photometric point in the far-IR, a higher fraction (with respect to a \textit{SPICA}-like mission) of sources with mid-IR spectra and providing R$=300$, $25-588\,\mu$m spectra for $\sim 50\,\%$ of all the AGN. 
 
\item The CT-AGN, for which the X-ray detection and characterization are challenging due to their high amount of obscuration, will be very well sampled by deep IR surveys. A \textit{SPICA}-like  DEEP survey should detect $86\,\%$ of the low-luminosity ($\log{(\text{L}_{\text{x}}/\text{erg/s})} < 43.5$) CT-AGN at $z \le 10$, while, \OST{} can provide with six mid- and far-IR photometric detections for a similar fraction of the AGN, and at least one band detection for more than $\sim 95\,\%$ of all the CT AGN at $z<10$. These kinds of missions would allow us to sample higher redshifts than what we are actually capable of at IR wavelengths, and detecting very obscured CT-AGN missed by X-ray surveys, to provide clues and possibly inputs to BHAD simulations. However, for most of these sources, the major contributor to the flux comes from the host galaxy emission, and the challenge will be to recognise them as AGN. In this context, pointed observations, such as those provided by \textit{SPICA SAFARI}-like or \OSS{} follow up, can be decisive in detecting the [Ne$\,\textsc{v}$]$24.3\mu$m and the [O$\,\textsc{iv}$]$25.9\mu$m lines, which would allow us to unambiguously classify the source as an AGN.

\item For high-luminosity CT-AGN, we will have X-ray detection with \textsl{Athena} up to $z \sim 3$. This will allow us to to ascertain the AGN nature of these obscured sources and will provide vital input for the selection of IR SED templates and for their spectroscopic characterization.

\item The use of mid- and far-IR photometric observations (such as those produced by SMI- and B-BOP-like instruments or by the \OST{} \OSS{} in R=4 photometric mode) can provide up to four (or six in the case of the \OSS) photometric data points, which will be unique to improve the reliability of SED-fitting techniques to estimate the AGN properties up and beyond the cosmic noon.

\item Mid-IR spectra are very helpful to recognize AGN and to separate their emission from the SF-related one. A \textit{SPICA}-like observatory can provide LR spectra (with the SMI-LR instrument) for the majority of the detected objects at $z~\le~2$, and up to $z\approx 4$ for the most luminous sources, as well as follow-up deep spectroscopy with SAFARI for galaxies up to and beyond $z \sim 4$. On the other hand the \OST{} \OSS{} can deliver R$=300$, $25-588\,\mu$m spectra for $\sim 50\,\%$ and $\sim 80\,\%$ of all the AGN with, respectively, its  WIDE and DEEP surveys, the last one up to and beyond $z \approx 4$. The main spectral features to focus on will be the host-galaxy PAH lines and the AGN $9.7\,\mu$m silicate feature. We showed that, for what concerns the PAHs, those can be used to quickly recover the source spectroscopic redshift, without requiring specific higher-resolution follow-ups, while the silicate feature, when it is in absorption, jointly with X-ray spectral information, will allow us to assess and quantify the amount of obscuration. An \textsl{Athena} DEEP (WIDE) survey will provide X-ray spectral characterization for $\approx 20\,\%$ ($\approx 6\,\%$) of all the AGN and up to $\approx 50\,\%$ ($\approx 21\,\%$) of the high-luminosity ones, and, with the help of a multi-wavelengths approach, we will be able to characterize the properties even for X-ray low-statistic sources, such as CT-AGN. 

\end{enumerate}

Mid- and far-IR deep surveys, such as those of a \textit{SPICA}-like mission or of the \OST{}, will allow us to fully exploit the potential of \textsl{Athena} for the measurement of the full Black-Hole Accretion History, back to an age of the Universe of $\sim 1\,$Gyr. This, along with the Star Formation History, will give us a much better understanding of galaxy evolution as compared to what we know today.

\begin{acknowledgements}
This paper is dedicated to the memory of Bruce Swinyard, who initiated the SPICA project in Europe, but unfortunately died on 22 May 2015 at the age of 52. He was ISO-LWS calibration scientist, \textit{Herschel-SPIRE} instrument scientist, first European PI of \textit{SPICA} and first design lead of SAFARI.\par
We acknowledge the whole \textit{SPICA} Collaboration Team.\par
FP, CV, CG, LB, and LS acknowledge financial support by the Agenzia Spaziale Italiana (ASI) under the research contract 2018-31-HH.0.
FJC acknowledges financial support from the Spanish Ministry MCIU under project RTI2018-096686-B-C21 (MCIU/AEI/FEDER/UE), cofunded by FEDER funds and from the Agencia Estatal de Investigaci\'on, Unidad de Excelencia Mar\'ia de Maeztu, ref. MDM-2017-0765.
MPS acknowledges support from the Comunidad de Madrid, Spain, through Atracci\'on de Talento Investigador Grant 2018-T1/TIC-11035 and PID2019-105423GA-I00 (MCIU/AEI/FEDER,UE).
AF acknowledges the support from grant PRIN MIUR2017-20173ML3WW\_001.\par
We also thank the \textit{SPICA} Science Study Team appointed by ESA and the \textit{SPICA} Galaxy Evolution Working Group. We should also thank here the Athena Science Study Team, also appointed by ESA, the Athena Science Working Groups Formation and growth of earliest SMBH and Understanding the build-up of SMBH and galaxies, and the Athena Community Office

\end{acknowledgements}

\bibliographystyle{aa.bst} 
\bibliography{biblio.bib} 

\begin{thebibliography}{142}
\expandafter\ifx\csname natexlab\endcsname\relax\def\natexlab#1{#1}\fi

\bibitem[{{Aird} {et~al.}(2015){Aird}, {Alexander}, {Ballantyne}, {Civano},
  {Del-Moro}, {Hickox}, {Lansbury}, {Mullaney}, {Bauer}, {Brand t}, {Comastri},
  {Fabian}, {Gandhi}, {Harrison}, {Luo}, {Stern}, {Treister}, {Zappacosta},
  {Ajello}, {Assef}, {Balokovi{\'c}}, {Boggs}, {Brightman}, {Christensen},
  {Craig}, {Elvis}, {Forster}, {Grefenstette}, {Hailey}, {Koss}, {LaMassa},
  {Madsen}, {Puccetti}, {Saez}, {Urry}, {Wik}, \& {Zhang}}]{aird15}
{Aird}, J., {Alexander}, D.~M., {Ballantyne}, D.~R., {et~al.} 2015, \apj, 815,
  66

\bibitem[{{Aird} {et~al.}(2010){Aird}, {Nandra}, {Laird}, {Georgakakis},
  {Ashby}, {Barmby}, {Coil}, {Huang}, {Koekemoer}, {Steidel}, \&
  {Willmer}}]{aird10}
{Aird}, J., {Nandra}, K., {Laird}, E.~S., {et~al.} 2010, \mnras, 401, 2531

\bibitem[{{Ananna} {et~al.}(2019){Ananna}, {Treister}, {Urry}, {Ricci},
  {Kirkpatrick}, {LaMassa}, {Buchner}, {Civano}, {Tremmel}, \&
  {Marchesi}}]{ananna19}
{Ananna}, T.~T., {Treister}, E., {Urry}, C.~M., {et~al.} 2019, \apj, 871, 240

\bibitem[{Andr\'e {et~al.}(2019)Andr\'e, Hughes, Guillet, Boulanger, Bracco,
  Ntormousi, Arzoumanian, Maury, Bernard, Bontemps, Ristorcelli, Girart, Motte,
  Tassis, Pantin, Montmerle, Johnstone, Gabici, Efstathiou, Basu, Béthermin,
  Beuther, Braine, Francesco, Falgarone, Ferrière, Fletcher, Galametz, Giard,
  Hennebelle, Jones, Kepley, Kwon, Lagache, Lesaffre, Levrier, Li, Li, Mao,
  Nakagawa, Onaka, Paladino, Peretto, Poglitsch, Revéret, Rodriguez, Sauvage,
  Soler, Spinoglio, Tabatabaei, Tritsis, van~der Tak, Ward-Thompson,
  Wiesemeyer, Ysard, \& Zhang}]{andre19}
Andr\'e, P., Hughes, A., Guillet, V., {et~al.} 2019, Probing the cold
  magnetized Universe with SPICA-POL (B-BOP)

\bibitem[{{Armus} {et~al.}(2006){Armus}, {Bernard-Salas}, {Spoon}, {Marshall},
  {Charmandaris}, {Higdon}, {Desai}, {Hao}, {Teplitz}, {Devost}, {Brandl},
  {Soifer}, \& {Houck}}]{armus06}
{Armus}, L., {Bernard-Salas}, J., {Spoon}, H.~W.~W., {et~al.} 2006, \apj, 640,
  204

\bibitem[{{Armus} {et~al.}(2007){Armus}, {Charmandaris}, {Bernard-Salas},
  {Spoon}, {Marshall}, {Higdon}, {Desai}, {Teplitz}, {Hao}, {Devost}, {Brandl},
  {Wu}, {Sloan}, {Soifer}, {Houck}, \& {Herter}}]{armus07}
{Armus}, L., {Charmandaris}, V., {Bernard-Salas}, J., {et~al.} 2007, \apj, 656,
  148

\bibitem[{{Arnaud}(1996)}]{xspec}
{Arnaud}, K.~A. 1996, in Astronomical Society of the Pacific Conference Series,
  Vol. 101, Astronomical Data Analysis Software and Systems V, ed. G.~H.
  {Jacoby} \& J.~{Barnes}, 17

\bibitem[{{Assef} {et~al.}(2013){Assef}, {Stern}, {Kochanek}, {Blain},
  {Brodwin}, {Brown}, {Donoso}, {Eisenhardt}, {Jannuzi}, {Jarrett}, {Stanford},
  {Tsai}, {Wu}, \& {Yan}}]{assef13}
{Assef}, R.~J., {Stern}, D., {Kochanek}, C.~S., {et~al.} 2013, \apj, 772, 26

\bibitem[{{Ballantyne} {et~al.}(2011){Ballantyne}, {Draper}, {Madsen}, {Rigby},
  \& {Treister}}]{ballantyne11}
{Ballantyne}, D.~R., {Draper}, A.~R., {Madsen}, K.~K., {Rigby}, J.~R., \&
  {Treister}, E. 2011, \apj, 736, 56

\bibitem[{{Barai} {et~al.}(2018){Barai}, {Gallerani}, {Pallottini}, {Ferrara},
  {Marconi}, {Cicone}, {Maiolino}, \& {Carniani}}]{Barai18}
{Barai}, P., {Gallerani}, S., {Pallottini}, A., {et~al.} 2018, \mnras, 473,
  4003

\bibitem[{{Barcons} {et~al.}(2017){Barcons}, {Barret}, {Decourchelle}, {den
  Herder}, {Fabian}, {Matsumoto}, {Lumb}, {Nandra}, {Piro}, {Smith}, \&
  {Willingale}}]{barcons17}
{Barcons}, X., {Barret}, D., {Decourchelle}, A., {et~al.} 2017, Astronomische
  Nachrichten, 338, 153

\bibitem[{{Barret} {et~al.}(2018){Barret}, {Lam Trong}, {den Herder}, {Piro},
  {Cappi}, {Houvelin}, {Kelley}, {Mas-Hesse}, {Mitsuda}, {Paltani}, {Rauw},
  {Rozanska}, {Wilms}, {Bandler}, {Barbera}, {Barcons}, {Bozzo}, {Ceballos},
  {Charles}, {Costantini}, {Decourchelle}, {den Hartog}, {Duband}, {Duval},
  {Fiore}, {Gatti}, {Goldwurm}, {Jackson}, {Jonker}, {Kilbourne}, {Macculi},
  {Mendez}, {Molendi}, {Orleanski}, {Pajot}, {Pointecouteau}, {Porter},
  {Pratt}, {Pr{\^e}le}, {Ravera}, {Sato}, {Schaye}, {Shinozaki}, {Thibert},
  {Valenziano}, {Valette}, {Vink}, {Webb}, {Wise}, {Yamasaki}, {Douchin},
  {Mesnager}, {Pontet}, {Pradines}, {Branduardi-Raymont}, {Bulbul}, {Dadina},
  {Ettori}, {Finoguenov}, {Fukazawa}, {Janiuk}, {Kaastra}, {Mazzotta},
  {Miller}, {Miniutti}, {Naze}, {Nicastro}, {Scioritino}, {Simonescu},
  {Torrejon}, {Frezouls}, {Geoffray}, {Peille}, {Aicardi}, {Andr{\'e}},
  {Daniel}, {Cl{\'e}net}, {Etcheverry}, {Gloaguen}, {Hervet}, {Jolly}, {Ledot},
  {Paillet}, {Schmisser}, {Vella}, {Damery}, {Boyce}, {Dipirro}, {Lotti},
  {Schwander}, {Smith}, {Van Leeuwen}, {van Weers}, {Clerc}, {Cobo}, {Dauser},
  {Kirsch}, {Cucchetti}, {Eckart}, {Ferrando}, \& {Natalucci}}]{barret18}
{Barret}, D., {Lam Trong}, T., {den Herder}, J.-W., {et~al.} 2018, in Society
  of Photo-Optical Instrumentation Engineers (SPIE) Conference Series, Vol.
  10699, Space Telescopes and Instrumentation 2018: Ultraviolet to Gamma Ray,
  106991G

\bibitem[{{Battersby} {et~al.}(2018){Battersby}, {Armus}, {Bergin}, {Kataria},
  {Meixner}, {Pope}, {Stevenson}, {Cooray}, {Leisawitz}, {Scott}, {Bauer},
  {Bradford}, {Ennico}, {Fortney}, {Kaltenegger}, {Melnick}, {Milam},
  {Narayanan}, {Padgett}, {Pontoppidan}, {Roellig}, {Sandstrom}, {Su},
  {Vieira}, {Wright}, {Zmuidzinas}, {Staguhn}, {Sheth}, {Benford}, {Mamajek},
  {Neff}, {Carey}, {Burgarella}, {De Beck}, {Gerin}, {Helmich}, {Moseley},
  {Sakon}, \& {Wiedner}}]{battersby18}
{Battersby}, C., {Armus}, L., {Bergin}, E., {et~al.} 2018, Nature Astronomy, 2,
  596

\bibitem[{{Bavdaz} {et~al.}(2018){Bavdaz}, {Wille}, {Ayre}, {Ferreira},
  {Shortt}, {Fransen}, {Collon}, {Vacanti}, {Barri{\`e}re}, {Landgraf},
  {Sforzini}, {Booysen}, {van Baren}, {Zuknik}, {Della Monica Ferreira},
  {Massahi}, {Christensen}, {Krumrey}, {M{\"u}ller}, {Burwitz}, {Pareschi},
  {Spiga}, {Valsecchi}, {Vernani}, {Oliver}, \& {Seidel}}]{bavdaz18}
{Bavdaz}, M., {Wille}, E., {Ayre}, M., {et~al.} 2018, in Society of
  Photo-Optical Instrumentation Engineers (SPIE) Conference Series, Vol. 10699,
  Space Telescopes and Instrumentation 2018: Ultraviolet to Gamma Ray, 106990X

\bibitem[{{Berta} {et~al.}(2013){Berta}, {Lutz}, {Santini}, {Wuyts}, {Rosario},
  {Brisbin}, {Cooray}, {Franceschini}, {Gruppioni}, {Hatziminaoglou}, {Hwang},
  {Le Floc'h}, {Magnelli}, {Nordon}, {Oliver}, {Page}, {Popesso}, {Pozzetti},
  {Pozzi}, {Riguccini}, {Rodighiero}, {Roseboom}, {Scott}, {Symeonidis},
  {Valtchanov}, {Viero}, \& {Wang}}]{berta13}
{Berta}, S., {Lutz}, D., {Santini}, P., {et~al.} 2013, \aap, 551, A100

\bibitem[{{Bisigello} {et~al.}(2020){Bisigello}, {Gruppioni}, {Feltre},
  {Calura}, {Pozzi}, {Vignali}, {Barchiesi}, {Rodighiero}, \&
  {Negrello}}]{bisigello20}
{Bisigello}, L., {Gruppioni}, C., {Feltre}, A., {et~al.} 2020, arXiv e-prints,
  arXiv:2011.07074

\bibitem[{{Bonato} {et~al.}(2019){Bonato}, {De Zotti}, {Leisawitz}, {Negrello},
  {Massardi}, {Baronchelli}, {Cai}, {Bradford}, {Pope}, {Murphy}, {Armus}, \&
  {Cooray}}]{bonato19}
{Bonato}, M., {De Zotti}, G., {Leisawitz}, D., {et~al.} 2019, \pasa, 36, e017

\bibitem[{{Brightman} \& {Nandra}(2011)}]{brightman11}
{Brightman}, M. \& {Nandra}, K. 2011, \mnras, 413, 1206

\bibitem[{{Brightman} \& {Ueda}(2012)}]{brightman12}
{Brightman}, M. \& {Ueda}, Y. 2012, \mnras, 423, 702

\bibitem[{{Burlon} {et~al.}(2011){Burlon}, {Ajello}, {Greiner}, {Comastri},
  {Merloni}, \& {Gehrels}}]{burlon11}
{Burlon}, D., {Ajello}, M., {Greiner}, J., {et~al.} 2011, \apj, 728, 58

\bibitem[{{Cash}(1979)}]{cash79}
{Cash}, W. 1979, \apj, 228, 939

\bibitem[{{Castell{\'o}-Mor} {et~al.}(2013){Castell{\'o}-Mor}, {Carrera},
  {Alonso-Herrero}, {Mateos}, {Barcons}, {Ranalli}, {P{\'e}rez-Gonz{\'a}lez},
  {Comastri}, {Vignali}, \& {Georgantopoulos}}]{castellomor13}
{Castell{\'o}-Mor}, N., {Carrera}, F.~J., {Alonso-Herrero}, A., {et~al.} 2013,
  \aap, 556, A114

\bibitem[{{Civano} {et~al.}(2016){Civano}, {Marchesi}, {Comastri}, {Urry},
  {Elvis}, {Cappelluti}, {Puccetti}, {Brusa}, {Zamorani}, {Hasinger},
  {Aldcroft}, {Alexand er}, {Allevato}, {Brunner}, {Capak}, {Finoguenov},
  {Fiore}, {Fruscione}, {Gilli}, {Glotfelty}, {Griffiths}, {Hao}, {Harrison},
  {Jahnke}, {Kartaltepe}, {Karim}, {LaMassa}, {Lanzuisi}, {Miyaji}, {Ranalli},
  {Salvato}, {Sargent}, {Scoville}, {Schawinski}, {Schinnerer}, {Silverman},
  {Smolcic}, {Stern}, {Toft}, {Trakhtenbrot}, {Treister}, \&
  {Vignali}}]{civano16}
{Civano}, F., {Marchesi}, S., {Comastri}, A., {et~al.} 2016, \apj, 819, 62

\bibitem[{{Comastri} {et~al.}(2015){Comastri}, {Gilli}, {Marconi}, {Risaliti},
  \& {Salvati}}]{comastri15}
{Comastri}, A., {Gilli}, R., {Marconi}, A., {Risaliti}, G., \& {Salvati}, M.
  2015, \aap, 574, L10

\bibitem[{{Comastri} {et~al.}(2011){Comastri}, {Ranalli}, {Iwasawa}, {Vignali},
  {Gilli}, {Georgantopoulos}, {Barcons}, {Brand t}, {Brunner}, {Brusa},
  {Cappelluti}, {Carrera}, {Civano}, {Fiore}, {Hasinger}, {Mainieri},
  {Merloni}, {Nicastro}, {Paolillo}, {Puccetti}, {Rosati}, {Silverman},
  {Tozzi}, {Zamorani}, {Balestra}, {Bauer}, {Luo}, \& {Xue}}]{comastri11}
{Comastri}, A., {Ranalli}, P., {Iwasawa}, K., {et~al.} 2011, \aap, 526, L9

\bibitem[{{Comastri} {et~al.}(1995){Comastri}, {Setti}, {Zamorani}, \&
  {Hasinger}}]{comastri95}
{Comastri}, A., {Setti}, G., {Zamorani}, G., \& {Hasinger}, G. 1995, \aap, 296,
  1

\bibitem[{{Del Moro} {et~al.}(2017){Del Moro}, {Alexander}, {Bauer}, {Daddi},
  {Kocevski}, {Stanley}, \& {McIntosh}}]{delmoro17}
{Del Moro}, A., {Alexander}, D.~M., {Bauer}, F.~E., {et~al.} 2017, Frontiers in
  Astronomy and Space Sciences, 4, 67

\bibitem[{{Del Moro} {et~al.}(2014){Del Moro}, {Mullaney}, {Alexander},
  {Comastri}, {Bauer}, {Treister}, {Stern}, {Civano}, {Ranalli}, {Vignali},
  {Aird}, {Ballantyne}, {Balokovi{\'c}}, {Boggs}, {Brandt}, {Christensen},
  {Craig}, {Gandhi}, {Gilli}, {Hailey}, {Harrison}, {Hickox}, {LaMassa},
  {Lansbury}, {Luo}, {Puccetti}, {Urry}, \& {Zhang}}]{delmoro14}
{Del Moro}, A., {Mullaney}, J.~R., {Alexander}, D.~M., {et~al.} 2014, \apj,
  786, 16

\bibitem[{{Delvecchio} {et~al.}(2014){Delvecchio}, {Gruppioni}, {Pozzi},
  {Berta}, {Zamorani}, {Cimatti}, {Lutz}, {Scott}, {Vignali}, {Cresci},
  {Feltre}, {Cooray}, {Vaccari}, {Fritz}, {Le Floc'h}, {Magnelli}, {Popesso},
  {Oliver}, {Bock}, {Carollo}, {Contini}, {Le F{\'e}vre}, {Lilly}, {Mainieri},
  {Renzini}, \& {Scodeggio}}]{delvecchio14}
{Delvecchio}, I., {Gruppioni}, C., {Pozzi}, F., {et~al.} 2014, \mnras, 439,
  2736

\bibitem[{{Delvecchio} {et~al.}(2015){Delvecchio}, {Lutz}, {Berta}, {Rosario},
  {Zamorani}, {Pozzi}, {Gruppioni}, {Vignali}, {Brusa}, {Cimatti}, {Clements},
  {Cooray}, {Farrah}, {Lanzuisi}, {Oliver}, {Rodighiero}, {Santini}, \&
  {Symeonidis}}]{delvecchio15}
{Delvecchio}, I., {Lutz}, D., {Berta}, S., {et~al.} 2015, \mnras, 449, 373

\bibitem[{{Di Matteo} {et~al.}(2005){Di Matteo}, {Springel}, \&
  {Hernquist}}]{dimatteo05}
{Di Matteo}, T., {Springel}, V., \& {Hernquist}, L. 2005, \nat, 433, 604

\bibitem[{Dowell {et~al.}(1998)Dowell, Hildebrand, Schleuning, Vaillancourt,
  Dotson, Novak, Renbarger, \& Houde}]{dowell98}
Dowell, C.~D., Hildebrand, R.~H., Schleuning, D.~A., {et~al.} 1998, The
  Astrophysical Journal, 504, 588

\bibitem[{{Duras} {et~al.}(2020){Duras}, {Bongiorno}, {Ricci}, {Piconcelli},
  {Shankar}, {Lusso}, {Bianchi}, {Fiore}, {Maiolino}, {Marconi}, {Onori},
  {Sani}, {Schneider}, {Vignali}, \& {La Franca}}]{duras20}
{Duras}, F., {Bongiorno}, A., {Ricci}, F., {et~al.} 2020, \aap, 636, A73

\bibitem[{{Elvis} {et~al.}(2009){Elvis}, {Civano}, {Vignali}, {Puccetti},
  {Fiore}, {Cappelluti}, {Aldcroft}, {Fruscione}, {Zamorani}, {Comastri},
  {Brusa}, {Gilli}, {Miyaji}, {Damiani}, {Koekemoer}, {Finoguenov}, {Brunner},
  {Urry}, {Silverman}, {Mainieri}, {Hasinger}, {Griffiths}, {Carollo}, {Hao},
  {Guzzo}, {Blain}, {Calzetti}, {Carilli}, {Capak}, {Ettori}, {Fabbiano},
  {Impey}, {Lilly}, {Mobasher}, {Rich}, {Salvato}, {Sand ers}, {Schinnerer},
  {Scoville}, {Shopbell}, {Taylor}, {Taniguchi}, \& {Volonteri}}]{elvis09}
{Elvis}, M., {Civano}, F., {Vignali}, C., {et~al.} 2009, \apjs, 184, 158

\bibitem[{{Fadda} {et~al.}(2010){Fadda}, {Yan}, {Lagache}, {Sajina}, {Lutz},
  {Wuyts}, {Frayer}, {Marcillac}, {Le Floc'h}, {Caputi}, {Spoon}, {Veilleux},
  {Blain}, \& {Helou}}]{fadda10}
{Fadda}, D., {Yan}, L., {Lagache}, G., {et~al.} 2010, \apj, 719, 425

\bibitem[{{Feltre} {et~al.}(2016){Feltre}, {Charlot}, \& {Gutkin}}]{feltre16}
{Feltre}, A., {Charlot}, S., \& {Gutkin}, J. 2016, \mnras, 456, 3354

\bibitem[{{Fern{\'a}ndez-Ontiveros} {et~al.}(2016){Fern{\'a}ndez-Ontiveros},
  {Spinoglio}, {Pereira-Santaella}, {Malkan}, {Andreani}, \&
  {Dasyra}}]{fernandez16}
{Fern{\'a}ndez-Ontiveros}, J.~A., {Spinoglio}, L., {Pereira-Santaella}, M.,
  {et~al.} 2016, \apjs, 226, 19

\bibitem[{{Ferrarese} \& {Merritt}(2000)}]{Ferrarese00}
{Ferrarese}, L. \& {Merritt}, D. 2000, \apjl, 539, L9

\bibitem[{{Feruglio} {et~al.}(2015){Feruglio}, {Fiore}, {Carniani},
  {Piconcelli}, {Zappacosta}, {Bongiorno}, {Cicone}, {Maiolino}, {Marconi},
  {Menci}, {Puccetti}, \& {Veilleux}}]{feruglio15}
{Feruglio}, C., {Fiore}, F., {Carniani}, S., {et~al.} 2015, \aap, 583, A99

\bibitem[{{Fiore} {et~al.}(2017){Fiore}, {Feruglio}, {Shankar}, {Bischetti},
  {Bongiorno}, {Brusa}, {Carniani}, {Cicone}, {Duras}, {Lamastra}, {Mainieri},
  {Marconi}, {Menci}, {Maiolino}, {Piconcelli}, {Vietri}, \&
  {Zappacosta}}]{Fiore17}
{Fiore}, F., {Feruglio}, C., {Shankar}, F., {et~al.} 2017, \aap, 601, A143

\bibitem[{Galitzki {et~al.}(2014)Galitzki, Ade, Angilè, Ashton, Beall, Becker,
  Bradford, Che, Cho, Devlin, Dober, Fissel, Fukui, Gao, Groppi, Hillbrand,
  Hilton, Hubmayr, Irwin, Klein, Van~Lanen, Li, Li, Lourie, Mani, Martin,
  Mauskopf, Nakamura, Novak, Pappas, Pascale, Pisano, Santos, Savini, Scott,
  Stanchfield, Tucker, Ullom, Underhill, Vissers, \&
  Ward-Thompson}]{galitzki14}
Galitzki, N., Ade, P. A.~R., Angilè, F.~E., {et~al.} 2014, Journal of
  Astronomical Instrumentation, 03, 1440001

\bibitem[{{Garc{\'\i}a-Bernete} {et~al.}(2016){Garc{\'\i}a-Bernete}, {Ramos
  Almeida}, {Acosta-Pulido}, {Alonso-Herrero}, {Gonz{\'a}lez-Mart{\'\i}n},
  {Hern{\'a}n-Caballero}, {Pereira-Santaella}, {Levenson}, {Packham},
  {Perlman}, {Ichikawa}, {Esquej}, \& {D{\'\i}az-Santos}}]{garcia16}
{Garc{\'\i}a-Bernete}, I., {Ramos Almeida}, C., {Acosta-Pulido}, J.~A.,
  {et~al.} 2016, \mnras, 463, 3531

\bibitem[{{Gebhardt} {et~al.}(2000){Gebhardt}, {Bender}, {Bower}, {Dressler},
  {Faber}, {Filippenko}, {Green}, {Grillmair}, {Ho}, {Kormendy}, {Lauer},
  {Magorrian}, {Pinkney}, {Richstone}, \& {Tremaine}}]{Gebhardt00}
{Gebhardt}, K., {Bender}, R., {Bower}, G., {et~al.} 2000, \apjl, 539, L13

\bibitem[{{Genzel} {et~al.}(1998){Genzel}, {Lutz}, {Sturm}, {Egami}, {Kunze},
  {Moorwood}, {Rigopoulou}, {Spoon}, {Sternberg}, {Tacconi-Garman}, {Tacconi},
  \& {Thatte}}]{genzel98}
{Genzel}, R., {Lutz}, D., {Sturm}, E., {et~al.} 1998, \apj, 498, 579

\bibitem[{{Georgantopoulos} {et~al.}(2013){Georgantopoulos}, {Comastri},
  {Vignali}, {Ranalli}, {Rovilos}, {Iwasawa}, {Gilli}, {Cappelluti}, {Carrera},
  {Fritz}, {Brusa}, {Elbaz}, {Mullaney}, {Castello-Mor}, {Barcons}, {Tozzi},
  {Balestra}, \& {Falocco}}]{georgantopoulos13}
{Georgantopoulos}, I., {Comastri}, A., {Vignali}, C., {et~al.} 2013, \aap, 555,
  A43

\bibitem[{{Giacconi} {et~al.}(2002){Giacconi}, {Zirm}, {Wang}, {Rosati},
  {Nonino}, {Tozzi}, {Gilli}, {Mainieri}, {Hasinger}, {Kewley}, {Bergeron},
  {Borgani}, {Gilmozzi}, {Grogin}, {Koekemoer}, {Schreier}, {Zheng}, \&
  {Norman}}]{giacconi02}
{Giacconi}, R., {Zirm}, A., {Wang}, J., {et~al.} 2002, \apjs, 139, 369

\bibitem[{{Gilli}(2013)}]{gilli13}
{Gilli}, R. 2013, \memsai, 84, 647

\bibitem[{{Gilli} {et~al.}(2007){Gilli}, {Comastri}, \& {Hasinger}}]{gilli07}
{Gilli}, R., {Comastri}, A., \& {Hasinger}, G. 2007, \aap, 463, 79

\bibitem[{{Gonz{\'a}lez} {et~al.}(2011){Gonz{\'a}lez}, {Labb{\'e}}, {Bouwens},
  {Illingworth}, {Franx}, \& {Kriek}}]{gonzalez11}
{Gonz{\'a}lez}, V., {Labb{\'e}}, I., {Bouwens}, R.~J., {et~al.} 2011, \apjl,
  735, L34

\bibitem[{{Goulding} {et~al.}(2012){Goulding}, {Alexander}, {Bauer}, {Forman},
  {Hickox}, {Jones}, {Mullaney}, \& {Trichas}}]{Goulding12}
{Goulding}, A.~D., {Alexander}, D.~M., {Bauer}, F.~E., {et~al.} 2012, \apj,
  755, 5

\bibitem[{{Griffin} {et~al.}(2010){Griffin}, {Abergel}, {Abreu}, {Ade},
  {Andr{\'e}}, {Augueres}, {Babbedge}, {Bae}, {Baillie}, {Baluteau}, {Barlow},
  {Bendo}, {Benielli}, {Bock}, {Bonhomme}, {Brisbin}, {Brockley-Blatt},
  {Caldwell}, {Cara}, {Castro-Rodriguez}, {Cerulli}, {Chanial}, {Chen},
  {Clark}, {Clements}, {Clerc}, {Coker}, {Communal}, {Conversi}, {Cox},
  {Crumb}, {Cunningham}, {Daly}, {Davis}, {de Antoni}, {Delderfield}, {Devin},
  {di Giorgio}, {Didschuns}, {Dohlen}, {Donati}, {Dowell}, {Dowell}, {Duband},
  {Dumaye}, {Emery}, {Ferlet}, {Ferrand}, {Fontignie}, {Fox}, {Franceschini},
  {Frerking}, {Fulton}, {Garcia}, {Gastaud}, {Gear}, {Glenn}, {Goizel},
  {Griffin}, {Grundy}, {Guest}, {Guillemet}, {Hargrave}, {Harwit}, {Hastings},
  {Hatziminaoglou}, {Herman}, {Hinde}, {Hristov}, {Huang}, {Imhof}, {Isaak},
  {Israelsson}, {Ivison}, {Jennings}, {Kiernan}, {King}, {Lange}, {Latter},
  {Laurent}, {Laurent}, {Leeks}, {Lellouch}, {Levenson}, {Li}, {Li},
  {Lilienthal}, {Lim}, {Liu}, {Lu}, {Madden}, {Mainetti}, {Marliani}, {McKay},
  {Mercier}, {Molinari}, {Morris}, {Moseley}, {Mulder}, {Mur}, {Naylor},
  {Nguyen}, {O'Halloran}, {Oliver}, {Olofsson}, {Olofsson}, {Orfei}, {Page},
  {Pain}, {Panuzzo}, {Papageorgiou}, {Parks}, {Parr-Burman}, {Pearce},
  {Pearson}, {P{\'e}rez-Fournon}, {Pinsard}, {Pisano}, {Podosek}, {Pohlen},
  {Polehampton}, {Pouliquen}, {Rigopoulou}, {Rizzo}, {Roseboom}, {Roussel},
  {Rowan-Robinson}, {Rownd}, {Saraceno}, {Sauvage}, {Savage}, {Savini},
  {Sawyer}, {Scharmberg}, {Schmitt}, {Schneider}, {Schulz}, {Schwartz},
  {Shafer}, {Shupe}, {Sibthorpe}, {Sidher}, {Smith}, {Smith}, {Smith},
  {Spencer}, {Stobie}, {Sudiwala}, {Sukhatme}, {Surace}, {Stevens}, {Swinyard},
  {Trichas}, {Tourette}, {Triou}, {Tseng}, {Tucker}, {Turner}, {Vaccari},
  {Valtchanov}, {Vigroux}, {Virique}, {Voellmer}, {Walker}, {Ward}, {Waskett},
  {Weilert}, {Wesson}, {White}, {Whitehouse}, {Wilson}, {Winter}, {Woodcraft},
  {Wright}, {Xu}, {Zavagno}, {Zemcov}, {Zhang}, \& {Zonca}}]{griffin10}
{Griffin}, M.~J., {Abergel}, A., {Abreu}, A., {et~al.} 2010, \aap, 518, L3

\bibitem[{{Gruppioni} {et~al.}(2016){Gruppioni}, {Berta}, {Spinoglio},
  {Pereira-Santaella}, {Pozzi}, {Andreani}, {Bonato}, {De Zotti}, {Malkan},
  {Negrello}, {Vallini}, \& {Vignali}}]{gruppioni16}
{Gruppioni}, C., {Berta}, S., {Spinoglio}, L., {et~al.} 2016, \mnras, 458, 4297

\bibitem[{{Gruppioni} {et~al.}(2020){Gruppioni}, {B{\'e}thermin}, {Loiacono},
  {Le F{\`e}vre}, {Capak}, {Cassata}, {Faisst}, {Schaerer}, {Silverman}, {Yan},
  {Bardelli}, {Boquien}, {Carraro}, {Cimatti}, {Dessauges-Zavadsky}, {Ginolfi},
  {Fujimoto}, {Hathi}, {Jones}, {Khusanova}, {Koekemoer}, {Lagache}, {Lemaux},
  {Oesch}, {Pozzi}, {Riechers}, {Rodighiero}, {Romano}, {Talia}, {Vallini},
  {Vergani}, {Zamorani}, \& {Zucca}}]{gruppioni20}
{Gruppioni}, C., {B{\'e}thermin}, M., {Loiacono}, F., {et~al.} 2020, \aap, 643,
  A8

\bibitem[{{Gruppioni} {et~al.}(2013){Gruppioni}, {Pozzi}, {Rodighiero},
  {Delvecchio}, {Berta}, {Pozzetti}, {Zamorani}, {Andreani}, {Cimatti},
  {Ilbert}, {Le Floc'h}, {Lutz}, {Magnelli}, {Marchetti}, {Monaco}, {Nordon},
  {Oliver}, {Popesso}, {Riguccini}, {Roseboom}, {Rosario}, {Sargent},
  {Vaccari}, {Altieri}, {Aussel}, {Bongiovanni}, {Cepa}, {Daddi},
  {Dom{\'\i}nguez-S{\'a}nchez}, {Elbaz}, {F{\"o}rster Schreiber}, {Genzel},
  {Iribarrem}, {Magliocchetti}, {Maiolino}, {Poglitsch}, {P{\'e}rez
  Garc{\'\i}a}, {Sanchez-Portal}, {Sturm}, {Tacconi}, {Valtchanov}, {Amblard},
  {Arumugam}, {Bethermin}, {Bock}, {Boselli}, {Buat}, {Burgarella},
  {Castro-Rodr{\'\i}guez}, {Cava}, {Chanial}, {Clements}, {Conley}, {Cooray},
  {Dowell}, {Dwek}, {Eales}, {Franceschini}, {Glenn}, {Griffin},
  {Hatziminaoglou}, {Ibar}, {Isaak}, {Ivison}, {Lagache}, {Levenson}, {Lu},
  {Madden}, {Maffei}, {Mainetti}, {Nguyen}, {O'Halloran}, {Page}, {Panuzzo},
  {Papageorgiou}, {Pearson}, {P{\'e}rez-Fournon}, {Pohlen}, {Rigopoulou},
  {Rowan-Robinson}, {Schulz}, {Scott}, {Seymour}, {Shupe}, {Smith}, {Stevens},
  {Symeonidis}, {Trichas}, {Tugwell}, {Vigroux}, {Wang}, {Wright}, {Xu},
  {Zemcov}, {Bardelli}, {Carollo}, {Contini}, {Le F{\'e}vre}, {Lilly},
  {Mainieri}, {Renzini}, {Scodeggio}, \& {Zucca}}]{gruppioni13}
{Gruppioni}, C., {Pozzi}, F., {Rodighiero}, G., {et~al.} 2013, \mnras, 432, 23

\bibitem[{{Harrison} {et~al.}(2016){Harrison}, {Aird}, {Civano}, {Lansbury},
  {Mullaney}, {Ballantyne}, {Alexander}, {Stern}, {Ajello}, {Barret}, {Bauer},
  {Balokovi{\'c}}, {Brandt}, {Brightman}, {Boggs}, {Christensen}, {Comastri},
  {Craig}, {Del Moro}, {Forster}, {Gandhi}, {Giommi}, {Grefenstette}, {Hailey},
  {Hickox}, {Hornstrup}, {Kitaguchi}, {Koglin}, {Luo}, {Madsen}, {Mao},
  {Miyasaka}, {Mori}, {Perri}, {Pivovaroff}, {Puccetti}, {Rana}, {Treister},
  {Walton}, {Westergaard}, {Wik}, {Zappacosta}, {Zhang}, \&
  {Zoglauer}}]{harrison16}
{Harrison}, F.~A., {Aird}, J., {Civano}, F., {et~al.} 2016, \apj, 831, 185

\bibitem[{{Heckman} \& {Best}(2014)}]{heckman14}
{Heckman}, T.~M. \& {Best}, P.~N. 2014, \araa, 52, 589

\bibitem[{{Hollenbach} \& {Tielens}(1999)}]{hollenbach99}
{Hollenbach}, D.~J. \& {Tielens}, A.~G.~G.~M. 1999, Reviews of Modern Physics,
  71, 173

\bibitem[{{Hopkins} {et~al.}(2006){Hopkins}, {Hernquist}, {Cox}, {Di Matteo},
  {Robertson}, \& {Springel}}]{hopkins06}
{Hopkins}, P.~F., {Hernquist}, L., {Cox}, T.~J., {et~al.} 2006, \apjs, 163, 1

\bibitem[{{Hopkins} {et~al.}(2007){Hopkins}, {Richards}, \&
  {Hernquist}}]{hopkins07}
{Hopkins}, P.~F., {Richards}, G.~T., \& {Hernquist}, L. 2007, \apj, 654, 731

\bibitem[{{Houck} {et~al.}(2005){Houck}, {Soifer}, {Weedman}, {Higdon},
  {Higdon}, {Herter}, {Brown}, {Dey}, {Jannuzi}, {Le Floc'h}, {Rieke}, {Armus},
  {Charmandaris}, {Brandl}, \& {Teplitz}}]{houck05}
{Houck}, J.~R., {Soifer}, B.~T., {Weedman}, D., {et~al.} 2005, \apjl, 622, L105

\bibitem[{{Hughes} {et~al.}(1998){Hughes}, {Serjeant}, {Dunlop},
  {Rowan-Robinson}, {Blain}, {Mann}, {Ivison}, {Peacock}, {Efstathiou}, {Gear},
  {Oliver}, {Lawrence}, {Longair}, {Goldschmidt}, \& {Jenness}}]{hughes98}
{Hughes}, D.~H., {Serjeant}, S., {Dunlop}, J., {et~al.} 1998, \nat, 394, 241

\bibitem[{{Jaacks} {et~al.}(2012){Jaacks}, {Choi}, {Nagamine}, {Thompson}, \&
  {Varghese}}]{jaacks12}
{Jaacks}, J., {Choi}, J.-H., {Nagamine}, K., {Thompson}, R., \& {Varghese}, S.
  2012, \mnras, 420, 1606

\bibitem[{{Kaneda} {et~al.}(2017){Kaneda}, {Ishihara}, {Oyabu}, {Yamagishi},
  {Wada}, {Armus}, {Baes}, {Charmandaris}, {Czerny}, {Efstathiou},
  {Fern{\'a}ndez-Ontiveros}, {Ferrara}, {Gonz{\'a}lez-Alfonso}, {Griffin},
  {Gruppioni}, {Hatziminaoglou}, {Imanishi}, {Kohno}, {Kwon}, {Nakagawa},
  {Onaka}, {Pozzi}, {Scott}, {Smith}, {Spinoglio}, {Suzuki}, {van der Tak},
  {Vaccari}, {Vignali}, \& {Wang}}]{kaneda17}
{Kaneda}, H., {Ishihara}, D., {Oyabu}, S., {et~al.} 2017, \pasa, 34, e059

\bibitem[{{Kormendy} \& {Richstone}(1995)}]{Kormendy95}
{Kormendy}, J. \& {Richstone}, D. 1995, \araa, 33, 581

\bibitem[{{La Caria} {et~al.}(2019){La Caria}, {Vignali}, {Lanzuisi},
  {Gruppioni}, \& {Pozzi}}]{lacaria19}
{La Caria}, M.~M., {Vignali}, C., {Lanzuisi}, G., {Gruppioni}, C., \& {Pozzi},
  F. 2019, \mnras, 487, 1662

\bibitem[{{Laigle} {et~al.}(2016){Laigle}, {McCracken}, {Ilbert}, {Hsieh},
  {Davidzon}, {Capak}, {Hasinger}, {Silverman}, {Pichon}, {Coupon}, {Aussel},
  {Le Borgne}, {Caputi}, {Cassata}, {Chang}, {Civano}, {Dunlop}, {Fynbo},
  {Kartaltepe}, {Koekemoer}, {Le F{\`e}vre}, {Le Floc'h}, {Leauthaud}, {Lilly},
  {Lin}, {Marchesi}, {Milvang-Jensen}, {Salvato}, {Sanders}, {Scoville},
  {Smolcic}, {Stockmann}, {Taniguchi}, {Tasca}, {Toft}, {Vaccari}, \&
  {Zabl}}]{laigle16}
{Laigle}, C., {McCracken}, H.~J., {Ilbert}, O., {et~al.} 2016, \apjs, 224, 24

\bibitem[{{Lamastra} {et~al.}(2013){Lamastra}, {Menci}, {Fiore}, {Santini},
  {Bongiorno}, \& {Piconcelli}}]{lamastra13}
{Lamastra}, A., {Menci}, N., {Fiore}, F., {et~al.} 2013, \aap, 559, A56

\bibitem[{{Lansbury} {et~al.}(2017{\natexlab{a}}){Lansbury}, {Alexander},
  {Aird}, {Gand hi}, {Stern}, {Koss}, {Lamperti}, {Ajello}, {Annuar}, {Assef},
  {Ballantyne}, {Balokovi{\'c}}, {Bauer}, {Brand t}, {Brightman}, {Chen},
  {Civano}, {Comastri}, {Del Moro}, {Fuentes}, {Harrison}, {Marchesi},
  {Masini}, {Mullaney}, {Ricci}, {Saez}, {Tomsick}, {Treister}, {Walton}, \&
  {Zappacosta}}]{lansbury17b}
{Lansbury}, G.~B., {Alexander}, D.~M., {Aird}, J., {et~al.} 2017{\natexlab{a}},
  \apj, 846, 20

\bibitem[{{Lansbury} {et~al.}(2017{\natexlab{b}}){Lansbury}, {Stern}, {Aird},
  {Alexander}, {Fuentes}, {Harrison}, {Treister}, {Bauer}, {Tomsick},
  {Balokovi{\'c}}, {Del Moro}, {Gandhi}, {Ajello}, {Annuar}, {Ballantyne},
  {Boggs}, {Brandt}, {Brightman}, {Chen}, {Christensen}, {Civano}, {Comastri},
  {Craig}, {Forster}, {Grefenstette}, {Hailey}, {Hickox}, {Jiang}, {Jun},
  {Koss}, {Marchesi}, {Melo}, {Mullaney}, {Noirot}, {Schulze}, {Walton},
  {Zappacosta}, \& {Zhang}}]{lansbury17a}
{Lansbury}, G.~B., {Stern}, D., {Aird}, J., {et~al.} 2017{\natexlab{b}}, \apj,
  836, 99

\bibitem[{{Lanzuisi} {et~al.}(2013){Lanzuisi}, {Civano}, {Elvis}, {Salvato},
  {Hasinger}, {Vignali}, {Zamorani}, {Aldcroft}, {Brusa}, {Comastri}, {Fiore},
  {Fruscione}, {Gilli}, {Ho}, {Mainieri}, {Merloni}, \&
  {Siemiginowska}}]{lanzuisi13}
{Lanzuisi}, G., {Civano}, F., {Elvis}, M., {et~al.} 2013, \mnras, 431, 978

\bibitem[{{Lanzuisi} {et~al.}(2017){Lanzuisi}, {Delvecchio}, {Berta}, {Brusa},
  {Comastri}, {Gilli}, {Gruppioni}, {Marchesi}, {Perna}, {Pozzi}, {Salvato},
  {Symeonidis}, {Vignali}, {Vito}, {Volonteri}, \& {Zamorani}}]{lanzuisi17}
{Lanzuisi}, G., {Delvecchio}, I., {Berta}, S., {et~al.} 2017, \aap, 602, A123

\bibitem[{{Lanzuisi} {et~al.}(2015){Lanzuisi}, {Ranalli}, {Georgantopoulos},
  {Georgakakis}, {Delvecchio}, {Akylas}, {Berta}, {Bongiorno}, {Brusa},
  {Cappelluti}, {Civano}, {Comastri}, {Gilli}, {Gruppioni}, {Hasinger},
  {Iwasawa}, {Koekemoer}, {Lusso}, {Marchesi}, {Mainieri}, {Merloni},
  {Mignoli}, {Piconcelli}, {Pozzi}, {Rosario}, {Salvato}, {Silverman},
  {Trakhtenbrot}, {Vignali}, \& {Zamorani}}]{lanzuisi15}
{Lanzuisi}, G., {Ranalli}, P., {Georgantopoulos}, I., {et~al.} 2015, \aap, 573,
  A137

\bibitem[{{Lapi} {et~al.}(2018){Lapi}, {Pantoni}, {Zanisi}, {Shi}, {Mancuso},
  {Massardi}, {Shankar}, {Bressan}, \& {Danese}}]{Lapi18}
{Lapi}, A., {Pantoni}, L., {Zanisi}, L., {et~al.} 2018, \apj, 857, 22

\bibitem[{{Leger} {et~al.}(1989){Leger}, {D'Hendecourt}, \&
  {Defourneau}}]{leger89}
{Leger}, A., {D'Hendecourt}, L., \& {Defourneau}, D. 1989, \aap, 216, 148

\bibitem[{{Lehmer} {et~al.}(2012){Lehmer}, {Xue}, {Brandt}, {Alexand er},
  {Bauer}, {Brusa}, {Comastri}, {Gilli}, {Hornschemeier}, {Luo}, {Paolillo},
  {Ptak}, {Shemmer}, {Schneider}, {Tozzi}, \& {Vignali}}]{lehmer12}
{Lehmer}, B.~D., {Xue}, Y.~Q., {Brandt}, W.~N., {et~al.} 2012, \apj, 752, 46

\bibitem[{Li {et~al.}(2008)Li, Dowell, Kirby, Novak, \& Vaillancourt}]{li08}
Li, H., Dowell, C.~D., Kirby, L., Novak, G., \& Vaillancourt, J.~E. 2008, Appl.
  Opt., 47, 422

\bibitem[{{Luo} {et~al.}(2008){Luo}, {Bauer}, {Brandt}, {Alexander}, {Lehmer},
  {Schneider}, {Brusa}, {Comastri}, {Fabian}, {Finoguenov}, {Gilli},
  {Hasinger}, {Hornschemeier}, {Koekemoer}, {Mainieri}, {Paolillo}, {Rosati},
  {Shemmer}, {Silverman}, {Smail}, {Steffen}, \& {Vignali}}]{luo08}
{Luo}, B., {Bauer}, F.~E., {Brandt}, W.~N., {et~al.} 2008, \apjs, 179, 19

\bibitem[{{Lusso} {et~al.}(2012){Lusso}, {Comastri}, {Simmons}, {Mignoli},
  {Zamorani}, {Vignali}, {Brusa}, {Shankar}, {Lutz}, {Trump}, {Maiolino},
  {Gilli}, {Bolzonella}, {Puccetti}, {Salvato}, {Impey}, {Civano}, {Elvis},
  {Mainieri}, {Silverman}, {Koekemoer}, {Bongiorno}, {Merloni}, {Berta}, {Le
  Floc'h}, {Magnelli}, {Pozzi}, \& {Riguccini}}]{lusso12}
{Lusso}, E., {Comastri}, A., {Simmons}, B.~D., {et~al.} 2012, Monthly Notices
  of the Royal Astronomical Society, 425, 623

\bibitem[{{Lusso} {et~al.}(2010){Lusso}, {Comastri}, {Vignali}, {Zamorani},
  {Brusa}, {Gilli}, {Iwasawa}, {Salvato}, {Civano}, {Elvis}, {Merloni},
  {Bongiorno}, {Trump}, {Koekemoer}, {Schinnerer}, {Le Floc'h}, {Cappelluti},
  {Jahnke}, {Sargent}, {Silverman}, {Mainieri}, {Fiore}, {Bolzonella}, {Le
  F{\`e}vre}, {Garilli}, {Iovino}, {Kneib}, {Lamareille}, {Lilly}, {Mignoli},
  {Scodeggio}, \& {Vergani}}]{lusso10}
{Lusso}, E., {Comastri}, A., {Vignali}, C., {et~al.} 2010, \aap, 512, A34

\bibitem[{{Lutz} {et~al.}(1998){Lutz}, {Spoon}, {Rigopoulou}, {Moorwood}, \&
  {Genzel}}]{lutz98}
{Lutz}, D., {Spoon}, H.~W.~W., {Rigopoulou}, D., {Moorwood}, A.~F.~M., \&
  {Genzel}, R. 1998, \apjl, 505, L103

\bibitem[{{Lutz} {et~al.}(2003){Lutz}, {Sturm}, {Genzel}, {Spoon}, {Moorwood},
  {Netzer}, \& {Sternberg}}]{lutz03}
{Lutz}, D., {Sturm}, E., {Genzel}, R., {et~al.} 2003, \aap, 409, 867

\bibitem[{{Madau} \& {Dickinson}(2014)}]{madau14}
{Madau}, P. \& {Dickinson}, M. 2014, \araa, 52, 415

\bibitem[{{Magnelli} {et~al.}(2013){Magnelli}, {Popesso}, {Berta}, {Pozzi},
  {Elbaz}, {Lutz}, {Dickinson}, {Altieri}, {Andreani}, {Aussel},
  {B{\'e}thermin}, {Bongiovanni}, {Cepa}, {Charmandaris}, {Chary}, {Cimatti},
  {Daddi}, {F{\"o}rster Schreiber}, {Genzel}, {Gruppioni}, {Harwit}, {Hwang},
  {Ivison}, {Magdis}, {Maiolino}, {Murphy}, {Nordon}, {Pannella}, {P{\'e}rez
  Garc{\'\i}a}, {Poglitsch}, {Rosario}, {Sanchez-Portal}, {Santini}, {Scott},
  {Sturm}, {Tacconi}, \& {Valtchanov}}]{magnelli13}
{Magnelli}, B., {Popesso}, P., {Berta}, S., {et~al.} 2013, \aap, 553, A132

\bibitem[{{Marchesi} {et~al.}(2016){Marchesi}, {Civano}, {Elvis}, {Salvato},
  {Brusa}, {Comastri}, {Gilli}, {Hasinger}, {Lanzuisi}, {Miyaji}, {Treister},
  {Urry}, {Vignali}, {Zamorani}, {Allevato}, {Cappelluti}, {Cardamone},
  {Finoguenov}, {Griffiths}, {Karim}, {Laigle}, {LaMassa}, {Jahnke}, {Ranalli},
  {Schawinski}, {Schinnerer}, {Silverman}, {Smolcic}, {Suh}, \&
  {Trakhtenbrot}}]{marchesi16}
{Marchesi}, S., {Civano}, F., {Elvis}, M., {et~al.} 2016, \apj, 817, 34

\bibitem[{{Mateos} {et~al.}(2012){Mateos}, {Alonso-Herrero}, {Carrera},
  {Blain}, {Watson}, {Barcons}, {Braito}, {Severgnini}, {Donley}, \&
  {Stern}}]{mateos12}
{Mateos}, S., {Alonso-Herrero}, A., {Carrera}, F.~J., {et~al.} 2012, \mnras,
  426, 3271

\bibitem[{{McCammon} {et~al.}(2002){McCammon}, {Almy}, {Apodaca}, {Bergmann
  Tiest}, {Cui}, {Deiker}, {Galeazzi}, {Juda}, {Lesser}, {Mihara},
  {Morgenthaler}, {Sanders}, {Zhang}, {Figueroa-Feliciano}, {Kelley},
  {Moseley}, {Mushotzky}, {Porter}, {Stahle}, \& {Szymkowiak}}]{mccammon02}
{McCammon}, D., {Almy}, R., {Apodaca}, E., {et~al.} 2002, \apj, 576, 188

\bibitem[{{Meidinger} {et~al.}(2018){Meidinger}, {Nandra}, \&
  {Plattner}}]{meidinger18}
{Meidinger}, N., {Nandra}, K., \& {Plattner}, M. 2018, in Society of
  Photo-Optical Instrumentation Engineers (SPIE) Conference Series, Vol. 10699,
  Space Telescopes and Instrumentation 2018: Ultraviolet to Gamma Ray, 106991F

\bibitem[{{Meixner} {et~al.}(2019){Meixner}, {Cooray}, {Leisawitz}, {Staguhn},
  {Armus}, {Battersby}, {Bauer}, {Bergin}, {Bradford}, {Ennico-Smith},
  {Fortney}, {Kataria}, {Melnick}, {Milam}, {Narayanan}, {Padgett},
  {Pontoppidan}, {Pope}, {Roellig}, {Sandstrom}, {Stevenson}, {Su}, {Vieira},
  {Wright}, {Zmuidzinas}, {Sheth}, {Benford}, {Mamajek}, {Neff}, {De Beck},
  {Gerin}, {Helmich}, {Sakon}, {Scott}, {Vavrek}, {Wiedner}, {Carey},
  {Burgarella}, {Moseley}, {Amatucci}, {Carter}, {DiPirro}, {Wu}, {Beaman},
  {Beltran}, {Bolognese}, {Bradley}, {Corsetti}, {D'Asto}, {Denis}, {Derkacz},
  {Earle}, {Fantano}, {Folta}, {Gavares}, {Generie}, {Hilliard}, {Howard},
  {Jamil}, {Jamison}, {Lynch}, {Martins}, {Petro}, {Ramspacher}, {Rao},
  {Sandin}, {Stoneking}, {Tompkins}, \& {Webster}}]{meixner19}
{Meixner}, M., {Cooray}, A., {Leisawitz}, D., {et~al.} 2019, arXiv e-prints,
  arXiv:1912.06213

\bibitem[{{Menci} {et~al.}(2008){Menci}, {Fiore}, {Puccetti}, \&
  {Cavaliere}}]{menci08}
{Menci}, N., {Fiore}, F., {Puccetti}, S., \& {Cavaliere}, A. 2008, \apj, 686,
  219

\bibitem[{{Mordini} {et~al.}(2021){Mordini}, {Spinoglio}, \&
  {Fern{\'a}ndez-Ontiveros}}]{mordini21}
{Mordini}, S., {Spinoglio}, L., \& {Fern{\'a}ndez-Ontiveros}, J.~A. 2021, arXiv
  e-prints, arXiv:2105.04584

\bibitem[{{Nakagawa} {et~al.}(1998){Nakagawa}, {Hayashi}, {Kawada},
  {Matsuhara}, {Matsumoto}, {Murakami}, {Okuda}, {Onaka}, {Shibai}, \&
  {Ueno}}]{nakagawa98}
{Nakagawa}, T., {Hayashi}, M., {Kawada}, M., {et~al.} 1998, Society of
  Photo-Optical Instrumentation Engineers (SPIE) Conference Series, Vol. 3356,
  {HII/L2 mission: future Japanese infrared astronomical mission}, ed. P.~Y.
  {Bely} \& J.~B. {Breckinridge}, 462--470

\bibitem[{{Nakagawa} {et~al.}(2014){Nakagawa}, {Shibai}, {Onaka}, {Matsuhara},
  {Kaneda}, {Kawakatsu}, \& {Roelfsema}}]{nakagawa14}
{Nakagawa}, T., {Shibai}, H., {Onaka}, T., {et~al.} 2014, Society of
  Photo-Optical Instrumentation Engineers (SPIE) Conference Series, Vol. 9143,
  {The next-generation infrared astronomy mission SPICA under the new
  framework}, 91431I

\bibitem[{{Nandra} {et~al.}(2013){Nandra}, {Barret}, {Barcons}, {Fabian}, {den
  Herder}, {Piro}, {Watson}, {Adami}, {Aird}, {Afonso}, {Alexander},
  {Argiroffi}, {Amati}, {Arnaud}, {Atteia}, {Audard}, {Badenes}, {Ballet},
  {Ballo}, {Bamba}, {Bhardwaj}, {Stefano Battistelli}, {Becker}, {De Becker},
  {Behar}, {Bianchi}, {Biffi}, {B{\^\i}rzan}, {Bocchino}, {Bogdanov}, {Boirin},
  {Boller}, {Borgani}, {Borm}, {Bouch{\'e}}, {Bourdin}, {Bower}, {Braito},
  {Branchini}, {Branduardi-Raymont}, {Bregman}, {Brenneman}, {Brightman},
  {Br{\"u}ggen}, {Buchner}, {Bulbul}, {Brusa}, {Bursa}, {Caccianiga},
  {Cackett}, {Campana}, {Cappelluti}, {Cappi}, {Carrera}, {Ceballos},
  {Christensen}, {Chu}, {Churazov}, {Clerc}, {Corbel}, {Corral}, {Comastri},
  {Costantini}, {Croston}, {Dadina}, {D'Ai}, {Decourchelle}, {Della Ceca},
  {Dennerl}, {Dolag}, {Done}, {Dovciak}, {Drake}, {Eckert}, {Edge}, {Ettori},
  {Ezoe}, {Feigelson}, {Fender}, {Feruglio}, {Finoguenov}, {Fiore}, {Galeazzi},
  {Gallagher}, {Gandhi}, {Gaspari}, {Gastaldello}, {Georgakakis},
  {Georgantopoulos}, {Gilfanov}, {Gitti}, {Gladstone}, {Goosmann}, {Gosset},
  {Grosso}, {Guedel}, {Guerrero}, {Haberl}, {Hardcastle}, {Heinz}, {Alonso
  Herrero}, {Herv{\'e}}, {Holmstrom}, {Iwasawa}, {Jonker}, {Kaastra}, {Kara},
  {Karas}, {Kastner}, {King}, {Kosenko}, {Koutroumpa}, {Kraft}, {Kreykenbohm},
  {Lallement}, {Lanzuisi}, {Lee}, {Lemoine-Goumard}, {Lobban}, {Lodato},
  {Lovisari}, {Lotti}, {McCharthy}, {McNamara}, {Maggio}, {Maiolino}, {De
  Marco}, {de Martino}, {Mateos}, {Matt}, {Maughan}, {Mazzotta}, {Mendez},
  {Merloni}, {Micela}, {Miceli}, {Mignani}, {Miller}, {Miniutti}, {Molendi},
  {Montez}, {Moretti}, {Motch}, {Naz{\'e}}, {Nevalainen}, {Nicastro}, {Nulsen},
  {Ohashi}, {O'Brien}, {Osborne}, {Oskinova}, {Pacaud}, {Paerels}, {Page},
  {Papadakis}, {Pareschi}, {Petre}, {Petrucci}, {Piconcelli}, {Pillitteri},
  {Pinto}, {de Plaa}, {Pointecouteau}, {Ponman}, {Ponti}, {Porquet}, {Pounds},
  {Pratt}, {Predehl}, {Proga}, {Psaltis}, {Rafferty}, {Ramos-Ceja}, {Ranalli},
  {Rasia}, {Rau}, {Rauw}, {Rea}, {Read}, {Reeves}, {Reiprich}, {Renaud},
  {Reynolds}, {Risaliti}, {Rodriguez}, {Rodriguez Hidalgo}, {Roncarelli},
  {Rosario}, {Rossetti}, {Rozanska}, {Rovilos}, {Salvaterra}, {Salvato}, {Di
  Salvo}, {Sanders}, {Sanz-Forcada}, {Schawinski}, {Schaye}, {Schwope},
  {Sciortino}, {Severgnini}, {Shankar}, {Sijacki}, {Sim}, {Schmid}, {Smith},
  {Steiner}, {Stelzer}, {Stewart}, {Strohmayer}, {Str{\"u}der}, {Sun}, {Takei},
  {Tatischeff}, {Tiengo}, {Tombesi}, {Trinchieri}, {Tsuru}, {Ud-Doula},
  {Ursino}, {Valencic}, {Vanzella}, {Vaughan}, {Vignali}, {Vink}, {Vito},
  {Volonteri}, {Wang}, {Webb}, {Willingale}, {Wilms}, {Wise}, {Worrall},
  {Young}, {Zampieri}, {In't Zand}, {Zane}, {Zezas}, {Zhang}, \&
  {Zhuravleva}}]{nandra13}
{Nandra}, K., {Barret}, D., {Barcons}, X., {et~al.} 2013, arXiv e-prints,
  arXiv:1306.2307

\bibitem[{{O'Dowd} {et~al.}(2009){O'Dowd}, {Schiminovich}, {Johnson}, {Treyer},
  {Martin}, {Wyder}, {Charlot}, {Heckman}, {Martins}, {Seibert}, \& {van der
  Hulst}}]{odowd09}
{O'Dowd}, M.~J., {Schiminovich}, D., {Johnson}, B.~D., {et~al.} 2009, \apj,
  705, 885

\bibitem[{{Page} {et~al.}(2004){Page}, {Stevens}, {Ivison}, \&
  {Carrera}}]{page04}
{Page}, M.~J., {Stevens}, J.~A., {Ivison}, R.~J., \& {Carrera}, F.~J. 2004,
  \apjl, 611, L85

\bibitem[{{Peca} {et~al.}(2021){Peca}, {Vignali}, {Gilli}, {Mignoli}, {Nanni},
  {Marchesi}, {Bolzonella}, {Brusa}, {Cappelluti}, {Comastri}, {Lanzuisi}, \&
  {Vito}}]{peca21}
{Peca}, A., {Vignali}, C., {Gilli}, R., {et~al.} 2021, \apj, 906, 90

\bibitem[{{Poglitsch} {et~al.}(2010){Poglitsch}, {Waelkens}, {Geis},
  {Feuchtgruber}, {Vandenbussche}, {Rodriguez}, {Krause}, {Renotte}, {van
  Hoof}, {Saraceno}, {Cepa}, {Kerschbaum}, {Agn{\`e}se}, {Ali}, {Altieri},
  {Andreani}, {Augueres}, {Balog}, {Barl}, {Bauer}, {Belbachir}, {Benedettini},
  {Billot}, {Boulade}, {Bischof}, {Blommaert}, {Callut}, {Cara}, {Cerulli},
  {Cesarsky}, {Contursi}, {Creten}, {De Meester}, {Doublier}, {Doumayrou},
  {Duband }, {Exter}, {Genzel}, {Gillis}, {Gr{\"o}zinger}, {Henning},
  {Herreros}, {Huygen}, {Inguscio}, {Jakob}, {Jamar}, {Jean}, {de Jong},
  {Katterloher}, {Kiss}, {Klaas}, {Lemke}, {Lutz}, {Madden}, {Marquet},
  {Martignac}, {Mazy}, {Merken}, {Montfort}, {Morbidelli}, {M{\"u}ller},
  {Nielbock}, {Okumura}, {Orfei}, {Ottensamer}, {Pezzuto}, {Popesso},
  {Putzeys}, {Regibo}, {Reveret}, {Royer}, {Sauvage}, {Schreiber}, {Stegmaier},
  {Schmitt}, {Schubert}, {Sturm}, {Thiel}, {Tofani}, {Vavrek}, {Wetzstein},
  {Wieprecht}, \& {Wiezorrek}}]{poglitsch10}
{Poglitsch}, A., {Waelkens}, C., {Geis}, N., {et~al.} 2010, \aap, 518, L2

\bibitem[{{Polletta} {et~al.}(2008){Polletta}, {Weedman}, {H{\"o}nig},
  {Lonsdale}, {Smith}, \& {Houck}}]{polletta08}
{Polletta}, M., {Weedman}, D., {H{\"o}nig}, S., {et~al.} 2008, \apj, 675, 960

\bibitem[{{Pozzi} {et~al.}(2007){Pozzi}, {Vignali}, {Comastri}, {Pozzetti},
  {Mignoli}, {Gruppioni}, {Zamorani}, {Lari}, {Civano}, {Brusa}, {Fiore},
  {Maiolino}, \& {La Franca}}]{pozzi07}
{Pozzi}, F., {Vignali}, C., {Comastri}, A., {et~al.} 2007, \aap, 468, 603

\bibitem[{{Ricci} {et~al.}(2015){Ricci}, {Ueda}, {Koss}, {Trakhtenbrot},
  {Bauer}, \& {Gandhi}}]{ricci15}
{Ricci}, C., {Ueda}, Y., {Koss}, M.~J., {et~al.} 2015, \apjl, 815, L13

\bibitem[{{Riechers} {et~al.}(2013){Riechers}, {Bradford}, {Clements},
  {Dowell}, {P{\'e}rez-Fournon}, {Ivison}, {Bridge}, {Conley}, {Fu}, {Vieira},
  {Wardlow}, {Calanog}, {Cooray}, {Hurley}, {Neri}, {Kamenetzky}, {Aguirre},
  {Altieri}, {Arumugam}, {Benford}, {B{\'e}thermin}, {Bock}, {Burgarella},
  {Cabrera-Lavers}, {Chapman}, {Cox}, {Dunlop}, {Earle}, {Farrah}, {Ferrero},
  {Franceschini}, {Gavazzi}, {Glenn}, {Solares}, {Gurwell}, {Halpern},
  {Hatziminaoglou}, {Hyde}, {Ibar}, {Kov{\'a}cs}, {Krips}, {Lupu}, {Maloney},
  {Martinez-Navajas}, {Matsuhara}, {Murphy}, {Naylor}, {Nguyen}, {Oliver},
  {Omont}, {Page}, {Petitpas}, {Rangwala}, {Roseboom}, {Scott}, {Smith},
  {Staguhn}, {Streblyanska}, {Thomson}, {Valtchanov}, {Viero}, {Wang},
  {Zemcov}, \& {Zmuidzinas}}]{riechers13}
{Riechers}, D.~A., {Bradford}, C.~M., {Clements}, D.~L., {et~al.} 2013, \nat,
  496, 329

\bibitem[{{Rodriguez} {et~al.}(2018){Rodriguez}, {Poglitsch}, {Aliane},
  {Martignac}, {Dubreuil}, {Dussopt}, {Rev{\'e}ret}, {Goudon}, {Bounissou},
  {Adami}, {Delisle}, {Gevin}, {De La Broise}, {Maffei}, \&
  {Sauvageot}}]{rodriguez18}
{Rodriguez}, L., {Poglitsch}, A., {Aliane}, A., {et~al.} 2018, Journal of Low
  Temperature Physics, 193, 449

\bibitem[{{Roelfsema} {et~al.}(2018){Roelfsema}, {Shibai}, {Armus}, {Arrazola},
  {Audard}, {Audley}, {Bradford}, {Charles}, {Dieleman}, {Doi}, {Duband},
  {Eggens}, {Evers}, {Funaki}, {Gao}, {Giard}, {di Giorgio}, {Gonz{\'a}lez
  Fern{\'a}ndez}, {Griffin}, {Helmich}, {Hijmering}, {Huisman}, {Ishihara},
  {Isobe}, {Jackson}, {Jacobs}, {Jellema}, {Kamp}, {Kaneda}, {Kawada},
  {Kemper}, {Kerschbaum}, {Khosropanah}, {Kohno}, {Kooijman}, {Krause}, {van
  der Kuur}, {Kwon}, {Laauwen}, {de Lange}, {Larsson}, {van Loon}, {Madden},
  {Matsuhara}, {Najarro}, {Nakagawa}, {Naylor}, {Ogawa}, {Onaka}, {Oyabu},
  {Poglitsch}, {Reveret}, {Rodriguez}, {Spinoglio}, {Sakon}, {Sato},
  {Shinozaki}, {Shipman}, {Sugita}, {Suzuki}, {van der Tak}, {Torres Redondo},
  {Wada}, {Wang}, {Wafelbakker}, {van Weers}, {Withington}, {Vandenbussche},
  {Yamada}, \& {Yamamura}}]{roelfsema18}
{Roelfsema}, P.~R., {Shibai}, H., {Armus}, L., {et~al.} 2018, \pasa, 35, e030

\bibitem[{{Rowan-Robinson} {et~al.}(1997){Rowan-Robinson}, {Mann}, {Oliver},
  {Efstathiou}, {Eaton}, {Goldschmidt}, {Mobasher}, {Serjeant}, {Sumner},
  {Danese}, {Elbaz}, {Franceschini}, {Egami}, {Kontizas}, {Lawrence},
  {McMahon}, {Norgaard-Nielsen}, {Perez-Fournon}, \&
  {Gonzalez-Serrano}}]{rowan-robinson97}
{Rowan-Robinson}, M., {Mann}, R.~G., {Oliver}, S.~J., {et~al.} 1997, \mnras,
  289, 490

\bibitem[{{Rowan-Robinson} {et~al.}(2016){Rowan-Robinson}, {Oliver}, {Wang},
  {Farrah}, {Clements}, {Gruppioni}, {Marchetti}, {Rigopoulou}, \&
  {Vaccari}}]{rowan-robinson16}
{Rowan-Robinson}, M., {Oliver}, S., {Wang}, L., {et~al.} 2016, \mnras, 461,
  1100

\bibitem[{{Schreiber} {et~al.}(2015){Schreiber}, {Pannella}, {Elbaz},
  {B{\'e}thermin}, {Inami}, {Dickinson}, {Magnelli}, {Wang}, {Aussel}, {Daddi},
  {Juneau}, {Shu}, {Sargent}, {Buat}, {Faber}, {Ferguson}, {Giavalisco},
  {Koekemoer}, {Magdis}, {Morrison}, {Papovich}, {Santini}, \&
  {Scott}}]{schreiber15}
{Schreiber}, C., {Pannella}, M., {Elbaz}, D., {et~al.} 2015, Astronomy and
  Astrophysics, 575, A74

\bibitem[{{Scoville} {et~al.}(2007){Scoville}, {Aussel}, {Brusa}, {Capak},
  {Carollo}, {Elvis}, {Giavalisco}, {Guzzo}, {Hasinger}, {Impey}, {Kneib},
  {LeFevre}, {Lilly}, {Mobasher}, {Renzini}, {Rich}, {Sanders}, {Schinnerer},
  {Schminovich}, {Shopbell}, {Taniguchi}, \& {Tyson}}]{scoville07}
{Scoville}, N., {Aussel}, H., {Brusa}, M., {et~al.} 2007, \apjs, 172, 1

\bibitem[{{Shankar} {et~al.}(2009){Shankar}, {Weinberg}, \&
  {Miralda-Escud{\'e}}}]{Shankar09}
{Shankar}, F., {Weinberg}, D.~H., \& {Miralda-Escud{\'e}}, J. 2009, \apj, 690,
  20

\bibitem[{{Shankar} {et~al.}(2013){Shankar}, {Weinberg}, \&
  {Miralda-Escud{\'e}}}]{shankar13}
{Shankar}, F., {Weinberg}, D.~H., \& {Miralda-Escud{\'e}}, J. 2013, \mnras,
  428, 421

\bibitem[{{Shi} {et~al.}(2006){Shi}, {Rieke}, {Hines}, {Gorjian}, {Werner},
  {Cleary}, {Low}, {Smith}, \& {Bouwman}}]{shi06}
{Shi}, Y., {Rieke}, G.~H., {Hines}, D.~C., {et~al.} 2006, \apj, 653, 127

\bibitem[{{Sijacki} {et~al.}(2015){Sijacki}, {Vogelsberger}, {Genel},
  {Springel}, {Torrey}, {Snyder}, {Nelson}, \& {Hernquist}}]{sijacki15}
{Sijacki}, D., {Vogelsberger}, M., {Genel}, S., {et~al.} 2015, \mnras, 452, 575

\bibitem[{{Silk} \& {Rees}(1998)}]{silk98}
{Silk}, J. \& {Rees}, M.~J. 1998, \aap, 331, L1

\bibitem[{{Simmonds} {et~al.}(2018){Simmonds}, {Buchner}, {Salvato}, {Hsu}, \&
  {Bauer}}]{simmonds18}
{Simmonds}, C., {Buchner}, J., {Salvato}, M., {Hsu}, L.~T., \& {Bauer}, F.~E.
  2018, \aap, 618, A66

\bibitem[{{Smith} {et~al.}(2007){Smith}, {Draine}, {Dale}, {Moustakas},
  {Kennicutt}, {Helou}, {Armus}, {Roussel}, {Sheth}, {Bendo}, {Buckalew},
  {Calzetti}, {Engelbracht}, {Gordon}, {Hollenbach}, {Li}, {Malhotra},
  {Murphy}, \& {Walter}}]{smith06}
{Smith}, J.~D.~T., {Draine}, B.~T., {Dale}, D.~A., {et~al.} 2007, \apj, 656,
  770

\bibitem[{{Smith} {et~al.}(2019){Smith}, {Tombesi}, {Veilleux}, {Lohfink}, \&
  {Luminari}}]{smith19}
{Smith}, R.~N., {Tombesi}, F., {Veilleux}, S., {Lohfink}, A.~M., \& {Luminari},
  A. 2019, \apj, 887, 69

\bibitem[{{Spergel} {et~al.}(2003){Spergel}, {Verde}, {Peiris}, {Komatsu},
  {Nolta}, {Bennett}, {Halpern}, {Hinshaw}, {Jarosik}, {Kogut}, {Limon},
  {Meyer}, {Page}, {Tucker}, {Weiland}, {Wollack}, \& {Wright}}]{spergel03}
{Spergel}, D.~N., {Verde}, L., {Peiris}, H.~V., {et~al.} 2003, \apjs, 148, 175

\bibitem[{{Spinoglio} {et~al.}(2017){Spinoglio}, {Alonso-Herrero}, {Armus},
  {Baes}, {Bernard-Salas}, {Bianchi}, {Bocchio}, {Bolatto}, {Bradford},
  {Braine}, {Carrera}, {Ciesla}, {Clements}, {Dannerbauer}, {Doi},
  {Efstathiou}, {Egami}, {Fern{\'a}ndez-Ontiveros}, {Ferrara}, {Fischer},
  {Franceschini}, {Gallerani}, {Giard}, {Gonz{\'a}lez-Alfonso}, {Gruppioni},
  {Guillard}, {Hatziminaoglou}, {Imanishi}, {Ishihara}, {Isobe}, {Kaneda},
  {Kawada}, {Kohno}, {Kwon}, {Madden}, {Malkan}, {Marassi}, {Matsuhara},
  {Matsuura}, {Miniutti}, {Nagamine}, {Nagao}, {Najarro}, {Nakagawa}, {Onaka},
  {Oyabu}, {Pallottini}, {Piro}, {Pozzi}, {Rodighiero}, {Roelfsema}, {Sakon},
  {Santini}, {Schaerer}, {Schneider}, {Scott}, {Serjeant}, {Shibai}, {Smith},
  {Sobacchi}, {Sturm}, {Suzuki}, {Vallini}, {van der Tak}, {Vignali}, {Yamada},
  {Wada}, \& {Wang}}]{spinoglio17}
{Spinoglio}, L., {Alonso-Herrero}, A., {Armus}, L., {et~al.} 2017, \pasa, 34,
  e057

\bibitem[{{Spinoglio} \& {Malkan}(1992)}]{spinoglio92}
{Spinoglio}, L. \& {Malkan}, M.~A. 1992, \apj, 399, 504

\bibitem[{{Spinoglio} {et~al.}(2021){Spinoglio}, {Mordini},
  {Fern{\'a}ndez-Ontiveros}, {Alonso-Herrero}, {Armus}, {Bisigello}, {Calura},
  {Carrera}, {Cooray}, {Dannerbauer}, {Decarli}, {Egami}, {Elbaz},
  {Franceschini}, {Gonz{\'a}lez Alfonso}, {Graziani}, {Gruppioni},
  {Hatziminaoglou}, {Kaneda}, {Kohno}, {Labiano}, {Magdis}, {Malkan},
  {Matsuhara}, {Nagao}, {Naylor}, {Pereira-Santaella}, {Pozzi}, {Rodighiero},
  {Roelfsema}, {Serjeant}, {Vignali}, {Wang}, \& {Yamada}}]{spinoglio21}
{Spinoglio}, L., {Mordini}, S., {Fern{\'a}ndez-Ontiveros}, J.~A., {et~al.}
  2021, \pasa, 38, e021

\bibitem[{{Sturm} {et~al.}(2006){Sturm}, {Hasinger}, {Lehmann}, {Mainieri},
  {Genzel}, {Lehnert}, {Lutz}, \& {Tacconi}}]{sturm06}
{Sturm}, E., {Hasinger}, G., {Lehmann}, I., {et~al.} 2006, \apj, 642, 81

\bibitem[{{Swinyard} {et~al.}(2009){Swinyard}, {Nakagawa}, {Merken}, {Royer},
  {Souverijns}, {Vandenbussche}, {Waelkens}, {Davis}, {Di Francesco},
  {Halpern}, {Houde}, {Johnstone}, {Joncas}, {Naylor}, {Plume}, {Scott},
  {Abergel}, {Bensammar}, {Braine}, {Buat}, {Burgarella}, {Cais}, {Dole},
  {Duband}, {Elbaz}, {Gerin}, {Giard}, {Goicoechea}, {Joblin}, {Jones},
  {Kneib}, {Lagache}, {Madden}, {Pons}, {Pajot}, {Rambaud}, {Ravera},
  {Ristorcelli}, {Rodriguez}, {Vives}, {Zavagno}, {Geis}, {Krause}, {Lutz},
  {Poglitsch}, {Raab}, {Stegmaier}, {Sturm}, {Tuffs}, {Lee}, {Koo}, {Im},
  {Pak}, {Han}, {Park}, {Nam}, {Jin}, {Lee}, {Yuk}, {Lee}, {Aikawa}, {Arimoto},
  {Doi}, {Enya}, {Fukagawa}, {Furusho}, {Hasegawa}, {Hayashi}, {Honda
  Kanagawa}, {Ida}, {Imanishi}, {Masatoshi}, {Inutsuka}, {Izumiura}, {Kamaya},
  {Kaneda}, {Kasuga}, {Kataza}, {Kawabata}, {Kawada}, {Kawakita}, {Kii},
  {Koda}, {Kodama}, {Kokubo}, {Komatsu}, {Matsuhara}, {Matsumoto}, {Matsuura},
  {Miyata}, {Miyata}, {Nagata}, {Nagata}, {Nakajima}, {Naoto}, {Nishi}, {Noda},
  {Okamoto}, {Okamoto}, {Omukai}, {Onaka}, {Ootsubo}, {Ouchi}, {Saito}, {Sato},
  {Sako}, {Sekiguchi}, {Shibai}, {Sugita}, {Sugitani}, {Susa}, {Pyo}, {Tamura},
  {Ueda}, {Ueno}, {Wada}, {Watanabe}, {Yamada}, {Yamamura}, {Yoshida},
  {Yoshimi}, {Yui}, {Benedettini}, {Cerulli}, {Di Giorgio}, {Molinari},
  {Orfei}, {Pezzuto}, {Piazzo}, {Saraceno}, {Spinoglio}, {de Graauw}, {de
  Korte}, {Helmich}, {Hoevers}, {Huisman}, {Shipman}, {van der Tak}, {van der
  Werf}, {Wild}, {Acosta-Pulido}, {Cernicharo}, {Herreros}, {Martin-Pintado},
  {Najarro}, {Perez-Fourmon}, {Ramon Pardo}, {Gomez}, {Castro Rodriguez},
  {Ade}, {Barlow}, {Clements}, {Ferlet}, {Fraser}, {Griffin}, {Griffin},
  {Hargrave}, {Isaak}, {Ivison}, {Mansour}, {Laniesse}, {Mauskopf}, {Morozov},
  {Oliver}, {Orlando}, {Page}, {Popescu}, {Serjeant}, {Sudiwala}, {Rigopoulou},
  {Walker}, {White}, {Viti}, {Winter}, {Bock}, {Bradford}, {Harwit}, \&
  {Holmes}}]{swinyard09}
{Swinyard}, B., {Nakagawa}, T., {Merken}, P., {et~al.} 2009, Experimental
  Astronomy, 23, 193

\bibitem[{{Teplitz} {et~al.}(2006){Teplitz}, {Armus}, {Soifer}, {Charmand
  aris}, {Marshall}, {Spoon}, {Lawrence}, {Hao}, {Higdon}, {Wu}, {Lacy},
  {Eisenhardt}, {Herter}, \& {Houck}}]{teplitz06}
{Teplitz}, H.~I., {Armus}, L., {Soifer}, B.~T., {et~al.} 2006, \apjl, 638, L1

\bibitem[{Thomas {et~al.}(2019)Thomas, Davé, Anglés-Alcázar, \&
  Jarvis}]{thomas19}
Thomas, N., Davé, R., Anglés-Alcázar, D., \& Jarvis, M. 2019, Monthly
  Notices of the Royal Astronomical Society, 487, 5764–5780

\bibitem[{{Tombesi} {et~al.}(2015){Tombesi}, {Mel{\'e}ndez}, {Veilleux},
  {Reeves}, {Gonz{\'a}lez-Alfonso}, \& {Reynolds}}]{tombesi15}
{Tombesi}, F., {Mel{\'e}ndez}, M., {Veilleux}, S., {et~al.} 2015, \nat, 519,
  436

\bibitem[{{Tommasin} {et~al.}(2010){Tommasin}, {Spinoglio}, {Malkan}, \&
  {Fazio}}]{tommasin10}
{Tommasin}, S., {Spinoglio}, L., {Malkan}, M.~A., \& {Fazio}, G. 2010, \apj,
  709, 1257

\bibitem[{{Tommasin} {et~al.}(2008){Tommasin}, {Spinoglio}, {Malkan}, {Smith},
  {Gonz{\'a}lez-Alfonso}, \& {Charmandaris}}]{tommasin08}
{Tommasin}, S., {Spinoglio}, L., {Malkan}, M.~A., {et~al.} 2008, \apj, 676, 836

\bibitem[{{Tozzi} {et~al.}(2006){Tozzi}, {Gilli}, {Mainieri}, {Norman},
  {Risaliti}, {Rosati}, {Bergeron}, {Borgani}, {Giacconi}, {Hasinger},
  {Nonino}, {Streblyanska}, {Szokoly}, {Wang}, \& {Zheng}}]{tozzi06}
{Tozzi}, P., {Gilli}, R., {Mainieri}, V., {et~al.} 2006, \aap, 451, 457

\bibitem[{{Treister} {et~al.}(2009){Treister}, {Urry}, \&
  {Virani}}]{treister09}
{Treister}, E., {Urry}, C.~M., \& {Virani}, S. 2009, \apj, 696, 110

\bibitem[{{Ueda} {et~al.}(2014){Ueda}, {Akiyama}, {Hasinger}, {Miyaji}, \&
  {Watson}}]{ueda14}
{Ueda}, Y., {Akiyama}, M., {Hasinger}, G., {Miyaji}, T., \& {Watson}, M.~G.
  2014, \apj, 786, 104

\bibitem[{{Vignali} {et~al.}(2011){Vignali}, {Piconcelli}, {Lanzuisi},
  {Feltre}, {Feruglio}, {Maiolino}, {Fiore}, {Fritz}, {La Parola}, {Mignoli},
  \& {Pozzi}}]{vignali11}
{Vignali}, C., {Piconcelli}, E., {Lanzuisi}, G., {et~al.} 2011, \mnras, 416,
  2068

\bibitem[{{Vignali} {et~al.}(2009){Vignali}, {Pozzi}, {Fritz}, {Comastri},
  {Gruppioni}, {Bellocchi}, {Fiore}, {Brusa}, {Maiolino}, {Mignoli}, {La
  Franca}, {Pozzetti}, {Zamorani}, \& {Merloni}}]{vignali09}
{Vignali}, C., {Pozzi}, F., {Fritz}, J., {et~al.} 2009, \mnras, 395, 2189

\bibitem[{{Vito} {et~al.}(2018){Vito}, {Brandt}, {Yang}, {Gilli}, {Luo},
  {Vignali}, {Xue}, {Comastri}, {Koekemoer}, {Lehmer}, {Liu}, {Paolillo},
  {Ranalli}, {Schneider}, {Shemmer}, {Volonteri}, \& {Wang}}]{vito18}
{Vito}, F., {Brandt}, W.~N., {Yang}, G., {et~al.} 2018, \mnras, 473, 2378

\bibitem[{{Vito} {et~al.}(2014){Vito}, {Gilli}, {Vignali}, {Comastri}, {Brusa},
  {Cappelluti}, \& {Iwasawa}}]{vito14}
{Vito}, F., {Gilli}, R., {Vignali}, C., {et~al.} 2014, \mnras, 445, 3557

\bibitem[{{Voit}(1992)}]{voit92}
{Voit}, G.~M. 1992, \mnras, 258, 841

\bibitem[{{Volonteri} {et~al.}(2016){Volonteri}, {Dubois}, {Pichon}, \&
  {Devriendt}}]{volonteri16}
{Volonteri}, M., {Dubois}, Y., {Pichon}, C., \& {Devriendt}, J. 2016, \mnras,
  460, 2979

\bibitem[{{Weedman} {et~al.}(2006){Weedman}, {Polletta}, {Lonsdale}, {Wilkes},
  {Siana}, {Houck}, {Surace}, {Shupe}, {Farrah}, \& {Smith}}]{weedman06}
{Weedman}, D., {Polletta}, M., {Lonsdale}, C.~J., {et~al.} 2006, \apj, 653, 101

\bibitem[{{Wu} {et~al.}(2009){Wu}, {Charmandaris}, {Huang}, {Spinoglio}, \&
  {Tommasin}}]{wu09}
{Wu}, Y., {Charmandaris}, V., {Huang}, J., {Spinoglio}, L., \& {Tommasin}, S.
  2009, \apj, 701, 658

\bibitem[{{Xue}(2017)}]{xue17}
{Xue}, Y.~Q. 2017, New Astronomy Review, 79, 59

\bibitem[{{Xue} {et~al.}(2011){Xue}, {Luo}, {Brandt}, {Bauer}, {Lehmer},
  {Broos}, {Schneider}, {Alexand er}, {Brusa}, {Comastri}, {Fabian}, {Gilli},
  {Hasinger}, {Hornschemeier}, {Koekemoer}, {Liu}, {Mainieri}, {Paolillo},
  {Rafferty}, {Rosati}, {Shemmer}, {Silverman}, {Smail}, {Tozzi}, \&
  {Vignali}}]{xue11}
{Xue}, Y.~Q., {Luo}, B., {Brandt}, W.~N., {et~al.} 2011, \apjs, 195, 10

\bibitem[{{Yan} {et~al.}(2007){Yan}, {Sajina}, {Fadda}, {Choi}, {Armus},
  {Helou}, {Teplitz}, {Frayer}, \& {Surace}}]{yan07}
{Yan}, L., {Sajina}, A., {Fadda}, D., {et~al.} 2007, \apj, 658, 778

\bibitem[{{Yang} {et~al.}(2020){Yang}, {Boquien}, {Buat}, {Burgarella},
  {Ciesla}, {Duras}, {Stalevski}, {Brandt}, \& {Papovich}}]{yang20}
{Yang}, G., {Boquien}, M., {Buat}, V., {et~al.} 2020, \mnras, 491, 740

\bibitem[{{Zappacosta} {et~al.}(2018){Zappacosta}, {Piconcelli}, {Duras},
  {Vignali}, {Valiante}, {Bianchi}, {Bongiorno}, {Fiore}, {Feruglio},
  {Lanzuisi}, {Maiolino}, {Mathur}, {Miniutti}, \& {Ricci}}]{zappacosta18}
{Zappacosta}, L., {Piconcelli}, E., {Duras}, F., {et~al.} 2018, \aap, 618, A28

\end{thebibliography}
\newpage
\begin{appendix}
\section{Affiliations}
\affil{$^1$Dipartimento di Fisica e Astronomia, Universit\`a degli Studi di Bologna, via P. Gobetti 93/2, 40129 Bologna, Italy}
\affil{$^2$INAF-OAS, Osservatorio di Astrofisica e Scienza dello Spazio di Bologna, via Gobetti 93/3, 40129 Bologna, Italy}
\affil{$^3$Instituto de F\'\i{}sica de Cantabria (CSIC-U. Cantabria), Avenida de los Castros, 39005 Santander, Spain}
\affil{$^4$Scuola Normale Superiore, Piazza dei Cavalieri 7,I-56126 Pisa, Italy}
\affil{$^{5}$Dipartimento di Fisica e Astronomia, Universit\`a di Firenze, via G. Sansone 1,50019 Sesto Fiorentino, Firenze, Italy}
\affil{$^{6}$INAF$-$Osservatorio Astrofisico di Arcetri, Largo Enrico Fermi 5, 50125 Firenze, Italy}
\affil{$^7$CEA, IRFU, DAp, AIM, Universit\'e Paris-Saclay, Universit\'e Paris Diderot, Sorbonne Paris Cit\'e, CNRS, F-91191 Gif-sur-Yvette, France}
\affil{$^8$INAF - Osservatorio Astronomico di Brera, via Brera 28, I-20121, Milano, Italy \& via Bianchi 46, I-23807, Merate, Italy}
\affil{$^9$School of Physics and Astronomy, Cardiff University, The Parade, Cardiff CF24 3AA, UK}
\affil{$^{10}$University of California, Irvine, Irvine, CA}
\affil{$^{11}$SISSA, Via Bonomea 265, 34136 Trieste, Italy}
\affil{$^{12}$Istituto di Astrofisica e Planetologia Spaziali - INAF, Rome, Via Fosso del Cavaliere 100, 00133, Roma, Italia}
\affil{$^{13}$Graduate School of Science, Nagoya University, Furo-cho, Chikusa-ku, Nagoya, Aichi 464-8602, Japan}
\affil{$^{14}$Institute of Liberal Arts and Sciences, Tokushima University, Minami Jousanjima-Machi 1-1, Tokushima, Tokushima 770-8502, Japan}
\affil{$^{15}$Centro de Astrobiolog\'ia (CSIC-INTA), Ctra. de Ajalvir, Km 4, 28850, Torrej\'on de Ardoz, Madrid, Spain}
\affil{$^{16}$Osservatorio Astronomico di Roma (INAF), Via Frascati 33, I-00040 Monte Porzio Catone (Roma), Italy}
\affil{$^{17}$N\'ucleo de Astronom\'ia de la Facultad de Ingenier\`ia, Universidad Diego Portales, Av. Ej\'ercito Libertador 441, Santiago, Chile}
\affil{$^{18}$Kavli Institute for Astronomy and Astrophysics, Peking University, Beijing 100871, People's Republic of China}
\affil{$^{19}$George Mason University, Department of Physics $\&$ Astronomy, MS 3F3, 4400 University Drive, Fairfax, VA 22030, USA}
\affil{$^{20}$Dipartimento di Fisica e Astronomia, Universit\'a di Padova, vicolo Osservatorio 3, 35122 Padova, Italy}
\affil{$^{21}$Department of Physics, University of Rome ``Tor Vergata'', Via della Ricerca Scientifica 1, I-00133 Rome, Italy}
\affil{$^{22}$Department of Astronomy, University of Maryland, College Park, MD 20742, USA}
\affil{$^{23}$Astrophysics Science Division, NASA Goddard Space Flight Center, Greenbelt, MD 20771, USA}

\end{appendix}

\end{document}